\documentclass[11pt,b4paper,superscriptaddress,floatfix]{article}
\usepackage{jheppub}
\usepackage{amsmath}
\usepackage{amssymb}

\usepackage{verbatim,graphicx}
\usepackage{mathtools}
\usepackage{color}
\usepackage[all]{xy}
\usepackage{simplewick}
\usepackage{hyperref}
\usepackage{setspace}
\usepackage{array}
\usepackage{tikz}
\usepackage{tkz-euclide}
\usepackage{amsthm}
\usepackage{enumitem} 
\usepackage{bbm}
\usepackage{bbold}
\usetikzlibrary{arrows}  
\usetikzlibrary{decorations.markings}

\newcounter{jvcc}

\newcounter{yjcc}

\newcommand{\one}{\mathbb{1}}
\newcommand{\Tr}{{\rm Tr}}

\newcommand{\eref}[1]{(\ref{#1})}
\newcommand{\nn}{\nonumber}

\newcommand{\be}{\begin{eqnarray}}
\newcommand{\ee}{\end{eqnarray}}
\newcommand{\vev}[1]{\langle #1\rangle}

\newcommand{\bmat}{\left ( \begin{array}{cc} }
\newcommand{\emat}{\end{array} \right ) }

\usetikzlibrary{shapes.misc}

\tikzset{cross/.style={cross out, draw=black, fill=none, minimum size=2*(#1-\pgflinewidth), inner sep=0pt, outer sep=0pt}, cross/.default={2pt}}

\def\Tr{\textrm{Tr}}

\usetikzlibrary{shapes.misc}

\newcommand{\beq}{\begin{equation}}
\newcommand{\beqs}{\begin{equation*}}
\newcommand{\eeq}{\end{equation}}
\newcommand{\eeqs}{\end{equation*}}

\allowdisplaybreaks

\begin{document}
\tikzset{%
          insert new path/.style={%
             insert path={%
                  node[midway,sloped]{\tikz \draw[#1,thick] (-.2pt,0) -- ++(.2 pt,0);}
                  }
             }
         }
\title{Chaos on the hypercube}
\author[a]{Yiyang Jia}
\author[a]{and Jacobus J. M. Verbaarschot}
\affiliation[a]{Department of Physics and Astronomy, Stony Brook University, Stony Brook, New York 11794, USA}
\emailAdd{yiyang.jia@stonybrook.edu}
\emailAdd{jacobus.verbaarschot@stonybrook.edu}

\abstract{We analyze the spectral properties of a $d$-dimensional HyperCubic (HC) lattice
  model originally introduced by Parisi. The U(1) gauge links of this model give rise
  to a magnetic flux  of constant magnitude $\phi$ but random orientation
  through the faces of the hypercube. The HC model,
which also can be written as a model of $2d$ interacting Majorana fermions,
has a spectral flow that is reminiscent of 
  Maldacena-Qi (MQ) model, and its spectrum at $\phi=0$, actually coincides with the
  coupling term of the MQ model.  As was already shown by Parisi, at leading order in $1/d$ , the spectral
  density of this model is given by the density function of the Q-Hermite
  polynomials, which is also the spectral density of the double-scaled Sachdev-Ye-Kitaev model. Parisi demonstrated this by mapping the moments of the HC model to Q-weighted sums on chord diagrams. We point out that the subleading moments of the HC model can also be mapped to weighted sums on chord diagrams, in a manner that descends from the leading moments.
  The HC model has a magnetic inversion symmetry that depends  on  both the magnitude and the
  orientation of the magnetic flux through the faces of the hypercube. The spectrum for fixed quantum number
  of this symmetry exhibits a transition from regular spectra at  $\phi=0$ to chaotic spectra with
  spectral statistics given by the Gaussian Unitary Ensembles (GUE) for larger values of $\phi$.
   For small magnetic flux, the ground state is gapped and is close to a  Thermofield Double (TFD) state.
}



\maketitle
\section{Introduction}

Many-body chaos has attracted a great deal of attention in recent years.
In particular, the study of the Sachdev-Kitaev-Ye (SYK) model 
\cite{sachdev1993,Kitaev2015} formerly known
as the two-body random ensemble \cite{mon1975}, has greatly improved our understanding
of the relationship between many-body chaos, 
disorder and spectral properties
of the underlying Hamiltonian (see \cite{brody1981,benet2003,Borgonovi:2016a,Borgonovi:2019mrk}
for reviews and recent work).
One of the main conclusions is that the
SYK model is a non-Fermi liquid with a many-body level density that increases
exponentially with the volume rather than a power of the volume for a
Fermi liquid.  A direct consequence is that the zero-temperature limit
of this model has an nonzero extensive entropy \cite{Georges_2001}. For the same reason, the SYK model can be
used to address questions related to understanding micro-states and entropy
of black holes \cite{sachdev2015}.

There are different ways to measure the chaotic properties of the SYK model.
The short-time behavior of the Out-of-Time-Order Correlator (OTOC), which in the
classical limit describes the exponential divergence of classical trajectories,
was shown \cite{Maldacena2016}
to saturate the chaos bound \cite{Maldacena2015}. This is also expected
to be the case for black holes, and was one of the main reasons for
the excitement for the SYK model. The paradigm of quantum chaos, though, is
that spectral correlations are given by Random Matrix Theory (RMT), which is
known as the Bohigas-Giannoni-Schmidt conjecture \cite{bohigas1984,Seligman1984}.
Indeed this was confirmed by numerical and analytical studies of
the SYK model \cite{You:2016ldz,Garcia-Garcia:2016mno,Cotler2016,Saad:2018bqo,Altland:2017eao,Jia:2019orl}.
One issue that has come forth in the study of the SYK model is to
what extent the disorder contributes to its chaotic properties. It has been
known for a long time \cite{Flores_2001} that level fluctuations at the
scale of many level spacings are dominated by fluctuations of the
width of the spectrum going from one disorder realization to the next. In the time domain, these fluctuations
\cite{Altland:2017eao,Garcia-Garcia:2018ruf,Saad:2018bqo,Gharibyan:2018jrp,Jia:2019orl}
give rise
to a peak at very short times
in the {\it connected}  spectral form factor. This
peak should not be confused with the peak due to the disconnected part
of the spectral form factor which is many orders of magnitude larger.
Fluctuations of other low-order moments also give significant contributions to the long-range spectral fluctuations. For an SYK  system of $N$ Majorana fermions,
the deviation from Random Matrix Theory are described by the covariance
matrix of the first $O(N)$ moments.\footnote{We do not claim certainty on the estimate $O(N)$, since it is inferred by observing limited numerics. In fact in \cite{Jia:2019orl} by the present authors, another estimate of $O(N\log N)$ was derived analytically, but that was also based on a crude estimate.} This gives an estimate of $2^{N/2}/N$
for the spectral range of RMT fluctuations or a time scale of $N 2^{-N/2}$
beyond which the spectral form factor is given by RMT.

It has been argued that the disorder is not important for the correlation
functions and thermodynamics of the SYK model \cite{Witten:2016iux} which also has been confirmed by melonic models which have similar properties in the absence of disorder
\cite{Klebanov:2016xxf,Klebanov:2019jup,Kim:2019upg,Krishnan:2017lra}. In this paper we study an
SYK-like model with much less disorder than the SYK model. This is the
hypercubic $U(1)$ lattice model in $d$ dimensions originally introduced by
Parisi \cite{Parisi:1994jg,Marinari:1995jwr}
as a model for an array of Josephson junctions.
This model has a magnetic flux of constant magnitude through each of the
faces of a $d$-dimensional hypercube, and only the sign of the flux through each face is random. 
In spite of the
$U(1)$ disorder on the links, the first six moments of the spectral density
do not depend on the disorder realization, and the scale fluctuations that limit
the agreement with random matrix theory are absent in this model. 
Experience with the 2+4-body SYK model shows \cite{Garcia-Garcia:2017bkg,Nosaka:2018iat,Nosaka:2019tcx} that although the two-body term is relevant, the model still remains chaotic, and also in the hypercubic
model we expect to find spectra correlated according to Random Matrix Theory.
We also note that Dirac spectral correlations of related gauge theories are described by Random Matrix Theory \cite{Halasz:1995vd,Gu:2019jub,Kieburg:2014eca,Kieburg:2017rrk}.

The original papers of Parisi and follow-ups \cite{Capelli:1997pm,Colomo:2001a,Colomo:2002rk} are mostly concerned with the thermodynamics of the hypercubic model, such as the free energy and heat capacity, which require the knowledge
of the average spectral density.   In this paper we are interested in the chaotic properties of Parisi's Hypercubic (HC) model, which require us to study the correlations among the energy levels. This in turn calls for a study of the symmetries of this model which leads to the discovery of a symmetry that was not known previously.
In the HC model, the magnitude of the flux (or equivalently the
Wilson loop) is parameterized by $\phi$. At $\phi =0$ the
Hamiltonian is given
by the adjacency matrix of the hypercube graph which is integrable and coincides with
the coupling term of the Maldacena-Qi model \cite{Maldacena:2018lmt}. The
spectral flow as a function of $\phi$ is also similar to that of the
Maldacena-Qi model, and exhibits an integrable-to-chaos transition. In addition, the hypercubic model has a previously
unknown discrete symmetry, which is a variant of the magnetic translation
symmetry \cite{Zak:1964,Rammal:1990,Wiegmann:1994js}, and is reminiscent of the discrete
symmetry of the Maldacena-Qi model. Understanding of the exact symmetries
is essential for a statistical analysis of the
spectral correlations of this model.

As was already noted by Parisi, the average
spectral density for large $d$ is well approximated by the Q-Hermite spectral density with $Q=\cos \phi$.
This also is the case for the double scaling limit with $q^2/N$ fixed ($Q=e^{-2q^2/N}$) for the
$q$-body SYK model of $N$ interacting Majorana fermions
\cite{Bagrets2016,Garcia-Garcia:2017pzl,Cotler2016,Jia:2018ccl,Berkooz:2018qkz,Berkooz:2018jqr}.
For $\phi > \pi/2$, $Q$ becomes negative
and spectrum splits into two bands, which also happens for the supercharge of
the supersymmetric SYK model \cite{Fu:2016vas,Garcia-Garcia:2018ruf,Kanazawa:2017dpd}.
The spectral fluctuations  of the HC model
from one realization to the next are quite different from those of the SYK model.
In the SYK model these fluctuations result from the covariance of
the first $O(N)$ moments, they decouple from the RMT fluctuations quite well,
and can be eliminated
\cite{Flores_2001,Altland:2017eao,Garcia-Garcia:2018ruf,Saad:2018bqo,Gharibyan:2018jrp,Jia:2019orl}.
For HC model, which can also formulated in terms
of gamma matrices in $2d$ dimensions, the fluctuations
due to the first six moments are absent, but higher moments  contribute significantly to the deviation from RMT level statistics. The scale of these fluctuations does not seem to separate well from the scale of the
RMT fluctuations.

The ground state of this  model has a gap that seems to remain in the thermodynamic limit for
$\phi < \pi/2$. Therefore the ground state entropy vanishes at zero temperature.
Since for zero flux the model coincides with the coupling Hamiltonian of
the Maldacena-Qi model, the ground state is also given by a ThermoField
Double (TFD) state. However, contrary to the Maldacena-Qi model, the overlap
with the TFD state  decreases considerably for nonzero magnetic flux.

This paper is organized as follows. In section \ref{sec:hc} we introduce
Parisi's hypercubic model which, as is explained in section \ref{sec:tensorRep},
can also be expressed as a sum of
tensor products of Pauli matrices. The novel discrete symmetry of this model
is discussed in section \ref{sec:symmetries}. In section \ref{sec:sumRules}
we show that the first six moments of this model do not depend on the
disorder realizations. Numerical results for the spectral density and
spectral correlations are presented in section \ref{sec:levelStat}.
Both the number variance and the spectral form factor are compared to
random matrix results. The ground state wave function is compared to
the TFD state in section \ref{sec:tfd} and concluding remarks are made
in section \ref{sec:conclusions}. Several technical results are worked
  out in two appendices. In
appendix
  \ref{app:m4m6Tensor} we calculate the fourth
  and sixth moments of the Hamiltonian in a tensor product representation,
  respectively. The connection between chord diagrams and the leading large $d$ moments of the Hamiltonian  is explained in appendix \ref{app:chords}, where we also demonstrate how subleading moments arise from chord diagram considerations.

\section{Parisi's hypercubic model}
\label{sec:hc}

Parisi \cite{Parisi:1994jg} studied a disordered $U(1)$ lattice gauge model
on a $d$-dimensional Euclidean hypercube. The lattice sites of this model are represented by
$d$-dimensional vectors $\vec x$ with components $x_\mu\in \{0,1\}$. The model considers a constant magnetic field such that the fluxes through all faces of the hypercube have the same magnitude $\phi$, but with random orientations. That is, we have the field strength tensor 
\begin{equation}
F_{\mu\nu} = \phi S_{\mu\nu},
\end{equation}
where $S_{\mu\nu}$ is an antisymmetric tensor with random entries $\pm 1$ with equal probabilities. Hence we are dealing with a finite ensemble with  $2^{\binom{d}{2}}$ disorder realizations. We can work in the axial gauge so that the link variables are given by
\begin{equation}
U_\mu (\vec x) = e^{i\phi \sum_{\nu=1}^{\mu-1} S_{\mu \nu}x_\nu },
\end{equation}
which is the phase we associate with the link emanating from site $\vec x$ along the $\mu$-th direction. Note the sum is over all the $\nu$'s with $\nu<\mu$, and if $\mu=1$ we define $U_1(\vec x)=1$. We wish to study a  Hamiltonian describing a particle hopping on the lattice sites through the lattice links, and picking up a phase of the corresponding link variable. In terms of matrix elements, the Hamiltonian $H$ has the form
\begin{equation}\label{Hdef}
H_{\vec x, \vec y} = \sum_{\mu}\left[\delta_{\vec x +\hat e^\mu, \vec y}\  U_\mu (\vec x) + \delta_{\vec x -\hat e^\mu, \vec y}\ U^*_\mu (\vec x) \right],
\end{equation} 
where $\hat e^\mu$ is the unit basis vector in the $\mu$-th direction. When $\phi=0$, this Hamiltonian becomes the adjacency matrix of the hypercube as a graph. We remark that Parisi was originally interested in the second quantized Hamiltonian 
\begin{equation}
\sum_{\vec x, \vec y}\varphi_{\vec{x}} \ H_{\vec x, \vec y}\ \varphi_{\vec{y}},
\end{equation}
where $\varphi$ is a scalar quantum field. However, in this paper we take a first quantized view and concern ourselves with the $H$ defined in equation \eqref{Hdef}, and the wave functions live in $\mathbb{C}^{2^d}$.  

Let us be very explicit on how to write the Hamiltonian matrix as a two-dimensional array of numbers: since $\vec x =(x_1, x_2,\ldots,x_d)$ is a string of $0$ and $1$'s of length $d$, we can naturally think of $\vec x$ as the binary representation of some integer between $0$ and $2^d-1$. Shifting this correspondence by one, we can represent any integer $m\in \{1,2,3, \ldots, 2^d\}$ through the relation 
\begin{equation}\label{eqn:base2To10}
[m-1]_2 = \overline{x_dx_{d-1}\ldots x_1},
\end{equation}
where $[m-1]_2$ denotes the number $m-1$ in the binary representation, and $\overline{x_dx_{d-1}\ldots x_1}$ denotes $x_dx_{d-1}\ldots x_1$ as a string of digits. We will use $m$ to index the matrix entries. Note we use the reverse order of $(x_1, x_2,\ldots,x_d) $ to represent binary digits because we wish the contributions from lower dimensions to appear as the upper-left of the matrix. For example, with these conventions we have 
\begin{equation}\label{eqn:Hexamples}
\begin{split}
H_{d=1} &= \begin{pmatrix} 
0&\ 1\\
1&\ 0
\end{pmatrix},\\
H_{d =2}&=  \begin{pmatrix} 
0&1&1&0\\
1&0&0&\ e^{i\phi S_{21}}\\
1&0&0&1\\
0&\ e^{-i\phi S_{21}}&1&0
\end{pmatrix},
\end{split}
\end{equation}
and so on.

The Hamiltonian can be obtained  recursively:
\be
H_{d} = \left ( \begin{array}{cc} H_{d-1} &  C_{d-1}\\
  (C_{d-1})^{-1} & H_{d-1} \end{array} \right ),
  \ee
 where $C_{d-1}$ is a diagonal unitary matrix with entries
  \begin{equation}
  (C_{d-1})_{k,k}= e^{i\phi \sum_{\nu=1}^{d-1} S_{d \nu}x_\nu(k)},
  \end{equation}
 where $x_\nu (k)$ is the $\nu$-th digit of $[k-1]_2$, as defined in equation \eqref{eqn:base2To10}.
We can verify that the following relation holds:
  \be
  \left( C_{d-1}\right)_{k,k}
  \left( C_{d-1} \right)_{2^{d-1}+1-k,\; 2^{d-1}+1-k}=e^{i \phi\sum_{\nu=1}^{d-1}S_{d\nu} },
\label{hrel}
  \ee
which will be useful for section \ref{sec:symmetries}.  For later convenience, we also introduce  the notation
  \be\label{eqn:Sdef}
  S_{\rho} :=\sum_{\nu=1}^{\rho-1} S_{\rho\nu},\quad S_1:= 0,
  \ee
 so that the right-hand side of equation \eqref{hrel} is simply $e^{i\phi S_d}$.

\section{Tensor product representation of the Hamiltonian}\label{sec:tensorRep}

Since the interaction between two lattice sites can be written in terms of the Pauli matrix $\sigma_1$, it is not surprising that the Hamiltonian can
be expressed in terms of tensor products of Pauli $\sigma$-like matrices. For $d=2$
it is clear from equation \eqref{eqn:Hexamples} that
\be
H_2 &=& \sigma_0 \otimes \sigma_1 + \sigma_{2,(0,0)}\otimes \rho_0
+ \sigma_{2,(1,0)}\otimes \rho_1 \nn\\
\ee
with 
\be
\label{rhodef}
\rho_0& = &\left ( \begin {array}{cc} 1& 0 \\  0&0\end{array}\right ),\qquad
   \rho_1 = \left (\begin{array}{cc} 0\ & 0\  \\0&1 \end{array}\right ),\qquad\sigma_0 = \left ( \begin{array}{cc} 1 & 0
\\ 0 & 1 \end{array} \right ),\nn\\
    \sigma_1 &= &\left ( \begin{array}{cc} 0 & 1
\\ 1 & 0 \end{array} \right ), \qquad \sigma_{d,\vec x}=\left ( \begin{array}{cc} 0 & e^{i\phi \sum_{\nu =1}^{d-1} x_\nu S_{d\nu}}
\\e^{-i\phi \sum_{\nu =1}^{d-1} x_\nu S_{d\nu}}& 0 \end{array} \right ),
\ee
where $\vec x=(x_1,\ldots,x_d)$. Notice that the definition of $\sigma_{d,\vec x}$ does not depend on the last component $x_d$ of $\vec x$, for example we have 
\begin{equation}
\sigma_{2,(0,0)}=\sigma_{2,(0,1)}=\begin{pmatrix}
0&1\\1&0
\end{pmatrix}=\sigma_1.
\end{equation}
For higher dimensions we have
\be
H_3 &=& \sigma_0 \otimes H_2 + \sum_{x_1,x_2} \sigma_{3,\vec x}\otimes \rho_{x_1} \otimes \rho_{x_2},\nn\\
H_4 &=& \sigma_0 \otimes H_3 + \sum_{x_1,x_2,x_3} \sigma_{4,\vec x}
\otimes \rho_{x_1} \otimes \rho_{x_2}\otimes \rho_{x_3},
\label{ht}
\ee
and in general we have
\be\label{eqn:tensorHami}
H_{d} = \sigma_0 \otimes H_{d-1} + \sum_{x_1,x_2,\cdots, x_{d-1}}\sigma_{d,\vec x}\otimes
\rho_{x_1}\otimes\rho_{x_2} \otimes \cdots \otimes \rho_{x_{d-1}},
\ee
where 
\begin{equation}
\sum_{x_1,x_2,\cdots, x_{d-1}} :=\sum_{x_1=0}^1 \sum_{x_2=0}^1\cdots\sum_{x_{d-1}=0}^1.
\end{equation}
\subsection{The Hamiltonian as a system of interacting Majorana fermions}

Since the Hamiltonian is a sum of tensor products of Pauli-like matrices, it is
natural to express the Hamiltonian as a sum of products of $\gamma$ matrices,
which then  can be interpreted as the Hamiltonian for a system of $2d$
Majorana fermions. The simplest case is  $\phi=0$. Then the Hamiltonian is just the adjacency matrix of a hypercube graph. In the tensor product representation it is given by
\be
H_d(\phi=0) =\sigma_0 \otimes H_{d-1}(\phi=0) +\sigma_1\otimes\overbrace{\sigma_0\otimes \cdots \otimes
  \sigma_0}^{d-1}, \quad  H_1(\phi=0) =\sigma_1.
\ee
If we introduce the gamma matrices
\be
\gamma_k^L &=& \overbrace{\sigma_1 \otimes\cdots\otimes\sigma_1}^{k-1}\otimes \sigma_3 \otimes \overbrace{\sigma_0 \otimes\cdots\otimes\sigma_0}^{d-k}
,\nn\\
\gamma_k^R &=& \overbrace{\sigma_1 \otimes\cdots\otimes\sigma_1}^{k-1}\otimes \sigma_2\otimes \overbrace{\sigma_0 \otimes\cdots\otimes\sigma_0}^{d-k}
,
\label{gamma-hc}
\ee
then the Hamiltonian can be written as
\be
H_d(\phi=0) = i\sum_{k=1}^{d} \gamma_k^L \gamma_k^R.
\ee
This is exactly the interaction term in the Maldacena-Qi model \cite{Maldacena:2018lmt}. This interaction term was shown \cite{Garcia-Garcia:2019poj} to have the spectrum 
\begin{equation}\label{eqn:zerophiSpectrum}
-d +2i, \ i=0,1,\ldots,d,
\end{equation}
with degeneracies 
\begin{equation}
\binom{d}{i},\ i=0,1,\ldots,d.
\end{equation}
Indeed this is also the well-known spectrum for the hypercube adjacency matrix. At $\phi\ne 0$ most other terms contributing to the Hamiltonian couple
the L and R spaces, which makes this model quite different from the
Maldacena-Qi model. In addition, interaction terms among any number of
$\gamma$ matrices appear in the Hamiltonian, which make the Hamiltonian look very complicated in a Majorana fermion representation.

\section{Symmetries}\label{sec:symmetries}
\subsection{Sublattice symmetry}\label{sec:sublattice}
Since the hypercube is a bi-partite lattice, the lattice links only connect one sublattice to the other, we conclude that the Hamiltonian \eqref{Hdef} has a sublattice symmetry
\be
\{\Gamma_5, H\} =0.
\ee
so that all eigenvalues appear in pairs $\pm \lambda_k$. In the tensor product representation described in section \ref{sec:tensorRep}, $\Gamma_5$ has the simple form of a tensor product of $\sigma_3$'s. Since each term contributing to $H_d$ in equation \eqref{eqn:tensorHami} contains exactly one off-diagonal $\sigma$ matrix
in the tensor product, we have
\be
\{\overbrace{\sigma_3\otimes \cdots \otimes \sigma_3}^d,H_d\} =0,
\ee
which proves the sublattice symmetry of the Hamiltonian.

\subsection{Magnetic inversion symmetry}
Since the field strength is constant in space and is a two-form, it is invariant under inversion
 \begin{equation}\label{eqn:spatialInversion}
 \vec x \to \vec{x}^{\,c}: =(1-x_1,1-x_2,\ldots,1-x_d).
 \end{equation} 
We choose the inverted coordinates to be $1-x_k$ instead of $-x_k$ so that the hypercube remains invariant too. Therefore, we expect a symmetry $A_d$ of the system acting on wave functions as
\begin{equation}
A_{d} \psi(\vec{x}) = \Omega_d(\vec x)\psi\left(\vec x^{\,c}\right),
\end{equation}
and we seek a position-dependent phase factor $\Omega_d(\vec{x})$ so that $[A_d,H_d]=0$. Its global phase is still ambiguous, which can be fixed by requiring $A_d^2=\one_{2^d\times 2^d}$ as a phase convention. We claim that the following choice  does the job:
    \begin{equation}\label{eqn:magneticPhase}
    \Omega_d(x)=  \exp\left(i\frac{\phi}{2}\sum_{\rho=2}^{d}S_\rho\right)\exp\left({i\phi \sum_{\nu>\mu} S_{\mu\nu}x_\nu}\right),
\end{equation}
where $S_\rho$ is defined as in equation \eqref{eqn:Sdef} and 
\begin{equation}
\sum_{\nu>\mu} := \sum_{\nu=2}^d\sum_{\mu=1}^{\nu-1}.
\end{equation}
Now equations \eqref{eqn:spatialInversion}--\eqref{eqn:magneticPhase} fix $A_d$ unambiguously.
We can write $A_d$  explicitly as a matrix through the recursion relation:
   \be
  A_{d} =
   \left ( \begin{array}{cc} 0 & e^{\frac i2\phi S_d} A_{d-1}\\
     e^{-\frac i2 \phi S_d}A_{d-1}&0 \end{array}
    \right ), \qquad A_1 =\sigma_1.
    \label{symRecursive}
\ee
 Note that  $A_d$ is a Hermitian anti-diagonal matrix. By induction we
    easily check that indeed $A_d^2=\one$ so that its eigenvalues can only be $\pm 1$. We will call $A_d$ the \textit{magnetic inversion}, because the operator implements a spatial inversion and is a function of the magnetic field. Let us remark that although we only wanted to implement an inversion in space, since we wrote down the Hamiltonian in a specific gauge (in our case the axial gauge along $x_1$ direction), spatial transformations may not always respect the gauge condition. The position-dependent phase factor is the price to pay to stay in the same gauge. This is analogous to the more familiar case of magnetic translations. In fact $A_d$ can be viewed as a magnetic translation if we view the inversion \eqref{eqn:spatialInversion} as a translation mod 2 along the longest body diagonal of the hypercube:
\begin{equation}
1-x_k = (1+x_k) \text{ mod 2}, \text{ for all }k=1,2,\ldots,d.
\end{equation}
Then the position-dependent phase factor in equation \eqref{eqn:magneticPhase} is exactly the standard phase factor for the corresponding magnetic translation \cite{Zak:1964,Rammal:1990,Wiegmann:1994js,Sekiguchi:2008kw}.

    We now prove $A_d$ is indeed a symmetry by induction. The commutator
$ [A_{d}, H_{d}]$ is given by
\be
   \begin{pmatrix}
     e^{\frac i2\phi S_{d}} A_{d-1} (C_{d-1})^{-1}-e^{-\frac i2 \phi S_{d}}C_{d-1}   A_{d-1} &
     e^{\frac i2 \phi S_{d}}[ A_{d-1}, H_{d-1}]  \\
     e^{-\frac i2\phi  S_{d}}[ A_{d-1}, H_{d-1}]  & e^{-\frac i2 \phi S_{d}} A_{d-1}  C_{d-1}
- e^{\frac i2 \phi S_{d}}  (C_{d-1})^{-1} A_{d-1}
   \end{pmatrix}. \quad
     \ee
     By induction hypothesis $[A_{d-1},H_{d-1}] =0$ which is satisfied for $d=2$ because $H_1=A_1=\sigma_1$, so we only have to worry about the diagonal blocks.
    We remind the readers that $C_{d-1}$ is diagonal and $A_{d-1}$ is anti-diagonal, so their product is anti-diagonal. So let us look at the only matrix elements that are possibly nonzero:
\be
\begin{split}\nn
  &\left[e^{\frac i2\phi S_{d}} A_{d-1} (C_{d-1})^{-1}\right]_{k,2^{d-1}+1-k}-\left[e^{-\frac i2 \phi S_{d}}C_{d-1}   A_{d-1}\right]_{k,2^{d-1}+1-k}\\
  =&e^{\frac i2\phi S_{d}} \left[ A_{d-1}\right]_{k,2^{d-1}+1-k} \left[C_{d-1}\right]_{2^{d-1}+1-k,2^{d-1}+1-k}^{-1}-e^{-\frac i2 \phi S_{d}}\left[C_{d-1}\right]_{kk}  \left[ A_{d-1}\right]_{k,2^{d-1}+1-k}\\
  =& e^{-\frac i2 \phi S_{d}}\left[ A_{d-1}\right]_{k,2^{d-1}+1-k} \left[C_{d-1}\right]_{2^{d-1}+1-k,2^{d-1}+1-k}^{-1}\left(e^{i \phi S_{d}}-\left[C_{d-1}\right]_{kk} \left[C_{d-1}\right]_{2^{d-1}+1-k,2^{d-1}+1-k}\right)\\
 =& 0,
  \end{split}
\ee
where for the last equality we used equation \eqref{hrel} and this completes the proof.

Since the symmetry operator is an anti-diagonal matrix, an orthogonal  set
of eigenvectors is given by $(0,\cdots, 0,b_k,0,\dots,0,\pm b_{2^{d}+1-k},0,\cdots,0)$,
where the $b_k$ are the anti-diagonal matrix elements. For the symmetry operator
$A_d$ we have that $b_{2^d+1-k}=b_k^*$. These eigenvectors can be used to
construct the unitary matrix that brings the Hamiltonian into a block-diagonal form where the two blocks
correspond to the $\pm 1$ eigenvalues of $A_d$. For our numerical results
to be discussed below, we block-diagonalize the Hamiltonian this way.

We can also discuss the magnetic inversion symmetry in the tensor product representation. If we define the unitary  Hermitian $2\times 2$ matrix
\be
K_d := \bmat 0 & e^{i\frac \phi 2 S_d}
  \\  e^{-i\frac \phi 2 S_d} &0\emat, \quad K_1 :=\sigma_1=\begin{pmatrix}
  0&1\\1&0
  \end{pmatrix},
    \ee
then it is clear from equation \eqref{symRecursive} that the magnetic inversion can be written as
\begin{equation}\label{eqn:magTransTensorRep}
A_d = K_d\otimes K_{d-1}\otimes\cdots\otimes K_1.
\end{equation}
We can check that
\be
K_d \sigma_{d,\vec{x}} K_d^{-1}&=& \sigma_{d,\vec x^c},\nn\\
K_d \rho_{x_k} K_d^{-1}&=& \rho_{x_k^c},
\ee
where $x_k^c = 1-x_k$ and $\vec x^{\,c} =(x_1^c,x_2^c \ldots,x_d^c)$, as defined in equation \eqref{eqn:spatialInversion}. Now we can prove $A_d H_d A_d^{-1}=H_d$ from induction again: the inductive hypothesis takes care of the first term on the right-hand side of equation \eqref{eqn:tensorHami}, and the second term becomes 
\begin{equation}
\sum_{x_1,x_2,\cdots, x_{d-1}}\sigma_{d,\vec x^c}\otimes
\rho_{x_1^c}\otimes\rho_{x_2^c} \otimes \cdots \otimes \rho_{x_{d-1}^c},
\end{equation} 
but we can freely re-index the summation as 
\begin{equation}
\sum_{x_1^c,x_2^c,\cdots, x_{d-1}^c}\sigma_{d,\vec x^c}\otimes
\rho_{x_1^c}\otimes\rho_{x_2^c} \otimes \cdots \otimes \rho_{x_{d-1}^c},
\end{equation} 
because both $x_k$ and $x_k^c$ sum over the same range, namely $\{0, 1\}$. Now it is clear that this term is indeed invariant by a simple change of dummy variables.

We end this section by noting the peculiarity of the situation: through its dependence on $S_{\mu\nu}$, the magnetic inversion symmetry $A_d$ depends on the disorder realization of the ensemble, hence the symmetry itself is disordered. This is exceptional in that the symmetries of most disordered systems do not depend on disorder realizations. However, the effects of this disordered symmetry are as real as the conventional cases. In particular, to study the level statistics we must focus on one block of the $H_d$ that is irreducible under $A_d$.

\section{Sum rules for the Hamiltonian}\label{sec:sumRules}

There are exact sum rules for the Hamiltonian that are valid even without taking the disorder average. They will account for some salient features of the level statistics we are going to see in section \ref{sec:levelStat}. The sum rules are consequences of the hypercubic geometry and the fact that the Hamiltonian \eqref{Hdef} has only nearest neighbor hoppings. The sum rules can  be evaluated in the tensor product representation \eqref{eqn:tensorHami} as well. Since this calculation for $\Tr H^4$ and $\Tr H^6$ in tensor product representation is rather
lengthy, we have moved it to  appendix \ref{app:m4m6Tensor}.

\subsection{$\Tr AH^p$}
In the study of level statistics, we analyze the energy eigenvalues in the same block under magnetic inversion symmetry $A$. So instead of the total moments $\Tr H^p$, what we really should be interested in is $\frac{1}{2}\Tr \left((\one + A)H^p\right)$. However in this section we will see 
\begin{equation}
\Tr A_dH_d^p=0 \quad\text{ for} \ p<d.
\end{equation}
So for low moments we might as well just study $\Tr H^p$.
\subsubsection*{Geometric picture}
The magnetic inversion $A_d$ has the physical meaning of particle hopping from one lattice site to the site sitting on the corresponding longest diagonal. The $H^p$ involves $p$-step hoppings connecting nearest neighbors. For the trace to be nonzero, we must have at least one hopping configuration that forms a loop. This means some of the $p$-step nearest-neighbor hops must reach  the longest diagonal to form a loop with the $A_d$ hopping. This is clearly impossible for $p<d$.
\subsubsection*{Tensor product picture}
Since $A_d$ is the tensor product of $d$ off-diagonal Pauli matrices (see equation \eqref{eqn:magTransTensorRep}),
while each factor contributing the Hamiltonian \eqref{eqn:tensorHami} contains only one off-diagonal Pauli matrix.  It is clear that $\Tr A_dH_d^p=0 $ for $p<d$,  because for $p<d$ every term in $H^p$ will have at least one diagonal $2\times 2$ matrix in the tensor product.
\subsection{$\Tr H^2$}

We wish to prove
\be
2^{-d} \Tr H_d^2 = d.
  \label{eqn:HsquareSum-rule}
  \ee
In fact, we will prove a stronger identity for the diagonal entries of $H^2$:
\begin{equation}  \label{eqn:HsquareDiagonals}
\left(H_d^2\right)_{ii} =d.
\end{equation}
 \subsubsection*{Geometric picture}
We note the diagonal elements $(H^2)_{ii}$ only receive contributions from 2-step loops. But a 2-step loop must be one step through some lattice link followed by one step back through the same link, and hence the phases cancel. We can choose the first step to be along any direction, thus in $d$ dimensions we have $d$ contributions, each being $1$. This gives \eqref{eqn:HsquareDiagonals}.

 \subsubsection*{Tensor product picture}
The tensor products in the Hamiltonian \eqref{eqn:tensorHami} involve both diagonal and off-diagonal two-by-two matrices. To contribute to $(H_d^2)_{ii}$, terms with the off-diagonal
Pauli matrices must be in the same position in both factors of $H$, so 
\be
(H_d^2)_{ii} = \left(\sigma_0 \otimes H_{d-1}\right)^2_{ii} + \left(\sum_{x_1,x_2,\cdots, x_{d-1}}\sigma_{d,\vec x}\otimes
\rho_{x_1}\otimes\rho_{x_2} \otimes \cdots \otimes \rho_{x_{d-1}}\right)^2_{ii},
\ee
and inside the second term, we have terms 
\be
\left(\sigma_{d,\vec x} \otimes \rho_{x_1} \otimes\cdots \otimes \rho_{x_{d-1}}\right)
\cdot \left(\sigma_{d,\vec x'} \otimes \rho_{x_1'} \otimes\cdots \otimes \rho_{x_{d-1}'}\right),
\label{terms-h2}
\ee 
which are only nonzero if $x_k = x_k'$ for all $1\leq k\leq d-1$. It is not hard to see then
the sum over $x_1,\ldots,x_{d-1}$ results 
  in a tensor product of $d$ identity matrices. The same argument applies
  to $ \left(\sigma_0 \otimes H_{d-1}\right)$ through the recursive definition of $H_{d-1}$. We can do this recursively all the way to $H_1$ in $d-1$ steps, and each step creates an identity matrix, so 
  \begin{equation}
  (H_d^2)_{ii} = \left(\sigma_0 \otimes\sigma_0\otimes\cdots\otimes H_1\right)^2_{ii} +(d-1) = d.
  \end{equation}

  \subsection{$H^2$ at $\pi$ flux}
  We just demonstrated that the diagonal entries of $(H_d^2)_{ii} =d$. We shall further show that at $\pi$ flux, 
  \begin{equation}\label{eqn:piFluxHsquare}
  H_d^2(\phi=\pi) = d \one.
\end{equation}   
Together with the sublattice symmetry described in section \ref{sec:sublattice}, this implies at $\pi$ flux $H_d$ has exactly half of its eigenvalues being $-\sqrt{d}$ and the other half being $\sqrt{d}$.
\subsubsection*{Geometric picture}
\begin{figure}
\begin{center}
\begin{tikzpicture}
\draw[fill=black] (0,0) circle (1.5pt);
\draw[fill=black] (0,2) circle (1.5pt);
\draw[fill=black] (2,0) circle (1.5pt);
\draw[fill=black] (2,2) circle (1.5pt);
\node at (-0.25,-0.25) {$i$};
\node at (2.25,2.25) {$j$};
\node at (1,1) {$\pi$};
\begin{scope}[thick,decoration={
    markings,
    mark=at position 0.5 with {\arrow{>}}}
    ] 
    \draw[postaction={decorate}] (0,0)--(2,0);
    \draw[postaction={decorate}] (2,0)--(2,2);
    \draw[postaction={decorate}] (0,2)--(2,2);
    \draw[postaction={decorate}] (0,0)--(0,2);
\end{scope}
\end{tikzpicture}
\end{center}
\caption{Two lattice paths that connect sites $i$ and $j$ with a $\pi$ flux.}\label{fig:PiFluxSquare}
\end{figure}
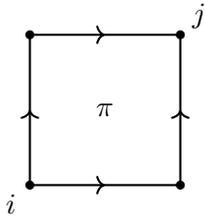

We only need to show 
\begin{equation}
\left(H_d^2\right)_{ij} = 0 \quad \text{if $i\neq j$.}
\end{equation}
Note $\left(H_d^2\right)_{ij} $ receives contributions from 2-step lattice paths that connect lattice sites $i$ and $j$.\footnote{Perhaps it is more precise to say ``sites represented by $i$ and $j$'', namely sites whose coordinates are $\vec{x}(i)$ and $\vec{x}(j)$, whose components are the binary digits of $i-1$ and $j-1$ in reverse order.} There are two scenarios for $i\neq j$:
\begin{enumerate}
\item There is no 2-step path from $i$ to $j$. For such pairs of $ij$ clearly $(H^2_d)_{ij}=0$.
\item Sites $i$ and $j$ can be connected by a 2-step path. If so then sites $i$ and $j$ must be siting on the diagonal of a face of the hypercube and there are exactly two paths connecting them, which form the four sides of the face, see figure \ref{fig:PiFluxSquare}. If the direction of one of the two paths is reversed, we will have a Wilson loop of flux $\pi$, and this means the two original paths give contributions that differ by a factor of $e^{i\pi}=-1$, so their sum vanishes. 
\end{enumerate} 
Hence equation \eqref{eqn:piFluxHsquare} is proven.

\subsubsection*{Tensor product picture}

We have seen in the last section that the diagonal entries of $H^2$ come from individual terms squared. Now we need to show the cross terms cancel out for $\phi=\pi$.  One such pair of  cross terms is an anticommutator
\begin{equation}
\left\{\sum_{x_1',\cdots, x_{d-2}'}\sigma_0 \otimes\sigma_{d-1,\vec x'}\otimes
\rho_{x_1'}\otimes \cdots \otimes \rho_{x_{d-2}'},\sum_{x_1,\cdots, x_{d-1}}\sigma_{d,\vec x}\otimes
\rho_{x_1}\otimes\rho_{x_2} \otimes \cdots \otimes \rho_{x_{d-1}}\right\}
\end{equation}
where the first factor is part of $\sigma_0\otimes H_{d-1}$ and the second factor is the second term in equation  \eqref{eqn:tensorHami}. Since $\rho_0\rho_1=0$, the product is only nonzero when
\begin{equation}
x_1'=x_2,\ x_2'=x_3, \ldots, x_{d-2}'=x_{d-1}.
\end{equation} 
So the sum reduces to\footnote{We remind the readers that $\sigma_{d-1,(x_2,x_3,\ldots,x_{d-1},x_{d-1}')}$ does not depend on the last coordinate $x_{d-1}'$.}

\begin{align}
&\sum_{x_1,x_2,\cdots, x_{d-1}}\sigma_{d,\vec x}\otimes
\left\{\sigma_{d-1,(x_2,x_3,\ldots,x_{d-1},x_{d-1}')},\rho_{x_1}\right\}\otimes\rho_{x_2} \otimes \cdots \otimes \rho_{x_{d-1}}\nn\\
=& \sum_{x_1,x_2,\cdots, x_{d-1}}\sigma_{d,\vec x}\otimes
\sigma_{d-1,(x_2,x_3,\ldots,x_{d-1},x_{d-1}')}\otimes\rho_{x_2} \otimes \cdots \otimes \rho_{x_{d-1}},\label{eqn:intermediateSum}
\end{align}
where we have used 
\begin{align}
\sigma_{d-1,(x_2,x_3,\ldots,x_{d-1},x_{d-1}')}\rho_{x_1}&= \rho_{x_1^c}\sigma_{d-1,(x_2,x_3,\ldots,x_{d-1},x_{d-1}')},\\
\rho_{x_1}+\rho_{x_1^c} &= \one_{2\times 2}.
\end{align}
Note only the first tensor factor in \eqref{eqn:intermediateSum} depends on $x_1$. Now we see the sum over $x_1$ already gives zero because 
\begin{equation}
\sigma_{d,(x_1,x_2,\ldots,x_d)} = -\sigma_{d,(x_1^c,x_2,\ldots,x_d)} \quad \text{when }\phi=\pi. 
\end{equation}
  The same argument can be applied to
  all other mixed terms. For $\phi = \pi$ we thus demonstrated $ H_d^2 = d\one$.

\subsection{$\Tr H^4$}\label{sec:fourthMomIndependence}

We wish to prove $\Tr H^4$ does not depend on disorder realizations of $S_{\mu\nu}$.
\subsubsection*{Geometric picture}
We need to consider all the 4-step loops on the hypercube. If the path is backtracking then the loop has zero area, so quite trivially they do not depend on flux realizations. The only other possibility for a 4-step loop is a one that travels the four sides of a hypercube face, and its contribution to the trace is its Wilson loop value $e^{i\phi S_{\mu\nu}}$. However, since each clockwise loop is accompanied by its counterclockwise counterpart, the contributions must be functions of $\cos \left(\phi S_{\mu\nu}\right) =\cos \phi$.\footnote{The crucial point is that a 4-step loop can at most loop around one face of the hypercube. For larger loops when several faces can be looped around, we generically have $\cos\left[\phi(S_{\mu_1\nu_1}+S_{\mu_2\nu_2}+\cdots)\right]$.} 
We see in both cases the contributions do not depend on the disorder realization of  $S_{\mu\nu}$.
\subsubsection*{Tensor product picture}
The fourth moment can also be worked out in the tensor representation of the Hamiltonian,
see appendix \ref{app:m4Tensor}. This allows us to obtain the exact result for the fourth
moment which is in agreement with that obtained in \cite{Parisi:1994jg}.

\subsection{$\Tr H^6$}

In this section we prove that $\Tr H^6$ does not depend on disorder realizations of $S_{\mu\nu}$. 
                        \subsubsection*{Geometric picture}
 A six-step loop can at most traverse three different dimensions.   Let us first think about $d=3$. As Parisi argued \cite{Parisi:1994jg}, in three dimensions the field strength tensor $\phi S_{\mu\nu}$ can be viewed as a vector, pointing along one of the longest diagonals of the 3-cube. Hence all possible realizations of the flux are related to each other by a spatial rotation in the cubic symmetry group, which implies their Hamiltonians all have the same spectrum independent of $S_{\mu\nu}$. The loops that contribute to $\Tr H^6$ can traverse one, two or three different dimensions. Those that traverse one and two dimensions are independent of $S_{\mu\nu}$ for reasons discussed in section \ref{sec:fourthMomIndependence}. This implies that for $d=3$ in particular, the sum of all Wilson loops that traverse three different dimensions is also independent of realizations of $S_{\mu\nu}$. Now let us consider general $d$. Since every three different dimensions uniquely define a 3-cube, it is evident that all loops that traverse three different dimensions can be partitioned into groups by the 3-cubes they reside in. By the argument just laid out, the sum of each group of such loops is independent of $S_{\mu\nu}$, and hence the total sum retains the  independence. It is important to separate the contributions of the loops that traverse three different dimensions from the rest for this argument to work, because  a loop that traverses one or two dimensions can reside in multiple 3-cubes.

\subsubsection*{Tensor product picture}
 For the calculation using the tensor representation we also have to distinguish several
 cases. Although the calculation is straightforward, the preponderance of indices makes
 this calculation rather cumbersome, and we have moved it to \ref{app:m6Tensor}.
 This calculation shows that the disorder independence of $\Tr H^6$ arises 
 because we have just enough terms in the expansion of $\Tr H^6$ to cancel the
 sine-dependent terms of the form $\sin (\phi S_{\mu\nu})$.
 However, the number of sine-dependent terms grows exponentially while the number of
 terms available for canceling sine-dependent terms does not grow as quickly,
 so for higher moments we cannot expect disorder independence.
 As it turns out the same calculation already fails for  $\Tr H^8$.

 \section{Chaos on the hypercube}\label{sec:levelStat}

\begin{figure}
\centerline{  \includegraphics[width=8cm]{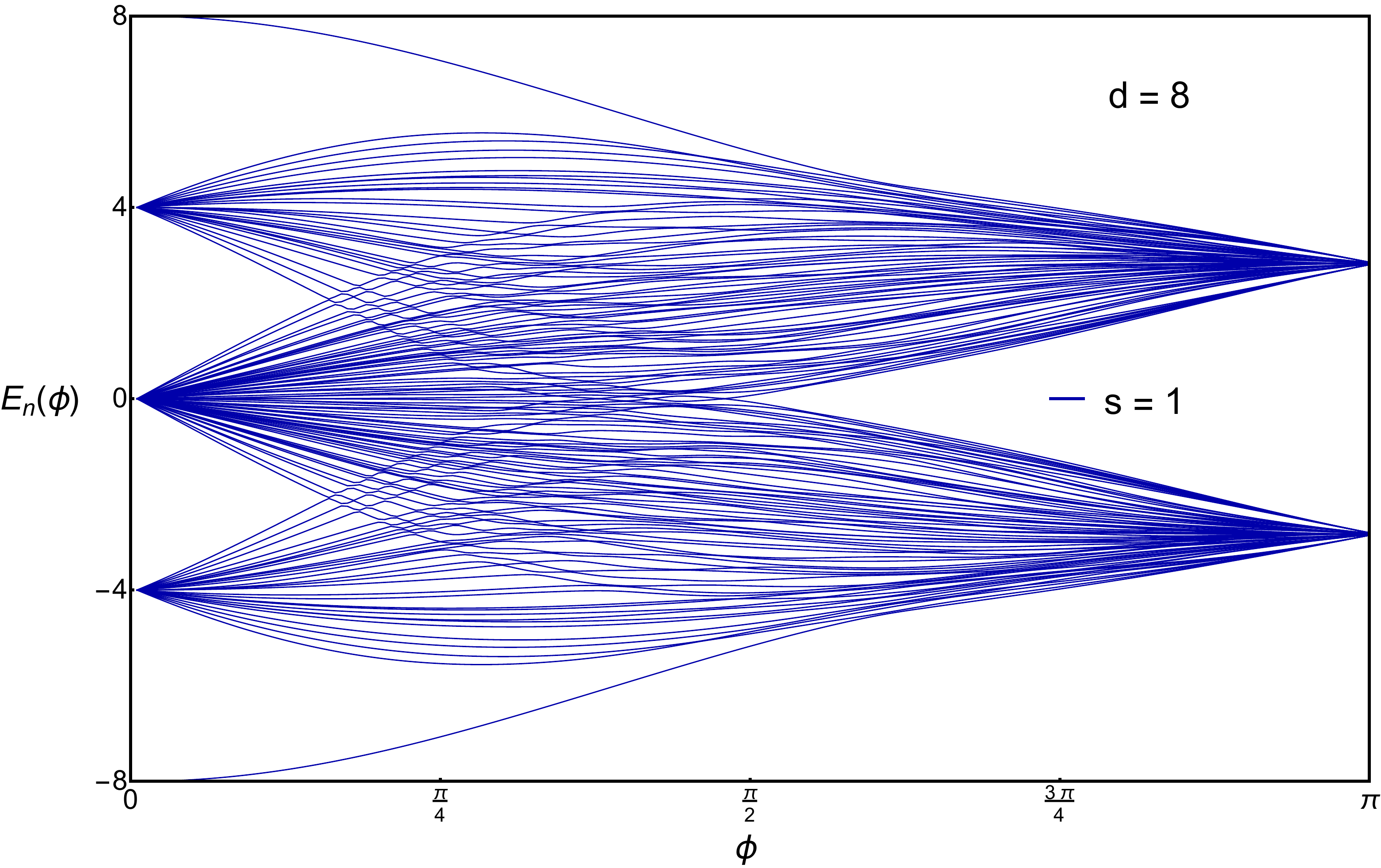}
  \includegraphics[width=8cm]{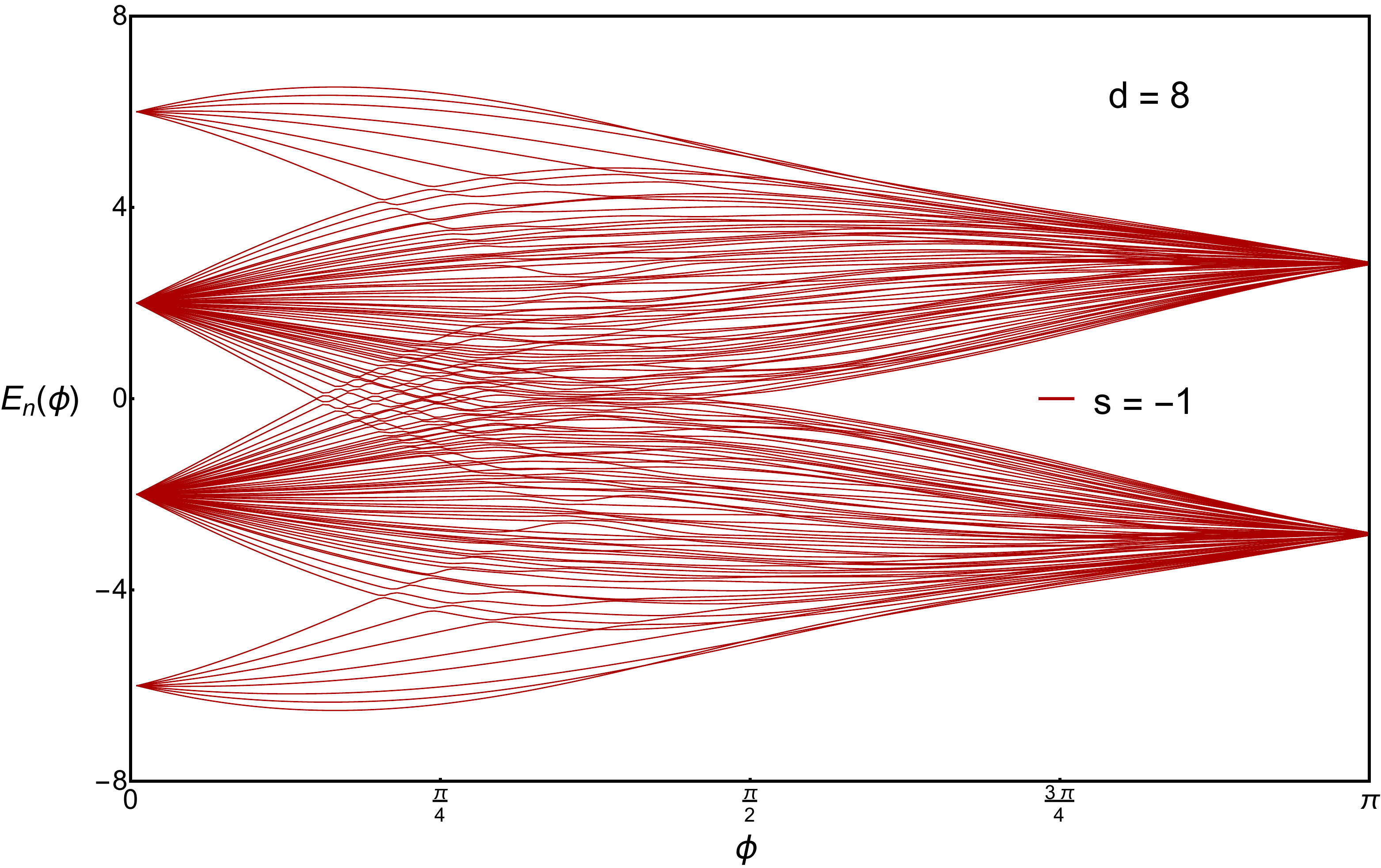}}
\caption{Spectral flow of the hypercubic Model as a function of the flux $\phi$. 
  for each of its two symmetry classes. To make individual curves visible, we show the
  results for $d=8$, but the features for larger values of $d$ are similar.}
\label{fig:flow}
  \end{figure}

For $\phi=0$ the model is integrable, and has a degenerate spectrum \eqref{eqn:zerophiSpectrum}. The degeneracies are lifted at nonzero $\phi$, but the eigenvalues will eventually flow to $\pm \sqrt d$
at $\phi = \pi$, as predicted by equation \eqref{eqn:piFluxHsquare} and the sublattice symmetry. A figure of the spectral flow as a function of $\phi$ is shown in figure \ref{fig:flow}
with the quantum number
of the magnetic inversion symmetry equal to $s=1$ in the left figure  and  $s=-1$ in the right figure.
The flow for $\phi<\pi/2$ is similar to the one of
the Maldacena-Qi model. At $\phi=0$ the spectrum and degeneracies are the same as for
the Maldacena-Qi model at infinite
coupling. The degeneracies are lifted at nonzero $\phi$, and at  $\phi=\pi/2$ the spectrum splits into two bands, a feature
that is not present in the MQ model. The ground state of the model is separated from the rest of the spectrum by a gap,
and our numerical results suggest that the gap likely remains finite for $\phi<\pi/2$ in the thermodynamical limit (see the left figure
of figure \ref{fig:flow}). 
We expect that the levels in each subsector become chaotic
as soon as the bands emanating from degenerate eigenvalues start overlapping
(at about $\phi=\pi/4$)
which will be studied in more detail below.\footnote{In fact, although bands are separate for very small $\phi$, the eigenvalues are repelled within each band (except for the lowest and highest energy states which are nondegenerate) for any small but nonzero $\phi$. A numerical analysis similar to that presented in section \ref{sec:levelStat} shows levels in each band are chaotic. In this sense the only integrable point of the HC model is at $\phi = 0$. }
The apparent crossings of the spectral flow lines are actually avoided crossings even though some are extremely close.

\subsection{Average spectral density}

It was already realized by Parisi (and more explicitly by Cappelli and  Colomo in \cite{Colomo:2001a}) that the spectral density of the large $d$
limit of the hypercube model is given by the ground state wave function density (wave function modulus squared) of the
Q-harmonic oscillator. The argument is essentially the same as in the case of
SYK model \cite{Erdos:2014a,Cotler2016,Garcia-Garcia:2017pzl,Berkooz:2018qkz,Berkooz:2018jqr}, and can be summarized
as follows (see appendix \ref{app:chords} for more details). The moments of the Hamiltonian $\vev {\Tr H^{2p}}$
can be written as a sum of Wilson loops on the lattice.
As is explained in appendix \ref{app:chords} paths
can be represented as chord diagrams, and in particular each loop is represented by a
chord diagram crossing. Each crossing gives rise to a factor of $q=\cos\phi$.\footnote{Note this $q$ is not the $q$ often used in the context of SYK model where it denotes the interaction order of Majorana fermions.}
For large $d$ the leading contributions are from Wilson loops traversing the
maximum number dimensions.
After ensemble averaging we thus obtain the $2p$-th moment:
\be\label{eqn:HCqHermite}
M_{2p}^{\text{HC}}:=\frac{ 2^{-d}\langle\Tr H^{2p}\rangle}{\left(2^{-d}\langle\Tr H^{2}\rangle\right)^p} = \sum_{k=0}^{d(d-1)/2}  a_k q^k + O(1/d),
\ee
where  $a_k$ is the number of chord diagrams with $k$ crossings. We have defined $M_{2p}^{\text{HC}}$ as a \textit{reduced moment} since we used $\Tr H^2$ in the denominator, but we will call $M_{2p}^{\text{HC}}$ ``moment'' when the context is free of confusion. In appendix \ref{app:chords} we lay out the arguments and derivations that lead to equation \eqref{eqn:HCqHermite} in more details, and discuss the subleading corrections.
 
\begin{figure}
  \centerline{\includegraphics[width=5cm]{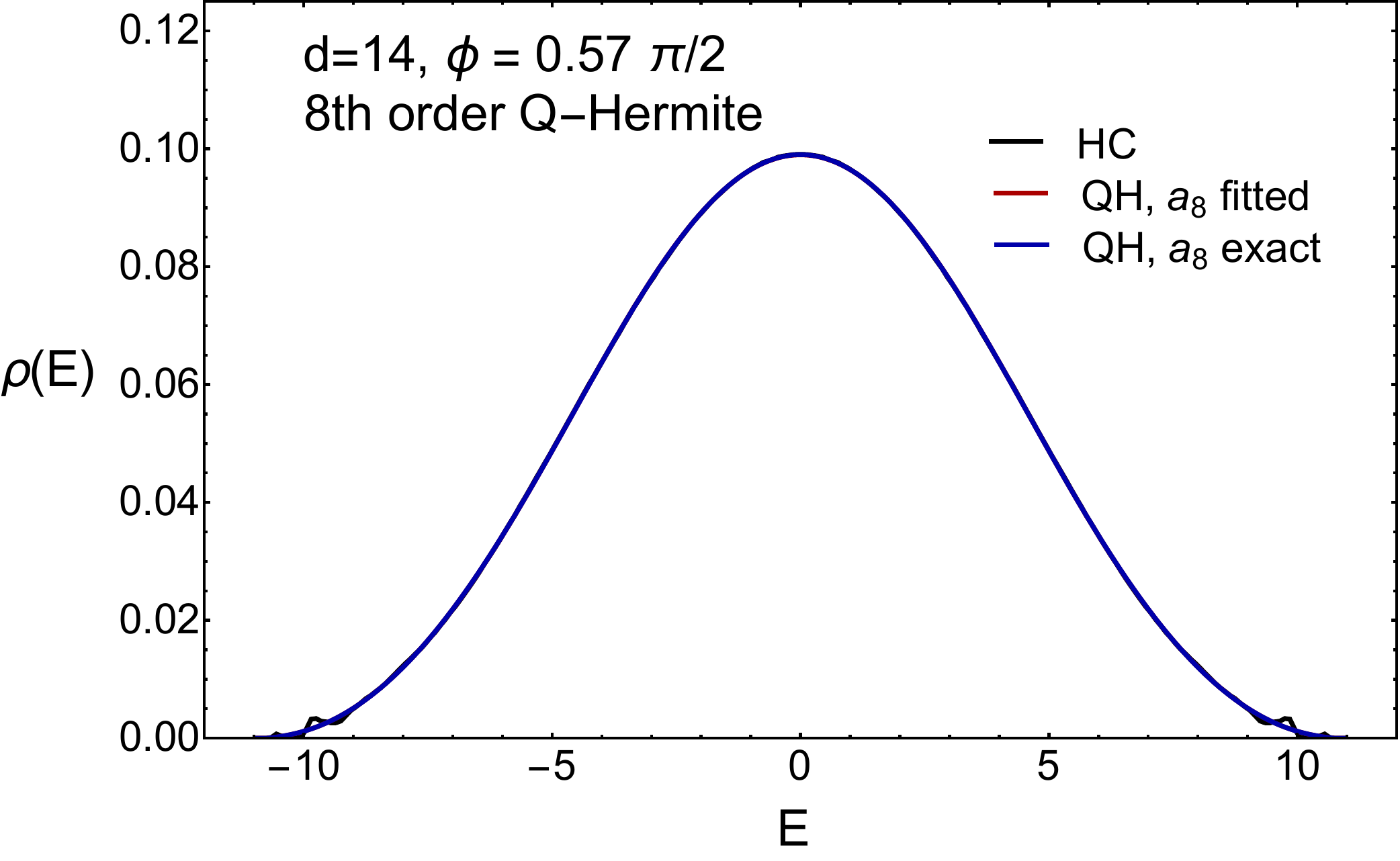}
    \includegraphics[width=5cm]{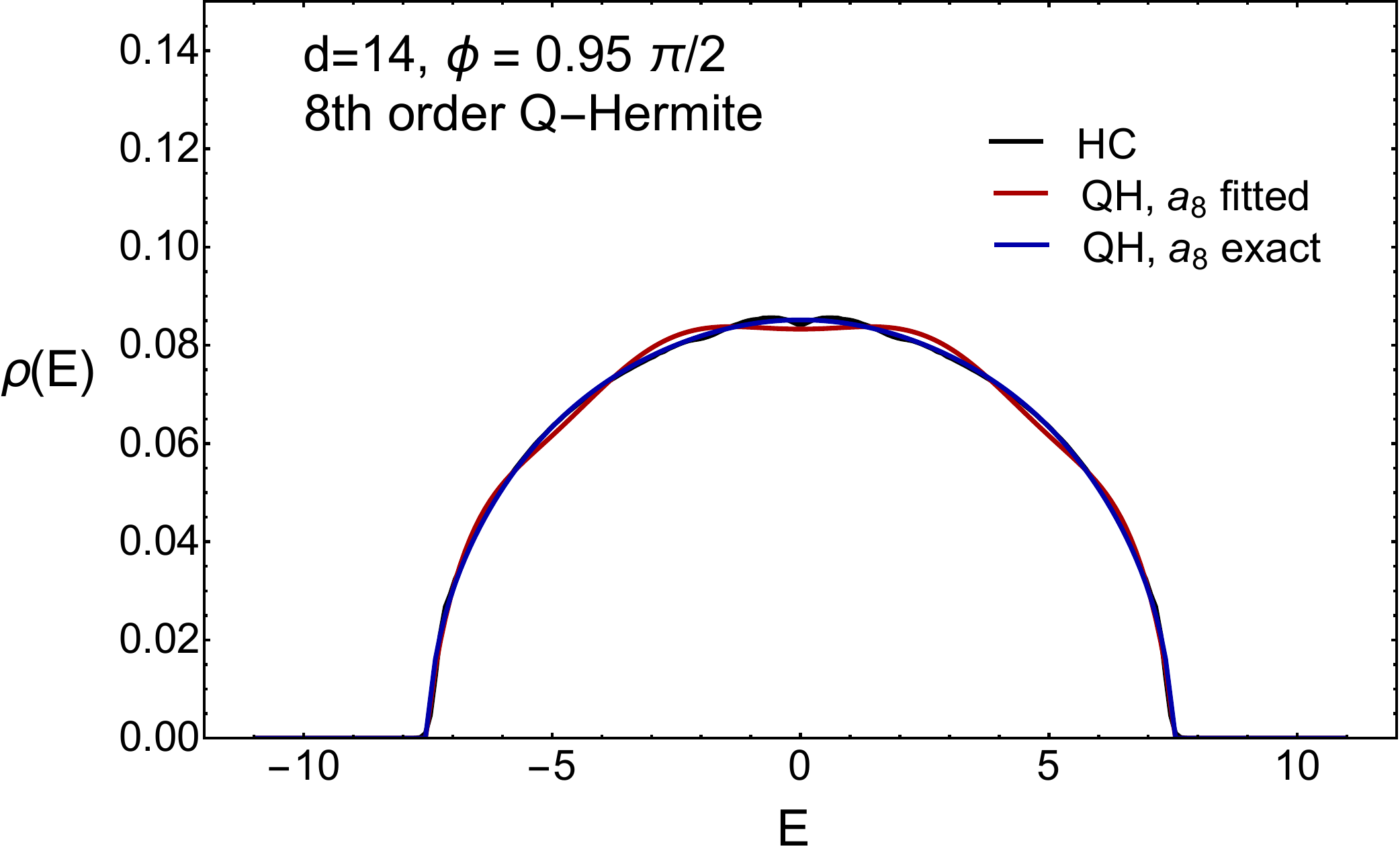}
  \includegraphics[width=5cm]{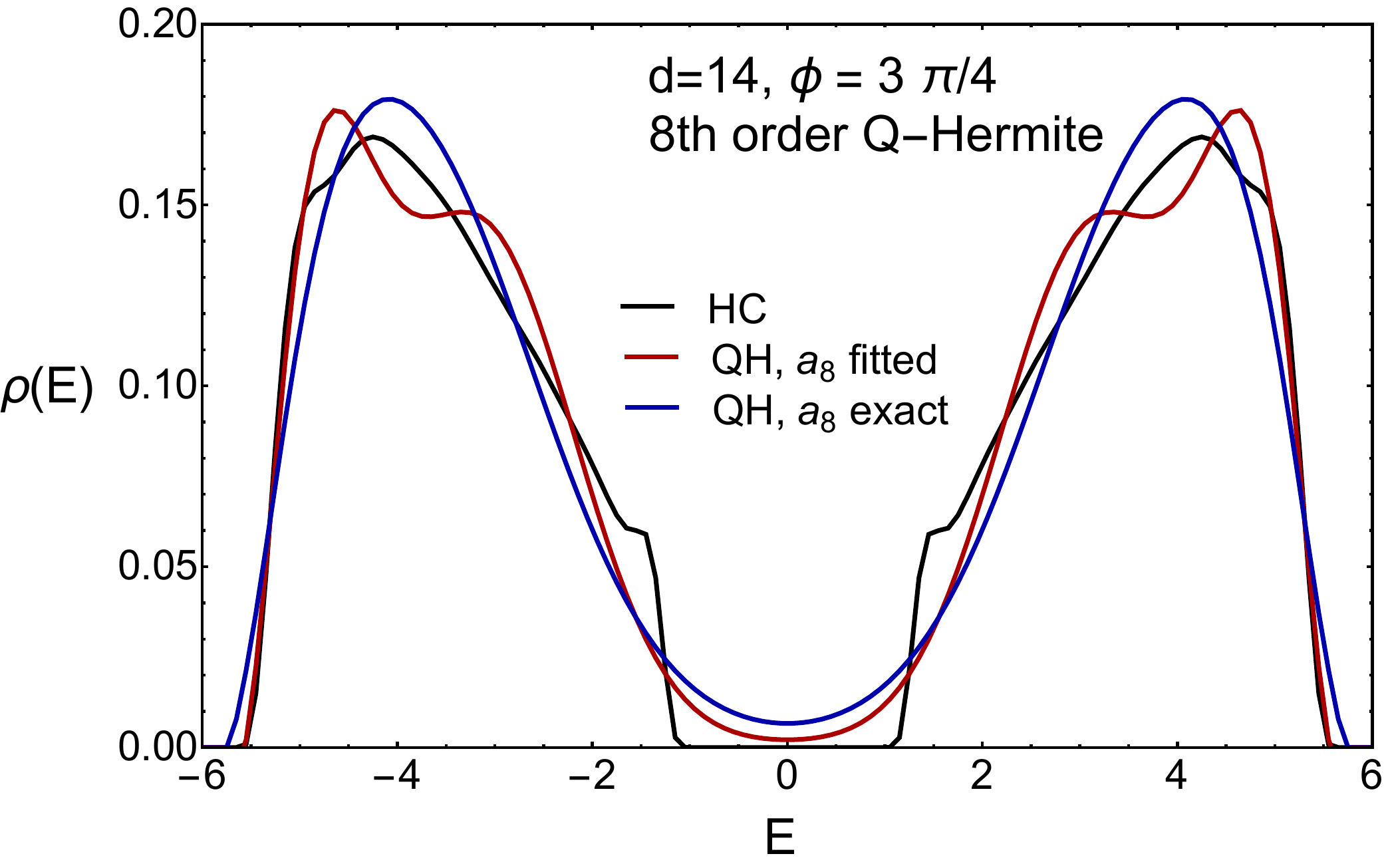}}
  \caption{The spectral density of the  hyper cubic model (black curve) compared to the
    eighth order Q-Hermite spectral density defined in \eqref{eqn:qhExpansion} for three different flux values as given in
    the legend of the figures.  In the left figure $\phi=0.57 \pi/2$,
    the curve resulted from a fitted $a_8$ (red curve) differs from the one resulted from the  $a_8$  that is calculated by equation \eqref{a68} (blue curve) by less than the line width of the curve's plot, so we do not see the red curve at all. The red curve and blue curve also agree very well in the middle figure where $\phi=0.95\pi/2$. The deviations are larger  for $\phi=\frac 34 \pi$ in the right figure. 
  }
\label{fig:spec57}
\end{figure}

\begin{figure}
  \centerline{\includegraphics[width=7cm]{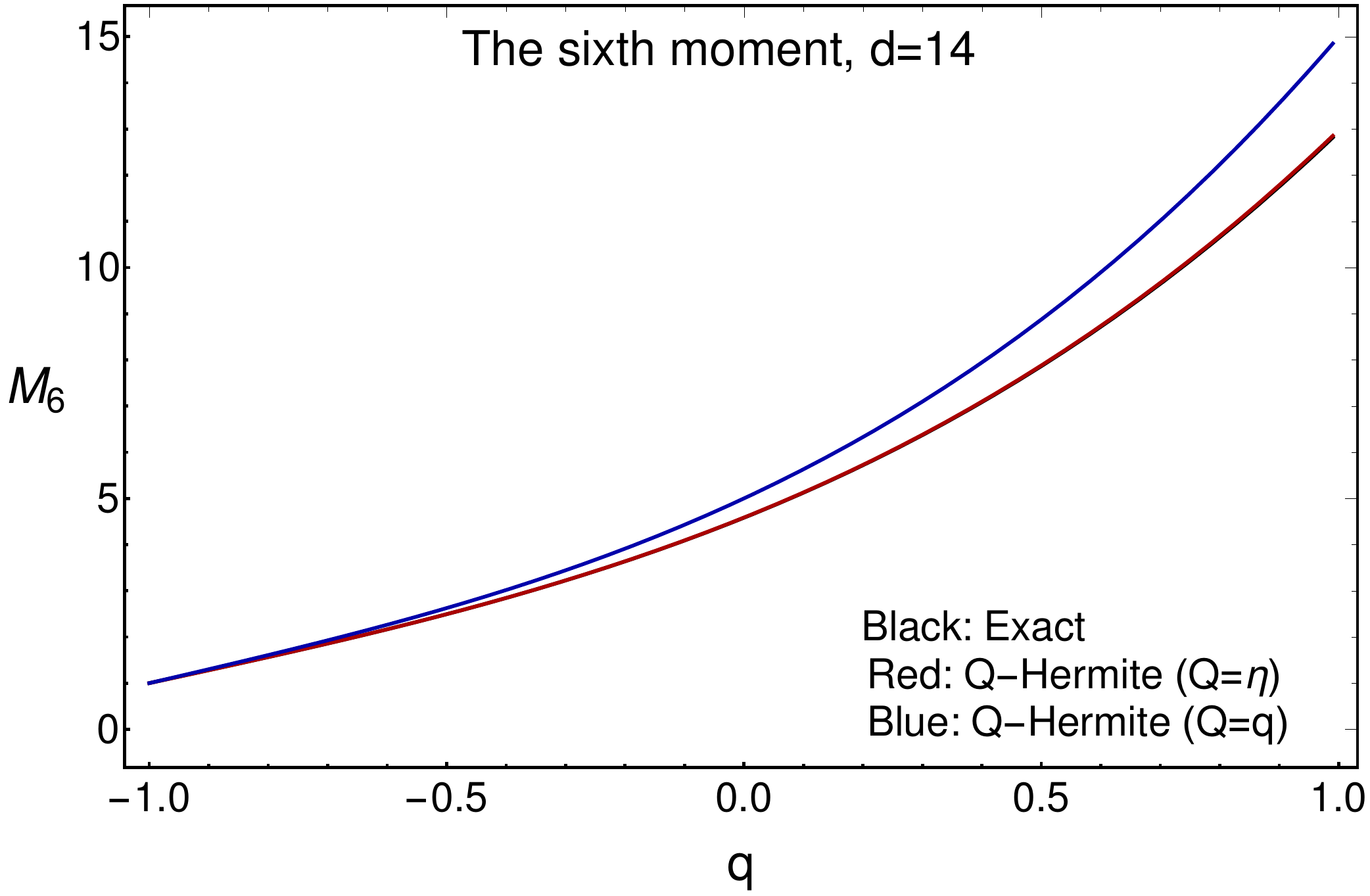}
    \includegraphics[width=7cm]{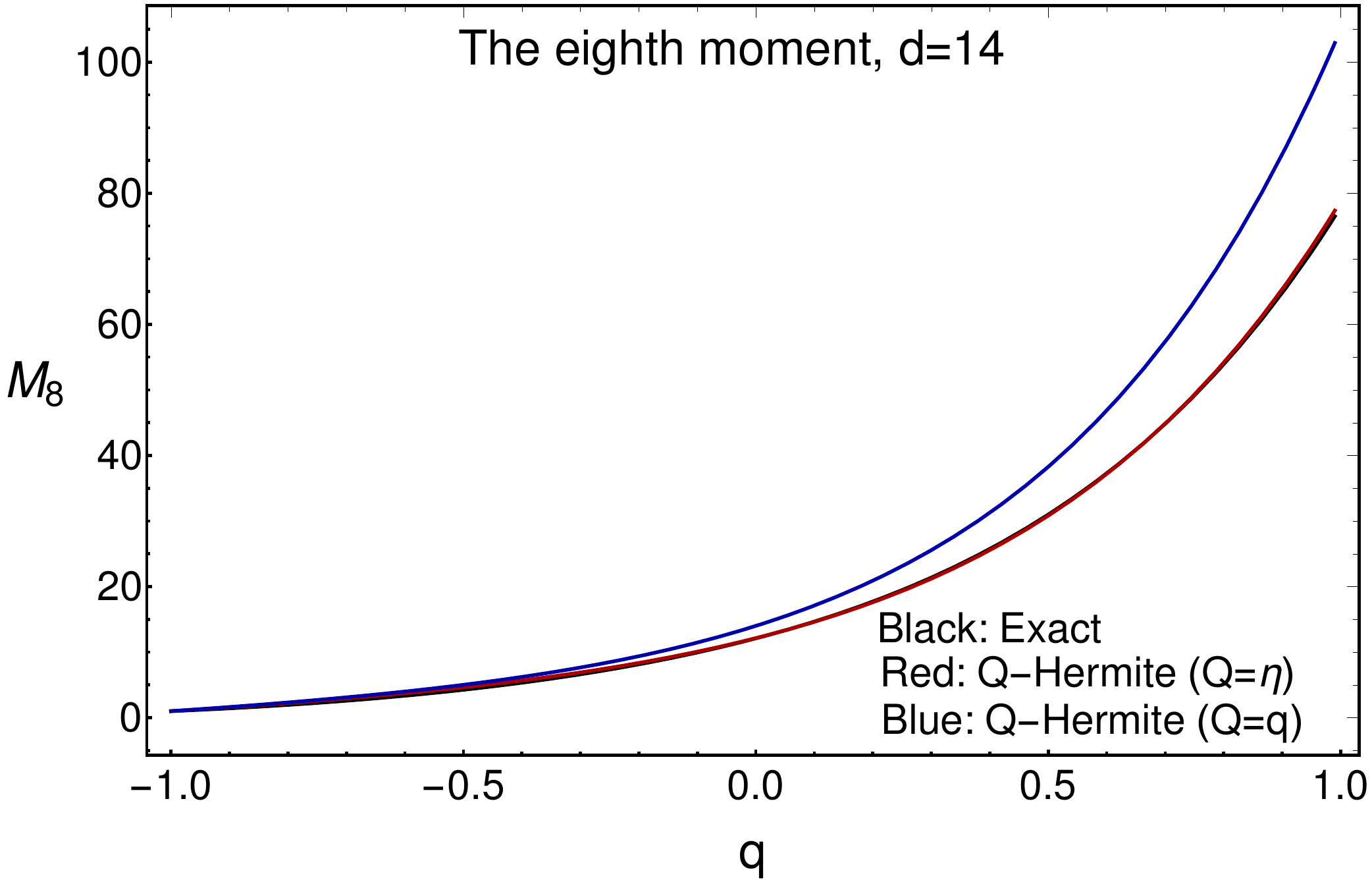}
  }
 \caption{The sixth (left) and the eighth (right) moments as functions of $q$ at $d=14$. The exact results (black) , the Q-Hermite (blue) results and the renormalized Q-Hermite (red) results are shown. We 
observe that renormalizing $q$ to $\eta$  greatly improves the accuracy at $d=14$. In fact, the renormalized results work so well that their curves can  barely be  distinguished from the exact results. }
\label{fig:momCompare}
\end{figure}

 The  moments given in equation \eref{eqn:HCqHermite} are the moments of  the
density function of the Q-Hermite polynomials:
\be
\rho^{\rm QH}(x) = \left(1-\frac{x^2}{e_0^2} \right )^{1/2}
\frac 2{\pi e_0}\prod_{k=1}^\infty\frac{1-Q^{2k}}{1-Q^{2k-1}}
\prod_{k=1}^\infty \left (1 - 4 \frac{x^2}{e_0^2}  \frac {Q^k}{(1 + Q^k)^2}\right),
\ee
with $e_0 = 2/\sqrt{1-Q}$ and $Q = q=\cos\phi$.  However, to include some of the finite-$d$ corrections we set $Q=\eta$, which is  a renormalized version of $q$, obtained by matching the fourth moment of $\rho^{\text{QH}}(x)$ and the fourth moment of the hypercube model exactly: 
\be
Q=\eta := M_4^{\rm HC}-2 =\cos \phi - \frac {\cos\phi  + 1}d.
\label{eta}
\ee
In addition, this renormalization absorbs the leading $1/d$ corrections of
the sixth moment, but not  higher moments.
It is clear $\eta\to q = \cos \phi$ in the large $d$ limit.
In figure \ref{fig:spec57} we show the average spectral densities for three different values of $\phi$ and compare the result with
the Q-Hermite spectral density with $Q=\eta$.
Renormalizing $q$ to $\eta$ improves the accuracy for finite $d$, but this is still not exact: the deviation will start to appear for the sixth and higher moments. We cite  \cite{Marinari:1995jwr} here for the exact results up to the eighth moment:
\begin{align}
M_4^{\text{HC}}&= \frac{d-1}{d}(2+q) +\frac{1}{d},\\
M_6^{\text{HC}}&= \frac{(d-1)(d-2)}{d^2}(5+6q+3q^2+q^3)+\frac{d-1}{d^2}(9+6q)+\frac{1}{d^2},\\
M_8^{\text{HC}}&= \frac{(d-1)(d-2)(d-3)}{d^3}(14+28q+28q^2+20q^3+10q^4+4q^5+q^6)\\
&\quad +\frac{(d-2)(d-1)}{d^2}(56+86q+52q^2+16q^3)+\frac{d-1}{d^3}(33+28q+2q^2)+\frac{1}{d^3}.\nn
\end{align}
Comparing with the exact results, we can see (in figure \ref{fig:momCompare}) that the renormalization indeed gives considerable improvements for the finite-$d$ results.
 In terms of $\rho^{\rm QH}_\eta(x)$ 
the spectral density (before ensemble averaging) can be expanded as 
\be\label{eqn:qhExpansion}
\rho^{\rm HC}(x)= \rho^{\rm QH}_\eta(x)(1 + a_6 H^\eta_6(x) + a_8 H^\eta_8(x)+\cdots),
\ee
where the coefficients $a_8,\ldots$ are random variables
(note that $a_6$ is determined by the sixth moment and does not depend on the disorder realization), and $H^\eta_n$ are the Q-Hermite polynomials defined by the recursion relation \cite{Viennot-1987}
 \be
 H_{n+1}^\eta(x) = x H_n^\eta(x) - \sum_{k=0}^{n-1}\eta^k H_{n-1}^\eta(x)
 \ee
 with the initial conditions
 \be
 H_0^\eta(x) = 1 \qquad {\rm and } \qquad H_1^\eta(x) = x.
 \ee
The Q-Hermite polynomials satisfy the orthogonality relation
 \be
 \int_{-\frac{2}{\sqrt{1-\eta}}}^{\frac{2}{\sqrt{1-\eta}}} dx \rho^{\text{QH}}_\eta(x) H_n^\eta\left (x \right )
 H_m^\eta\left (x \right )=\delta_{nm}n_\eta!,
 \ee
 where $n_\eta!$ is the Q-factorial (with Q $=\eta$) defined as
 \be
 n_\eta! = \prod_{k=1}^{n-1}\
 \left(\sum_{s=0}^k\eta^s\right).
 \ee
 Note that for the choice of $\eta$ in \eref{eta} the coefficients of $H^\eta_2(x)$ and $H^\eta_4(x)$ vanish since $\rho^{\text{QH}}_\eta$ already gives the exact results for $M^{\text{HC}}_2$ and $M^{\text{HC}}_4$. We stress that they vanish not just after averaging but also realization by realization, this is because in section \ref{sec:sumRules} we have proven $\Tr H^2$  and  $\Tr H^4$ are independent of disorder realizations. The coefficients $a_6$
 and $ a_8$ (after ensemble averaging) are given by (in the normalization where $M_2 =1$)
\be
a_6 &=&\frac 1{ 6_\eta!} (M_6^{\rm HC}- M_6^{{\rm QH}, \eta}),\nn\\
\vev {a_8} &=& \frac 1{ 8_\eta!} (M_8^{\rm HC}- M_8^{{\rm QH}, \eta})-\frac{(7+6\eta+5\eta^2+4\eta^3+3\eta^4+2\eta^5 +\eta^6)a_6 6_\eta!}{8_\eta !},
\label{a68}
\ee
where we note again $\vev {a_6} = a_6$ because $\Tr H^6$ is independent of disorder realizations, which is not true for $a_8$. 
This is not a good expansion for negative $\eta$ when $n_\eta!$ becomes small, see table \ref{tab:one}. For example, the
large value of $a_8$ for $\phi = 3\pi/4$ is due to the smallness of $8_\eta! \approx 0.01$. The expansion
diverges for $\phi \to \pi$. The reason is that
\be
(2p)_{(-1+x)}! = p! x^p + O(x^{p+1}), 
\ee
while
\be
M_8^{\rm HC}- M_8^{{\rm QH}, q-(q+1)/d} \sim -\frac{(q+1)^2}{3d},
\ee
so that $a_8$ diverges as $1/(q+1)^2$ for $q \to -1$.  This explains why in the left two figures of figure \ref{fig:spec57} the
fitted values of $a_8$ are close to the calculated values of $a_8$, whereas the in the right figure the agreement is not as good. For $a_6$ we are in a better position:
\be
M_6^{\rm HC}- M_6^{{\rm QH}, q-(q+1)/d} \sim \frac{(q+1)^3}{d^2}
\ee
so that $a_6 \sim 1/d^2$. This also explains why $a_6 <<1 $ for $d=14$, see table \ref{tab:one}.
For a given realization, the expansion coefficient $a_8$ is also given
by equation \eref{a68} but with $M_8^{\text{HC}}$ replaced by the eighth moment of that
realization. 

\subsection{Spectral correlations}

\begin{figure}[t!]
\centerline{  \includegraphics[width=8cm]{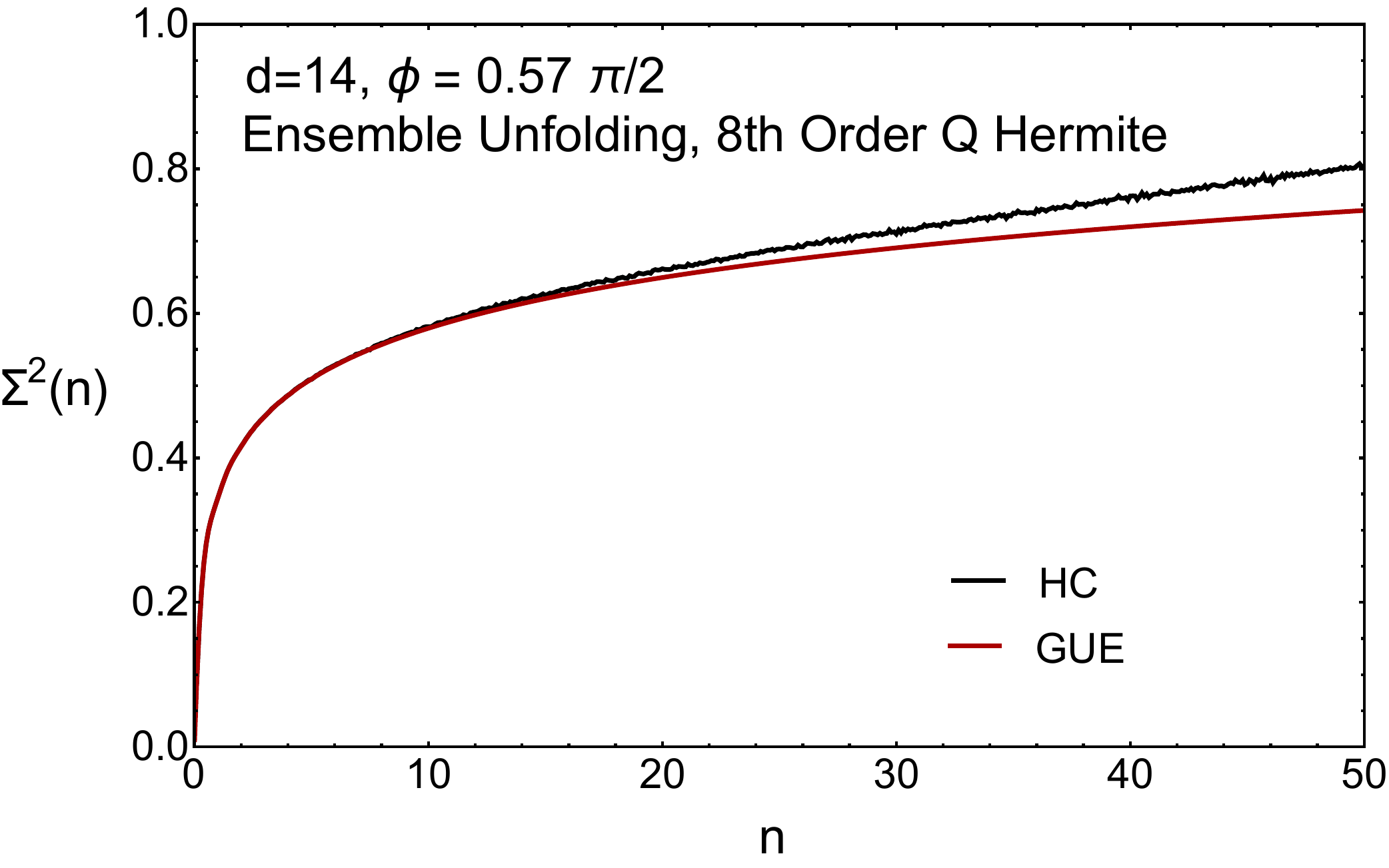}
  \includegraphics[width=8cm]{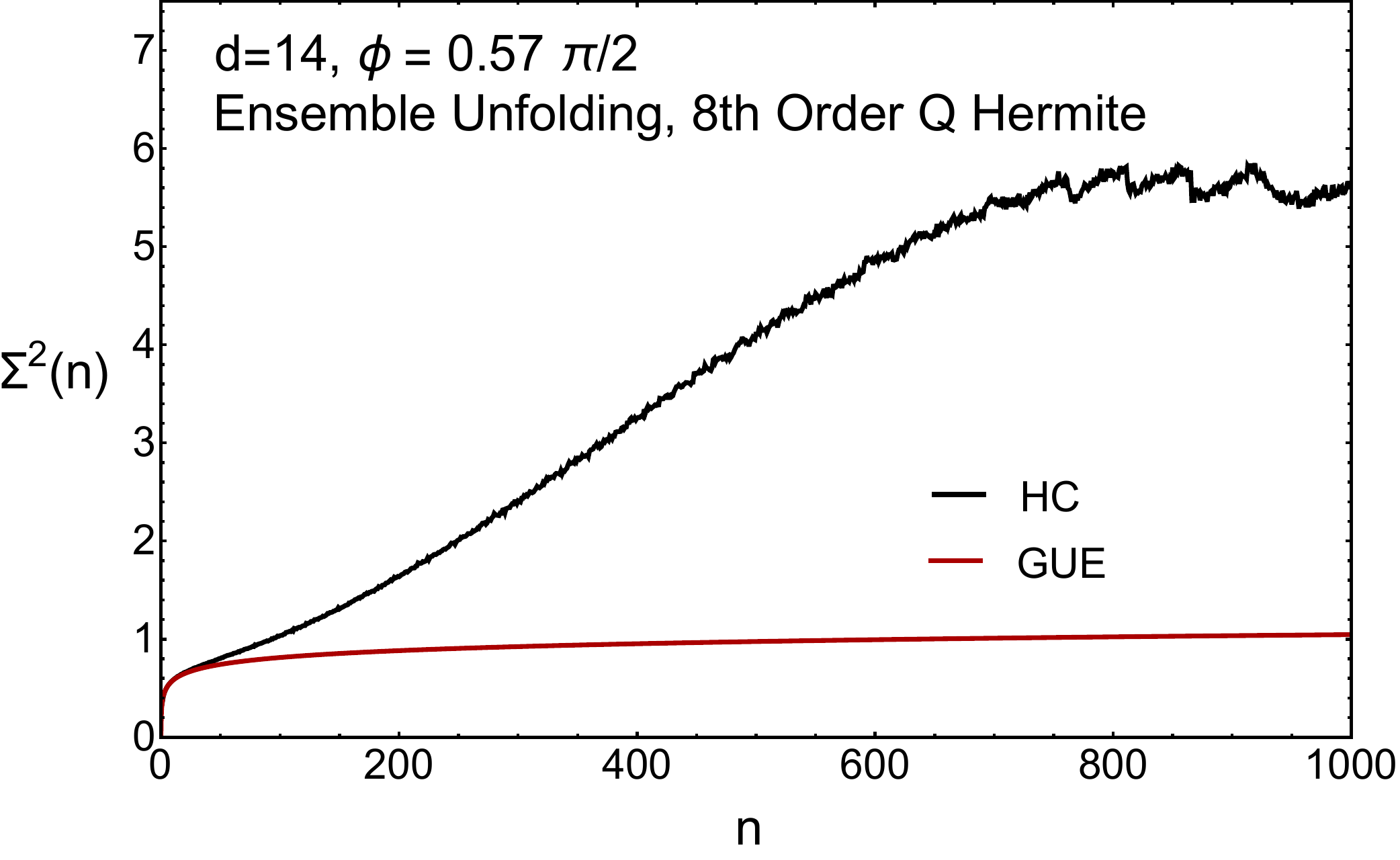}}
\centerline{  \includegraphics[width=8cm]{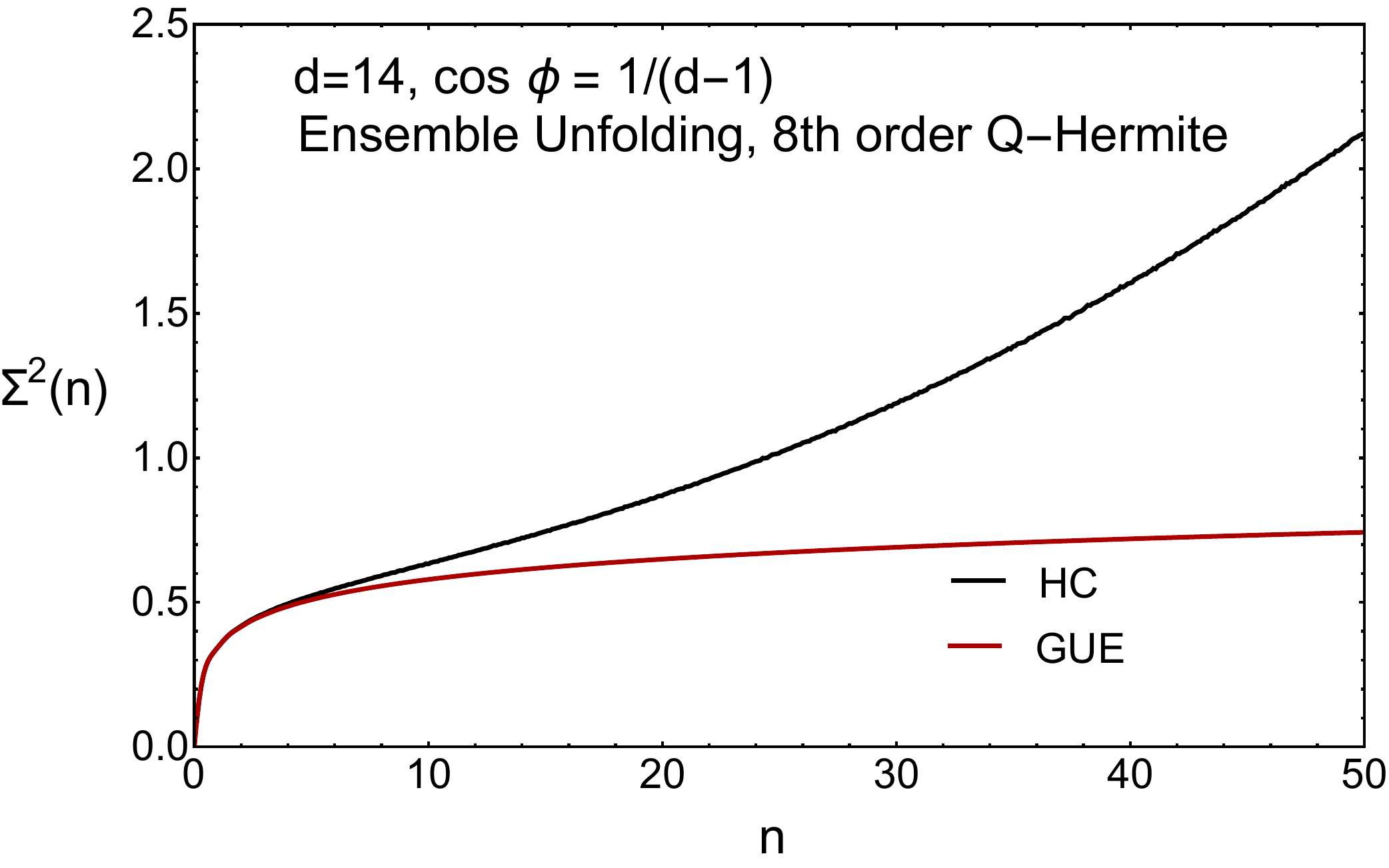}
  \includegraphics[width=8cm]{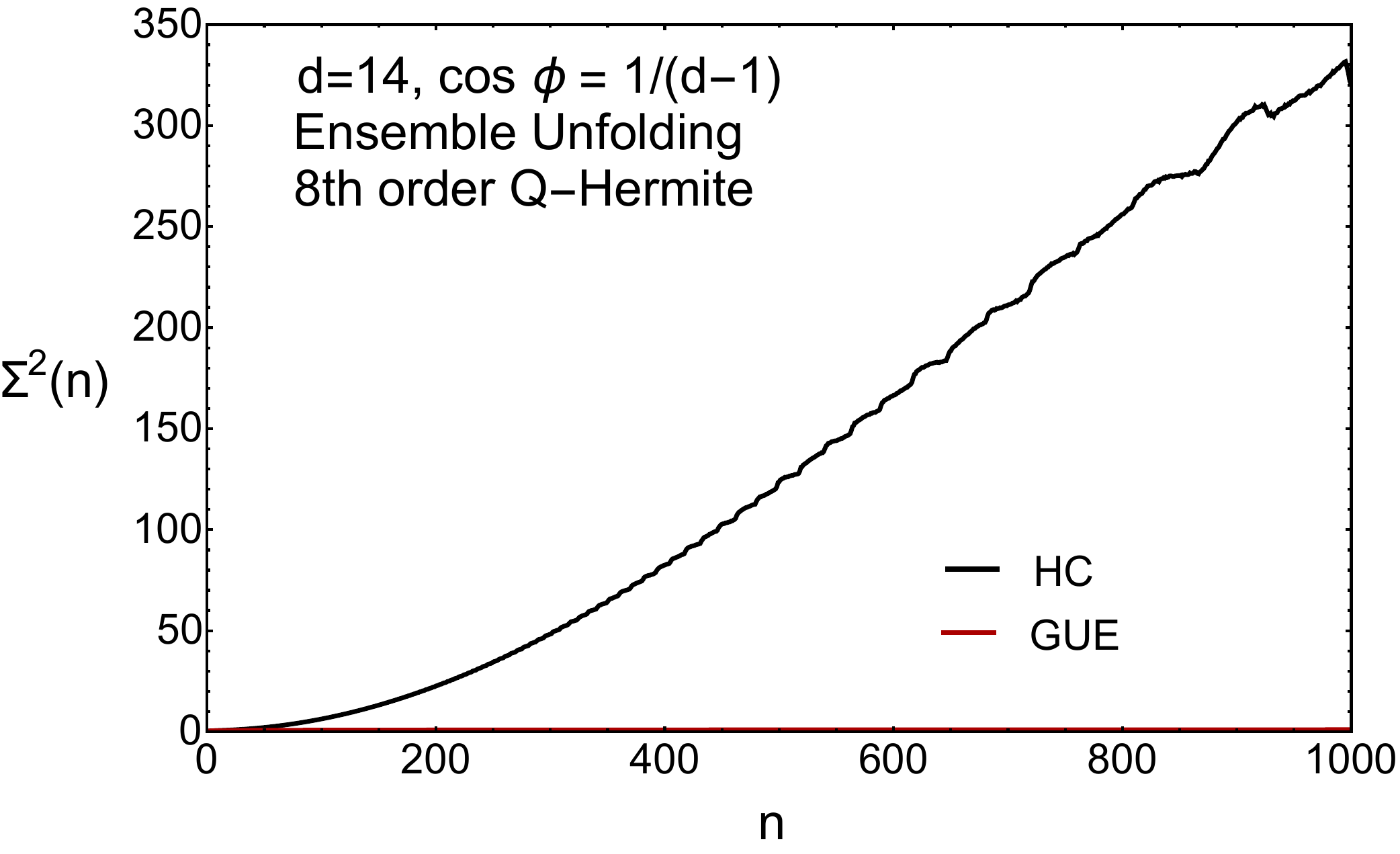}}
\centerline{  \includegraphics[width=8cm]{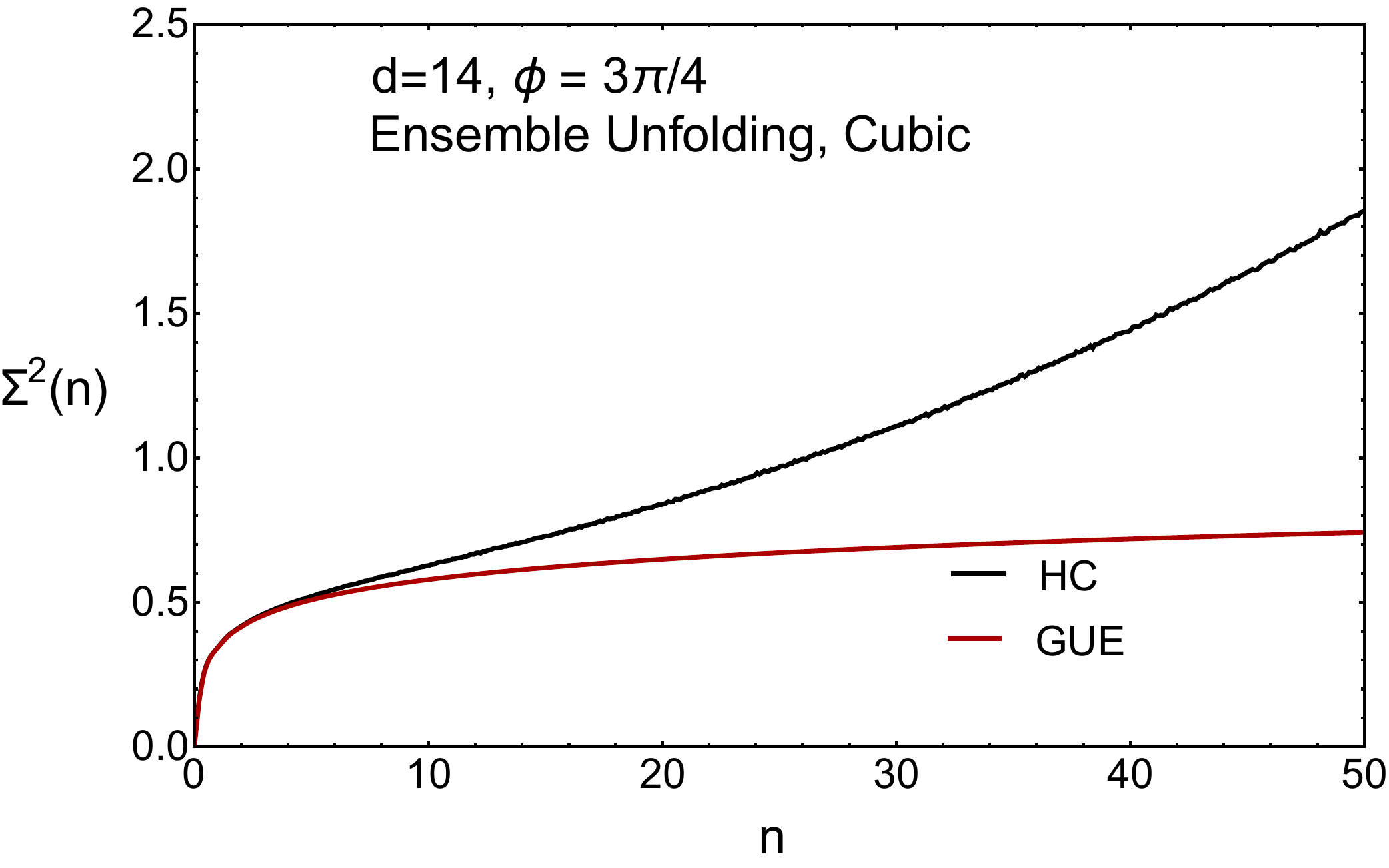}
 \includegraphics[width=8cm]{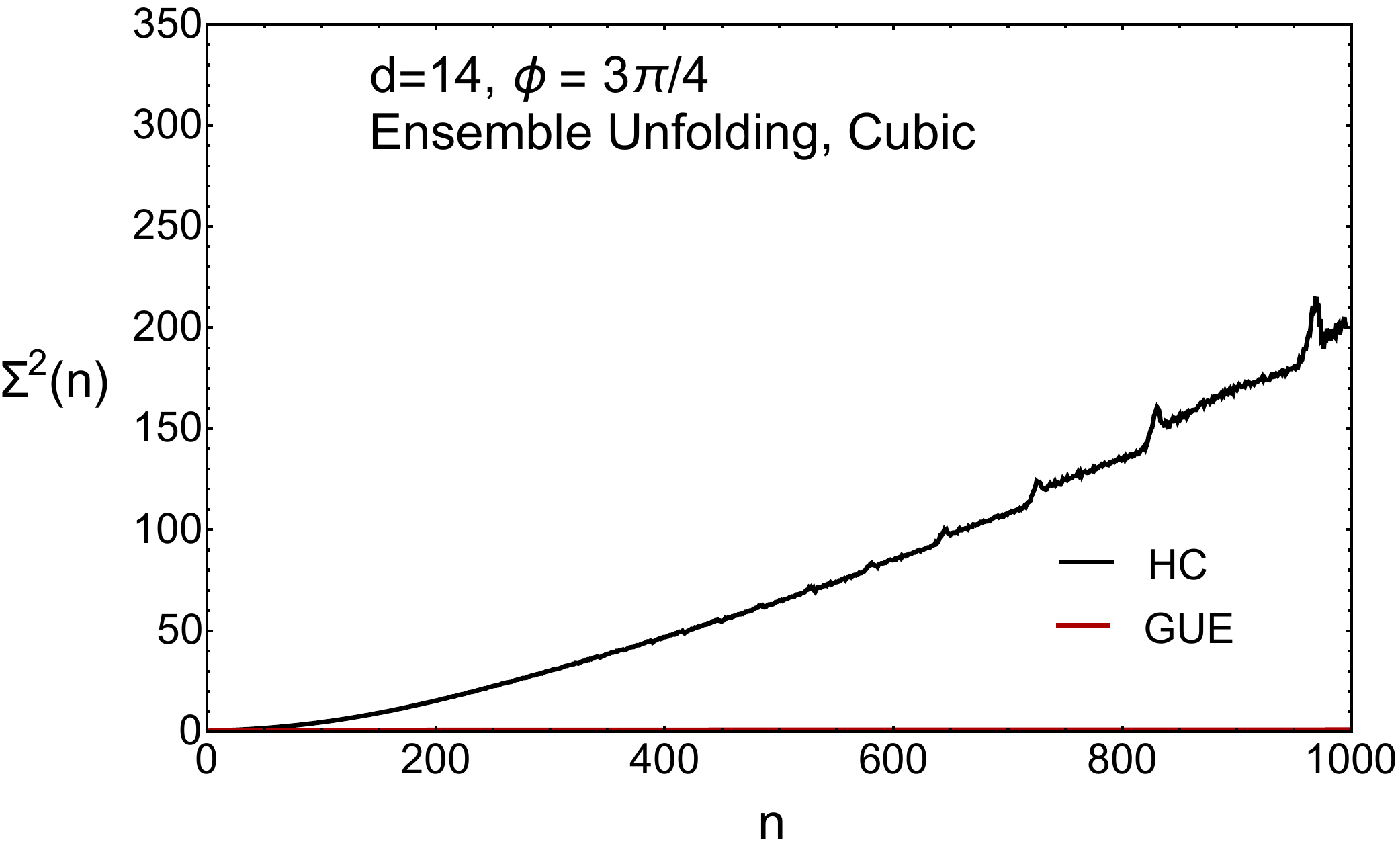}
}
\caption{The number variance $\Sigma^2(n)$ versus $n$ for $\phi = 0.57 \pi/2$,
  $\phi=0.95 \pi/2$ and $\phi =\frac 34 \pi$. The spectra have been unfolded
  using the ensemble average of the spectral density.
  The right figures show the number variance for larger values of $n$.}
\label{fig:numvar57}
\end{figure}
\begin{figure}[t!]
 \centerline{ \includegraphics[width=8cm]{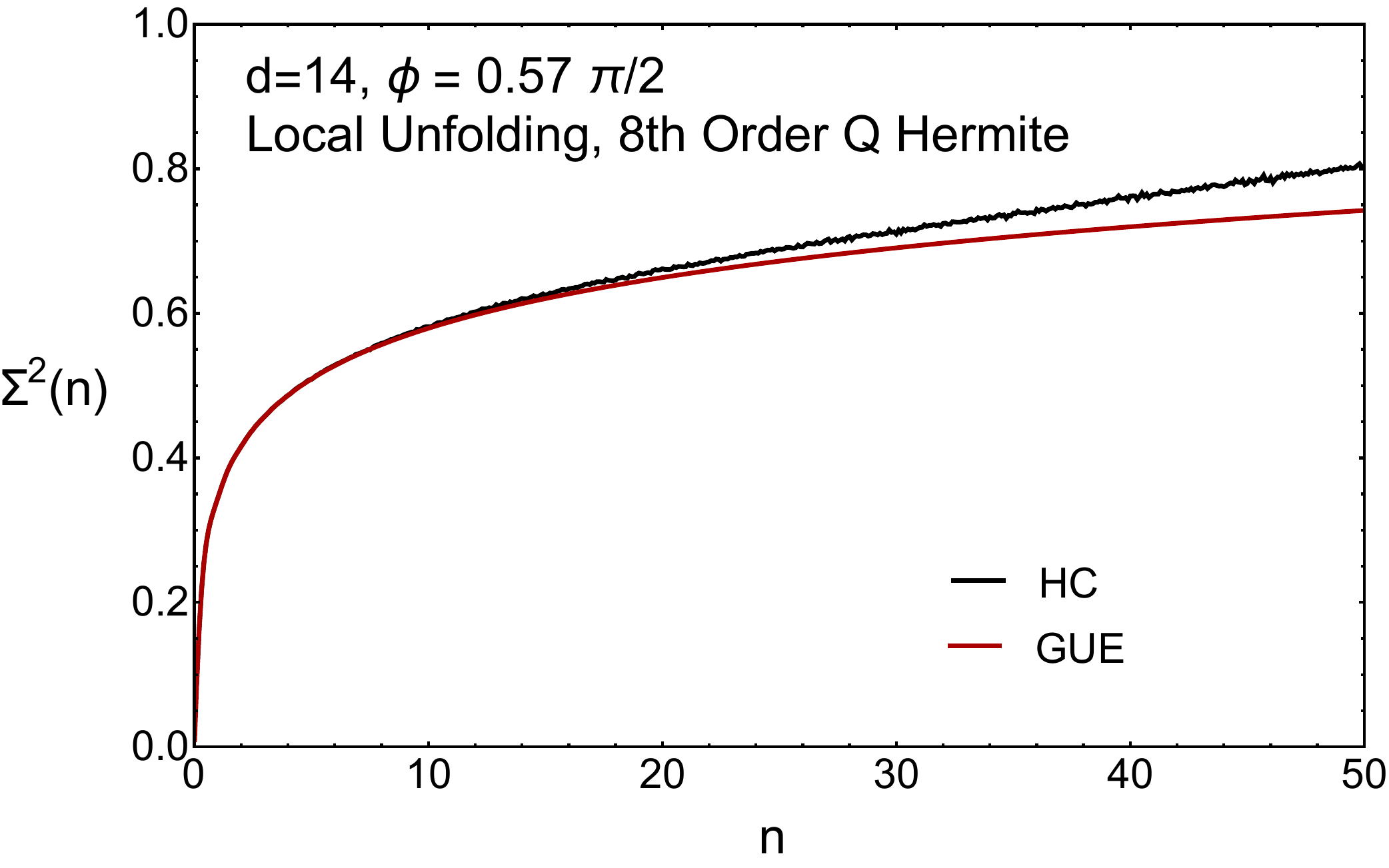}
  \includegraphics[width=8cm]{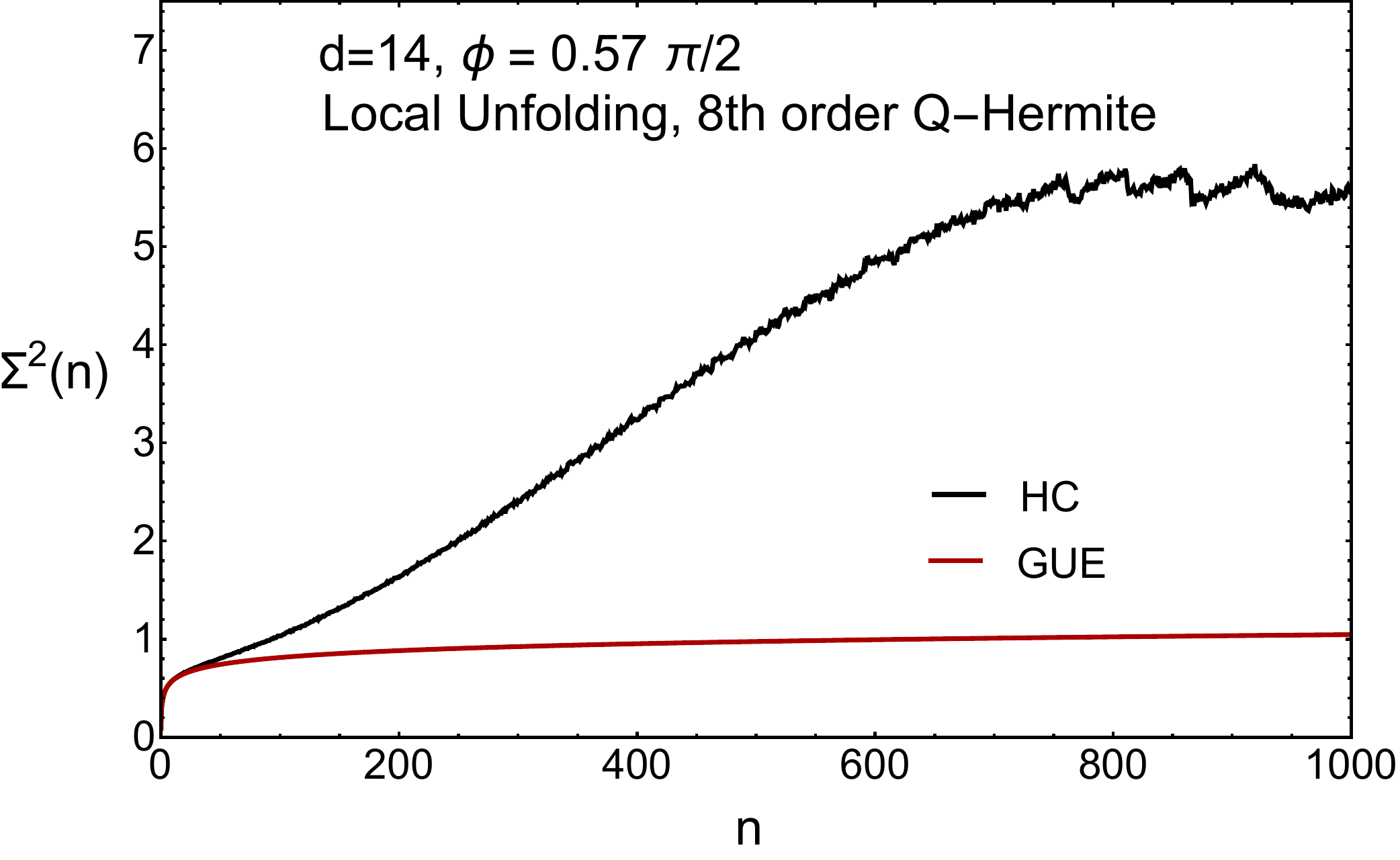}}
 \centerline{\includegraphics[width=8cm]{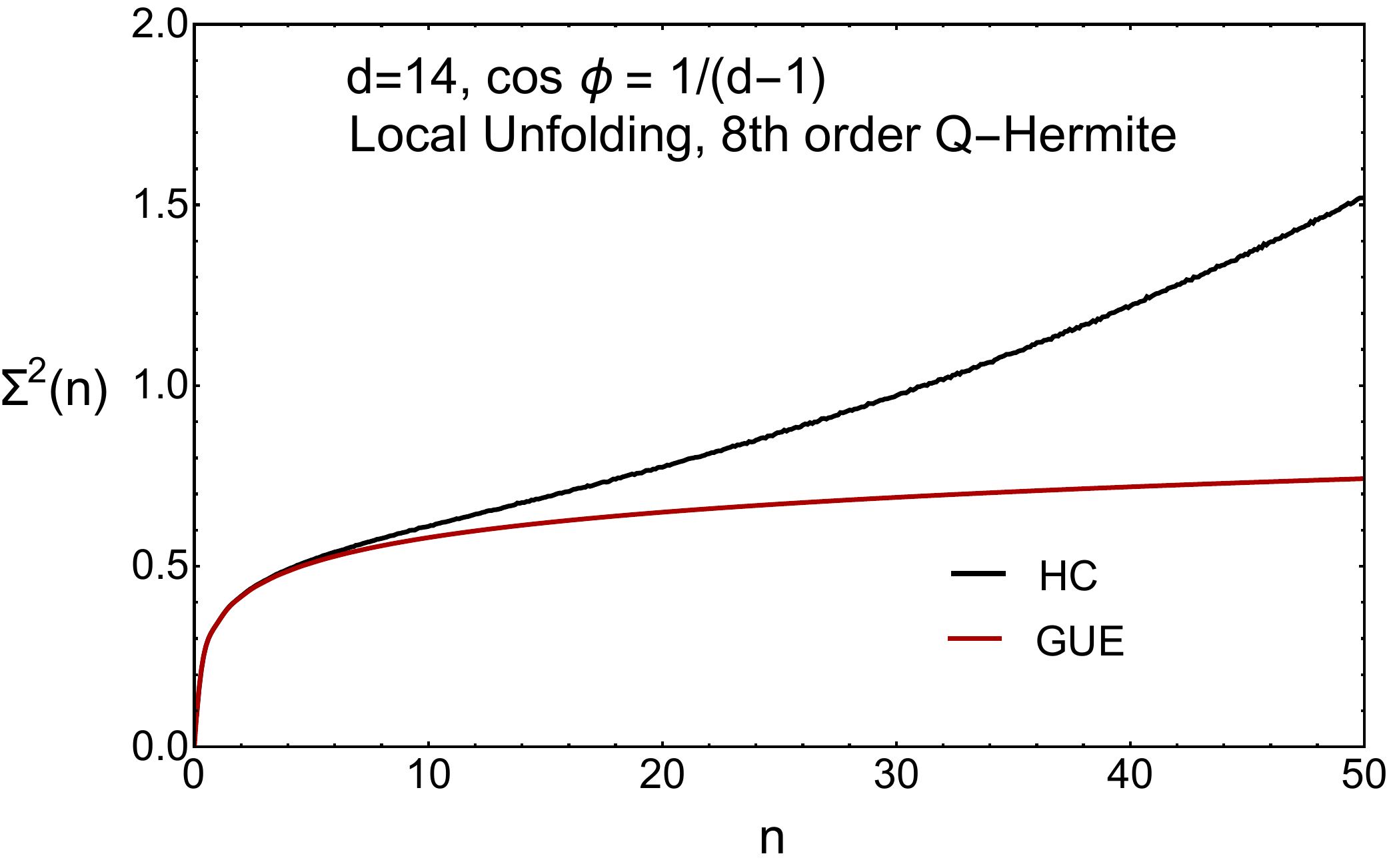}
  \includegraphics[width=8cm]{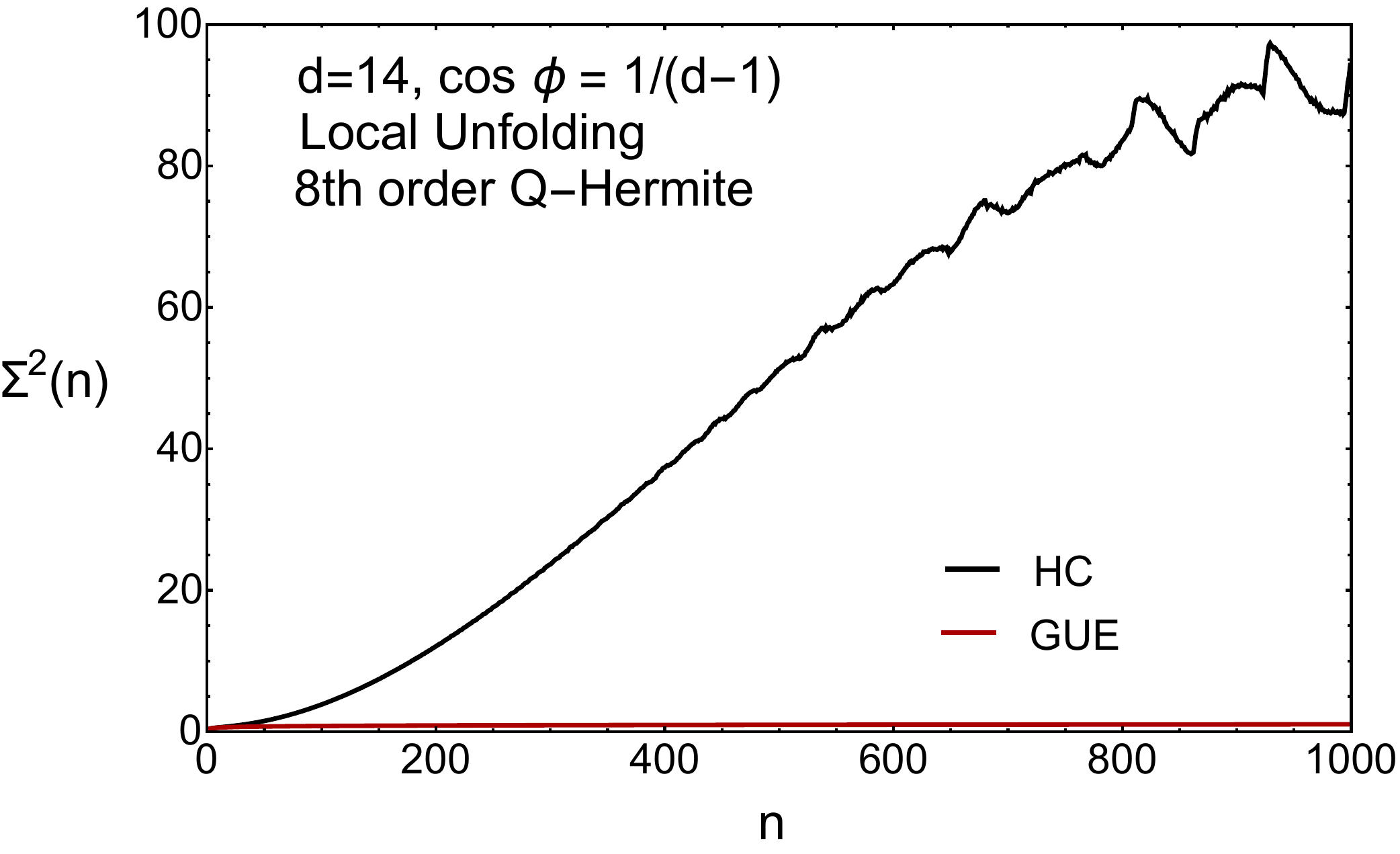}}
\centerline{ \includegraphics[width=8cm]{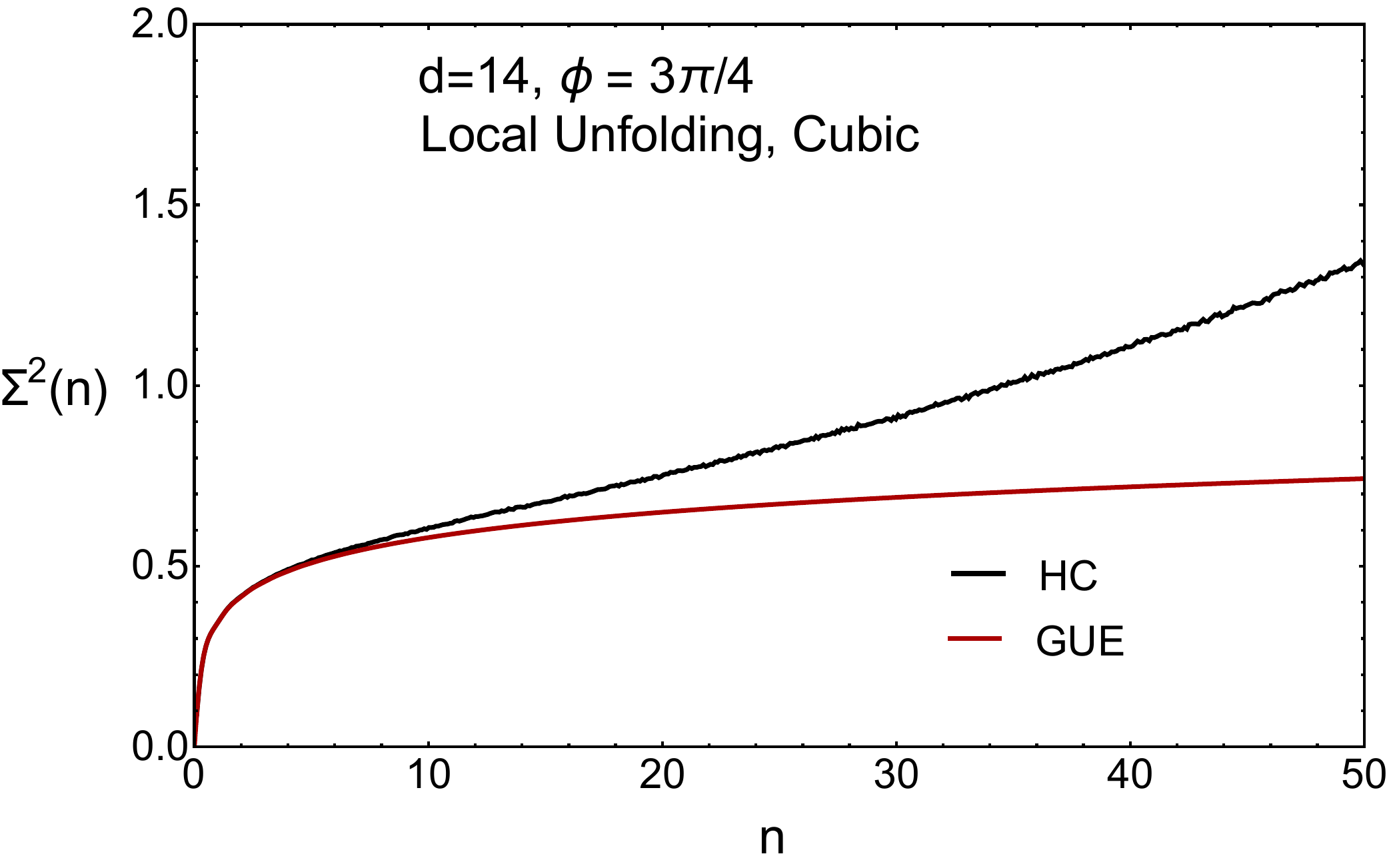}
  \includegraphics[width=8cm]{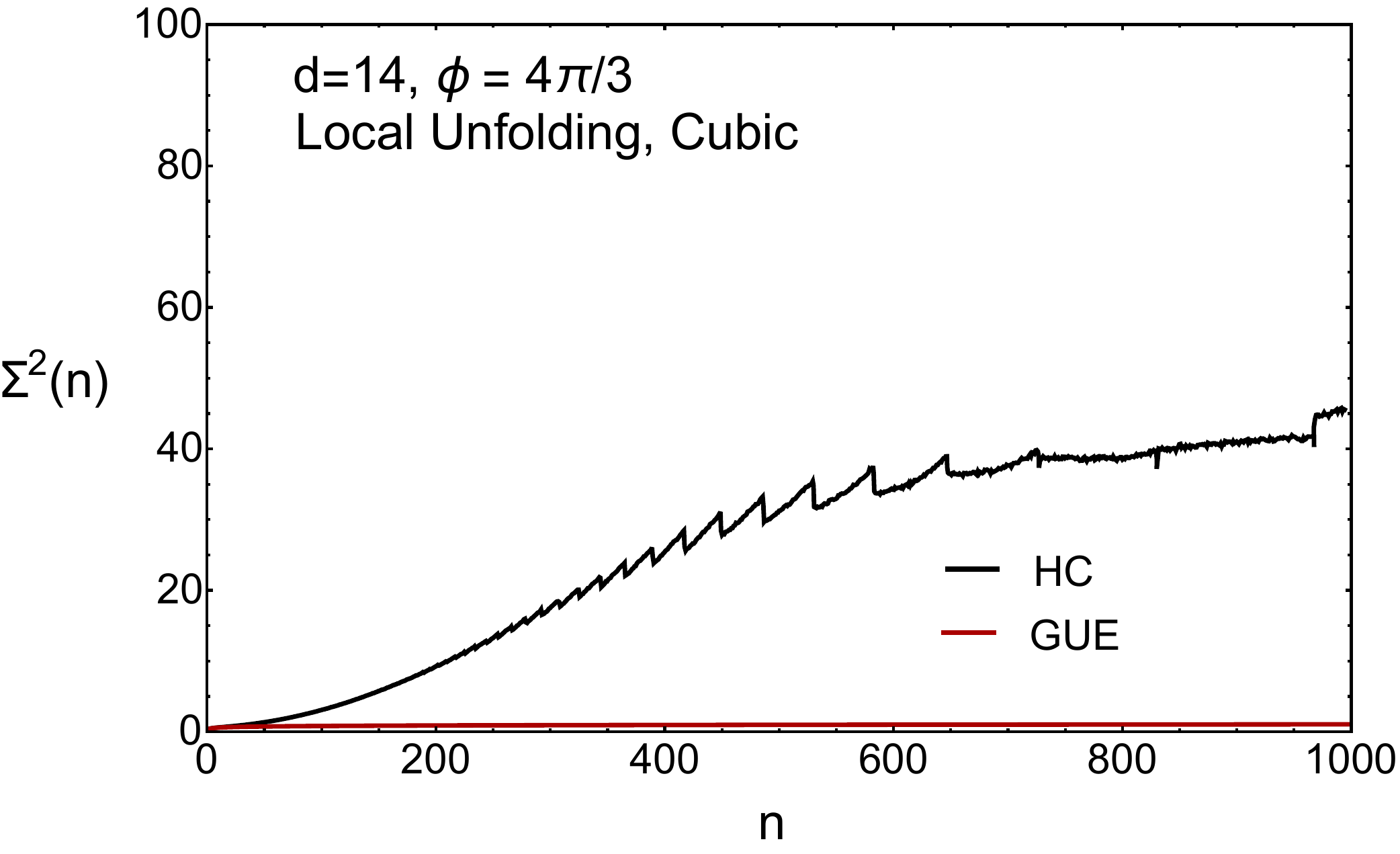}}
  
  \caption{The number variance $\Sigma^2(n)$ versus the average number of eigenvalues $n$ in
    the intervals for $d=14$. The values of the fluxes are indicated in the legends of the figures.
    The results have been obtained by unfolding the spectral for each realization
    separately (local unfolding). For $\phi = 0.57 \pi/2$, the curves are indistinguishable
    from the results for ensemble unfolding.
  The left figures give the same curves as the right figures but for a smaller range of $n$.}\label{fig:num-local}
 \end{figure}

In the SYK model the spectral correlations show agreement with random matrix theory for a distance of about $2^{N/2}/N$
level spacings if the fluctuations from one realization to the next one are eliminated. If we include those
fluctuations, the range of agreement is reduced to $O(N^2)$ which can be easily understood by analyzing the effect
of overall scale fluctuations due to the fact that the number of independent random variables is only of order
$N^4$ \cite{Flores_2001,Altland:2017eao,Garcia-Garcia:2018ruf,Gharibyan:2018jrp,Jia:2019orl} while the number of eigenvalues is $2^{N/2}/2$.
In the hypercubic model, the first six moments are independent of the realizations,
and  fluctuations of
the overall scale and low-order moments are mostly absent.
The sixth order Q-Hermite result already gives a very
accurate description of the average spectral density for values of $ \frac \pi 4 <
\phi < \pi/2$.
Indeed for $\phi=0.57 \pi/2$, there is very little difference in the statistical spectral
observables between local unfolding, where the spectral density of each realization is fitted to a smooth curve,
and unfolding with the ensemble-averaged spectral density.
In the left column of figure \ref{fig:numvar57} we show the number variance $\Sigma^2(n)$ versus the average number of levels $n$ in an interval for $n$ up to 50, and in the right column  (black curves)
up to 1000. In figure \ref{fig:num-local} we show the same quantities but with local unfolding.
We compare these results to the analytical expression for the Gaussian Unitary Ensemble (red curve). Deviations from the universal random
matrix curve start at $n \approx d$. This is in agreement with the observation that the
hypercubic Hamiltonian is determined by $O(d^2)$ random variables so  that the relative
fluctuations in $a_8$ and higher order expansion coefficients are of order $1/d$.
 \begin{table}
   \begin{center}
  \begin{tabular}{|c|c|c|c|c|c|}
    \hline$\cos \phi $ & $\eta $ & $a_6$ &$ \langle a_8 \rangle $  &
    $  \langle (\delta a_8)^2\rangle^{1/2} $ &$\langle a_8\rangle^{\text{fit}}$\\
    \hline
    0.6252 &0.509   &-0.0010 & $-4.16\times10^{-4}$ & $2.66\times10^{-4}$&
    $-3.72\times 10^{-4}$\\
    \hline
    1/13 & 0& -0.0059& -0.00237 & 0.0273& $-3.68\times 10^{-4}$\\
    \hline
   $ -1/\sqrt 2$ & -0.728&  -0.0038 & -2.003 &0.086&-1.80\\
   \hline
  \end{tabular}
  \end{center}
  \caption{Collective spectral fluctuations as measured by the coefficient
    $a_8$ in the expansion of the spectral density in Q-Hermite polynomials. The
  fitted value of $a_8$ is within the range of these fluctuations.}
    \label{tab:one}
  \end{table}
The
fluctuations of the number of levels in an interval containing $n$ levels on average is thus
$\delta n/n \sim O(1/d)$ resulting in a correction to the number variance that behaves as
$n^2/d^2$.
The results for $\phi= 0.57 \pi/2$
are significantly closer to the random matrix result than those for the other
values of $\phi$. For the first ($\phi=0.57\pi/2$) and second row ($\phi =\arccos (1/13)$) of figure \ref{fig:numvar57}  we used the ensemble average of
the eighth order Q-Hermite
result to unfold the spectral density, while for the third row ($\phi=\frac 34 \pi$) a third
order polynomial fit to the ensemble average of the spectral density
was used to unfold the bulk of the spectrum.

\begin{figure}[t!]
  \centerline{\includegraphics[width=8cm]{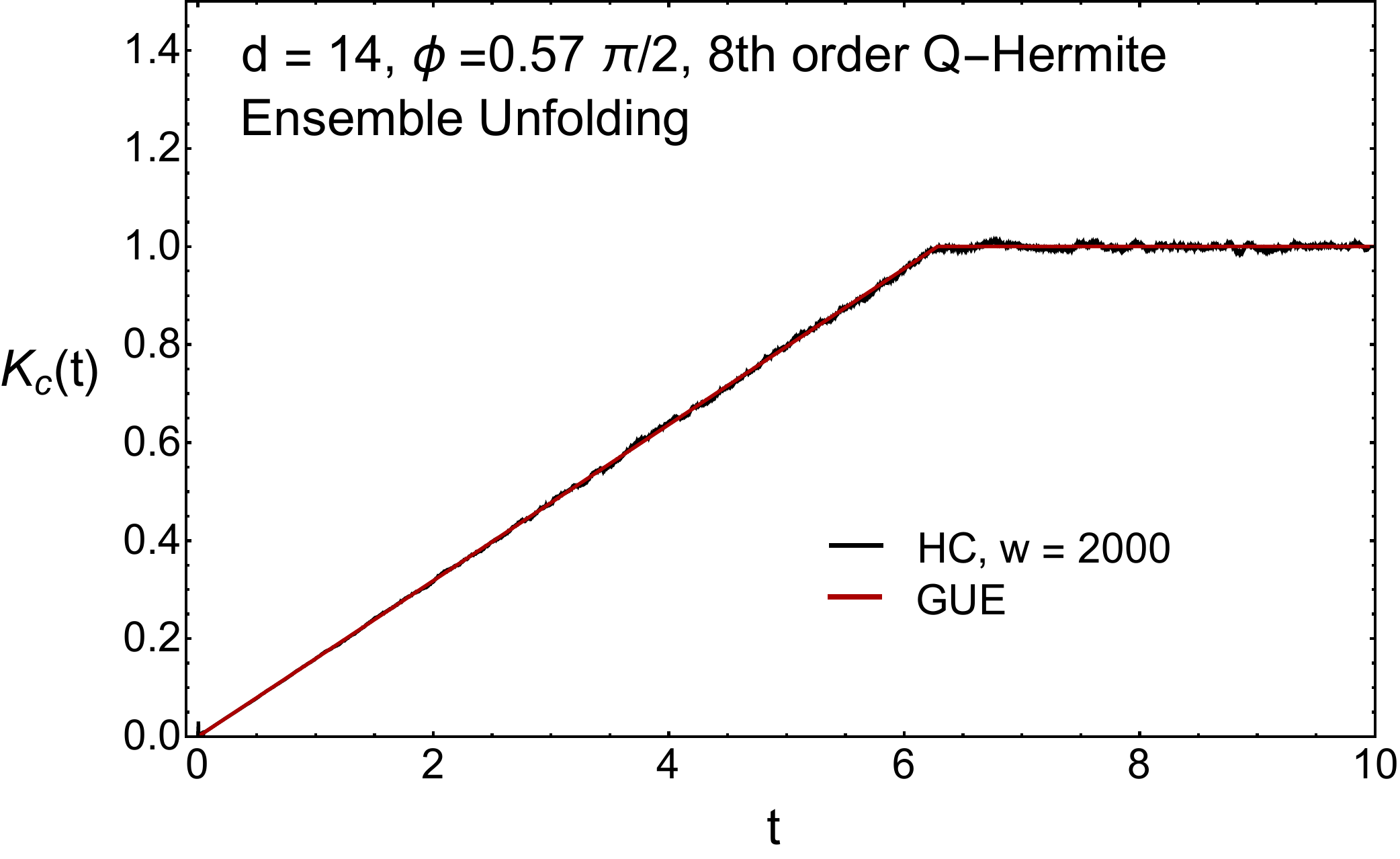}
    \includegraphics[width=8cm]{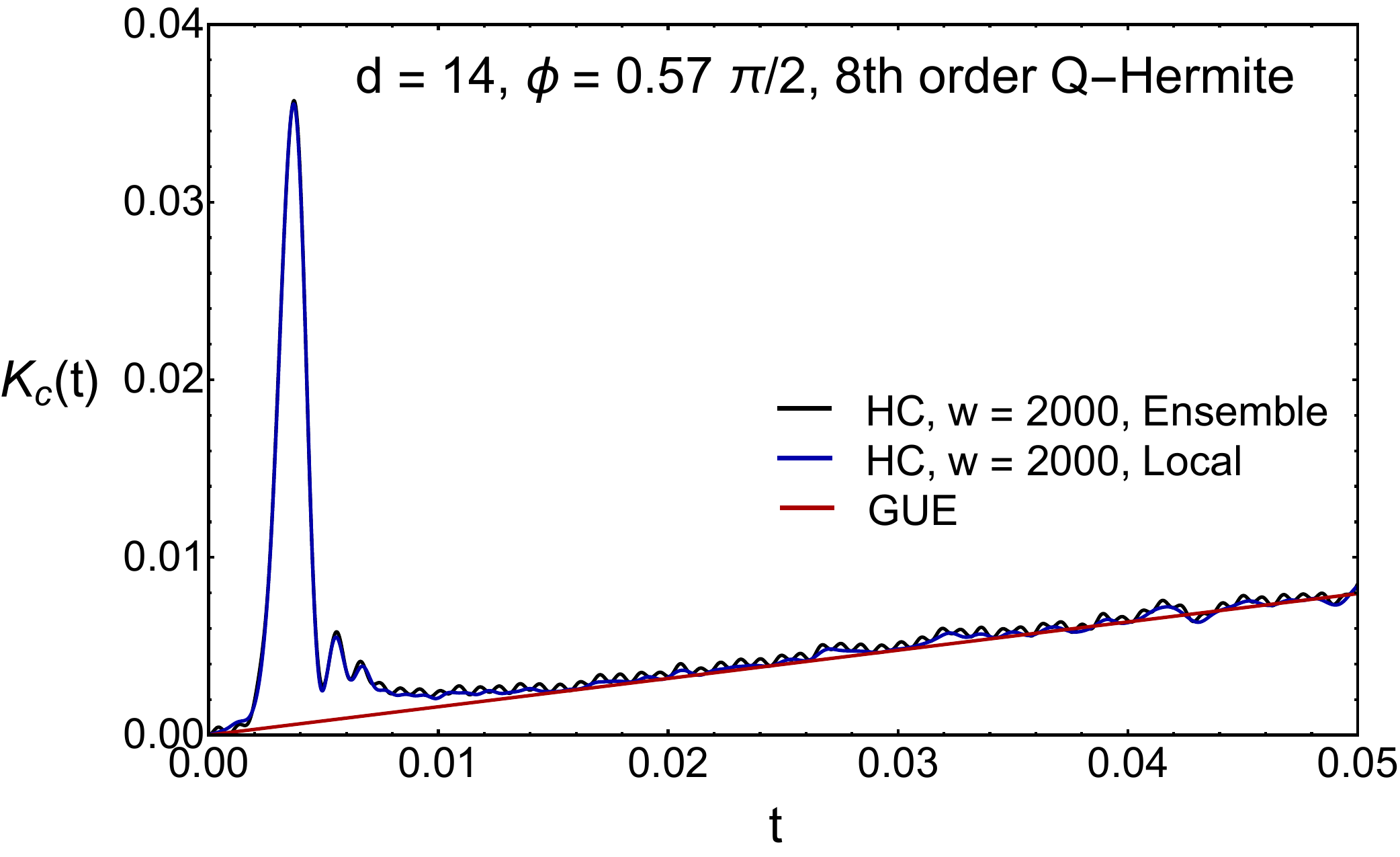}}
   \centerline {\includegraphics[width=8cm]{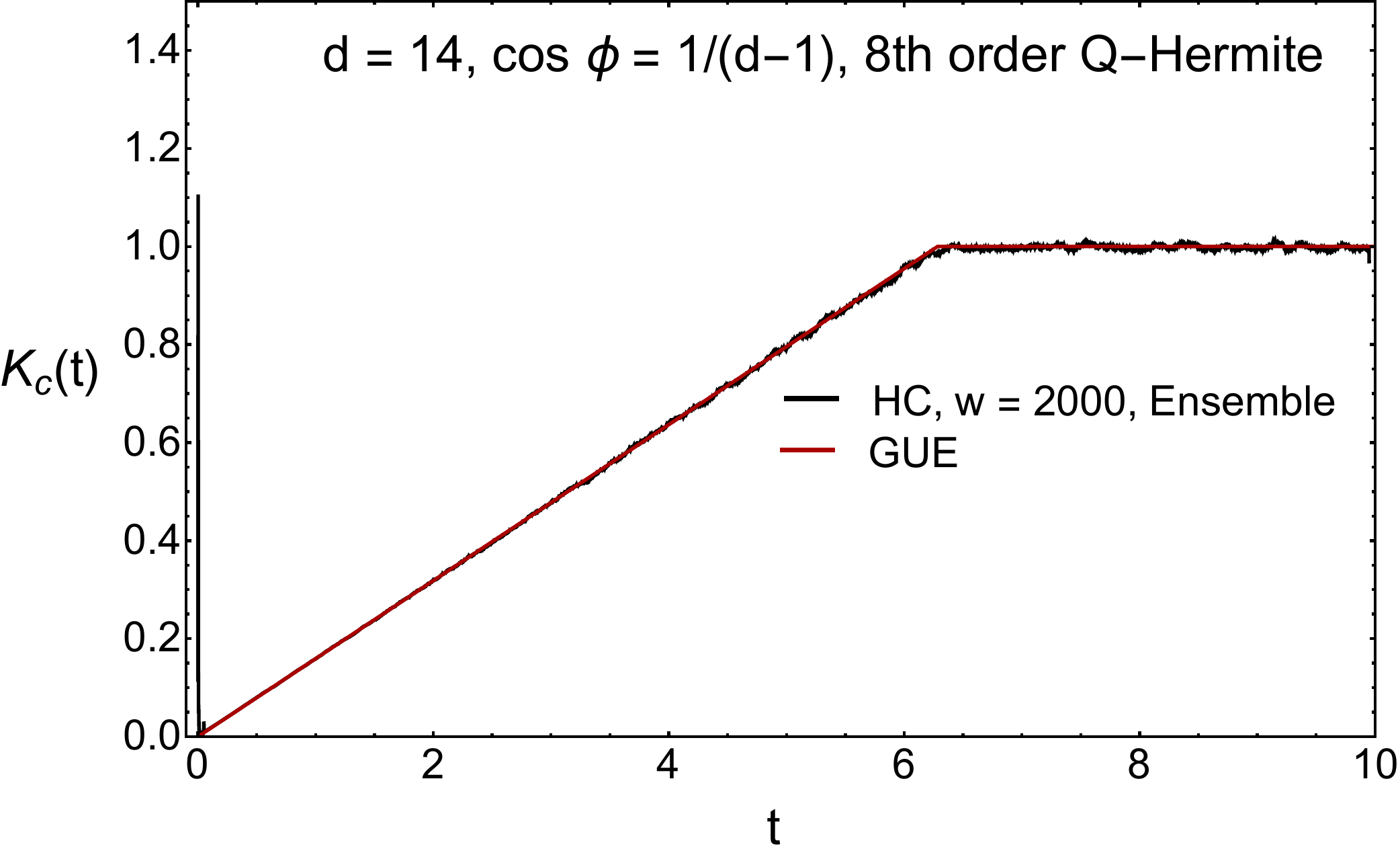}
   \includegraphics[width=8cm]{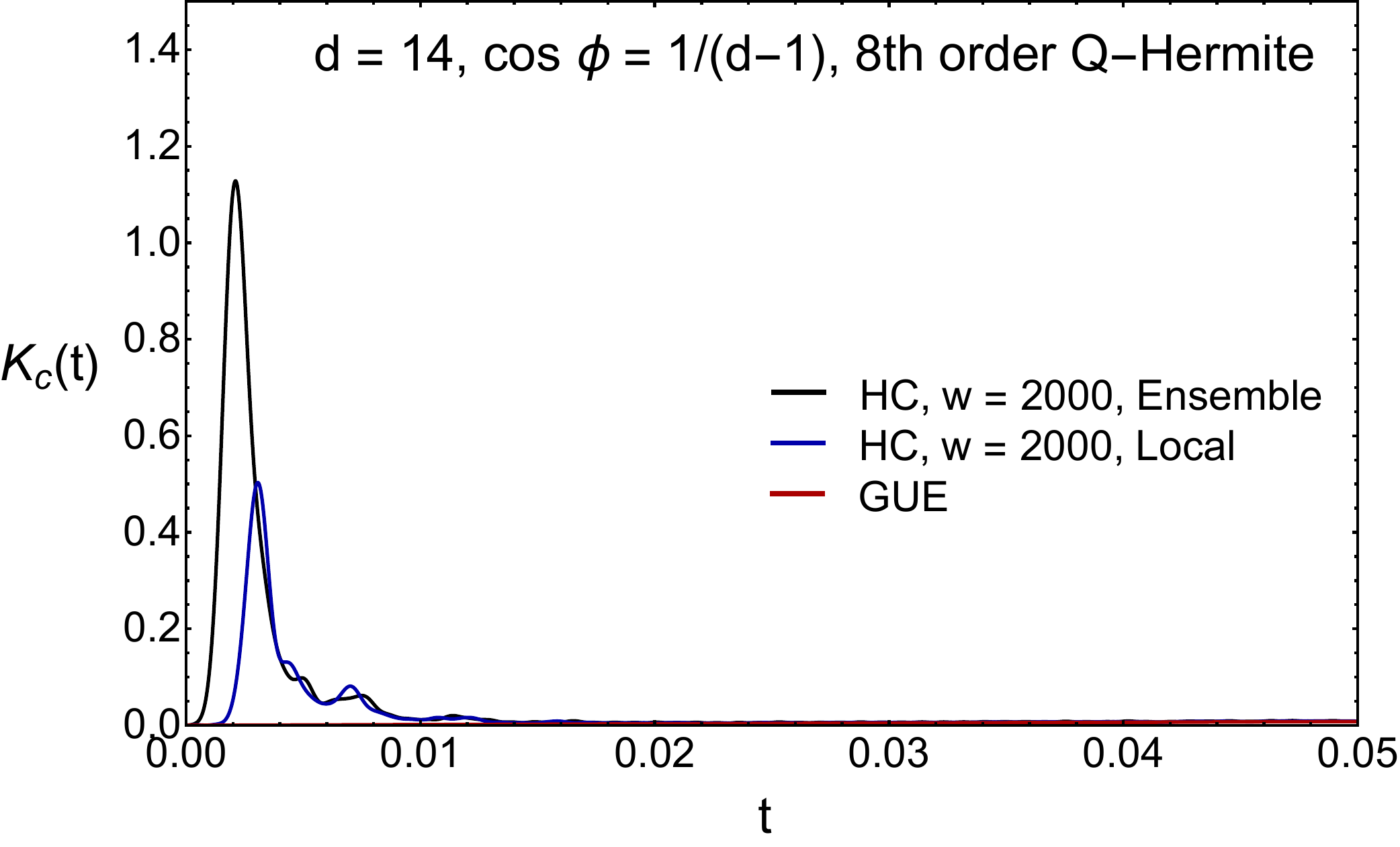}}
\centerline {\includegraphics[width=8cm]{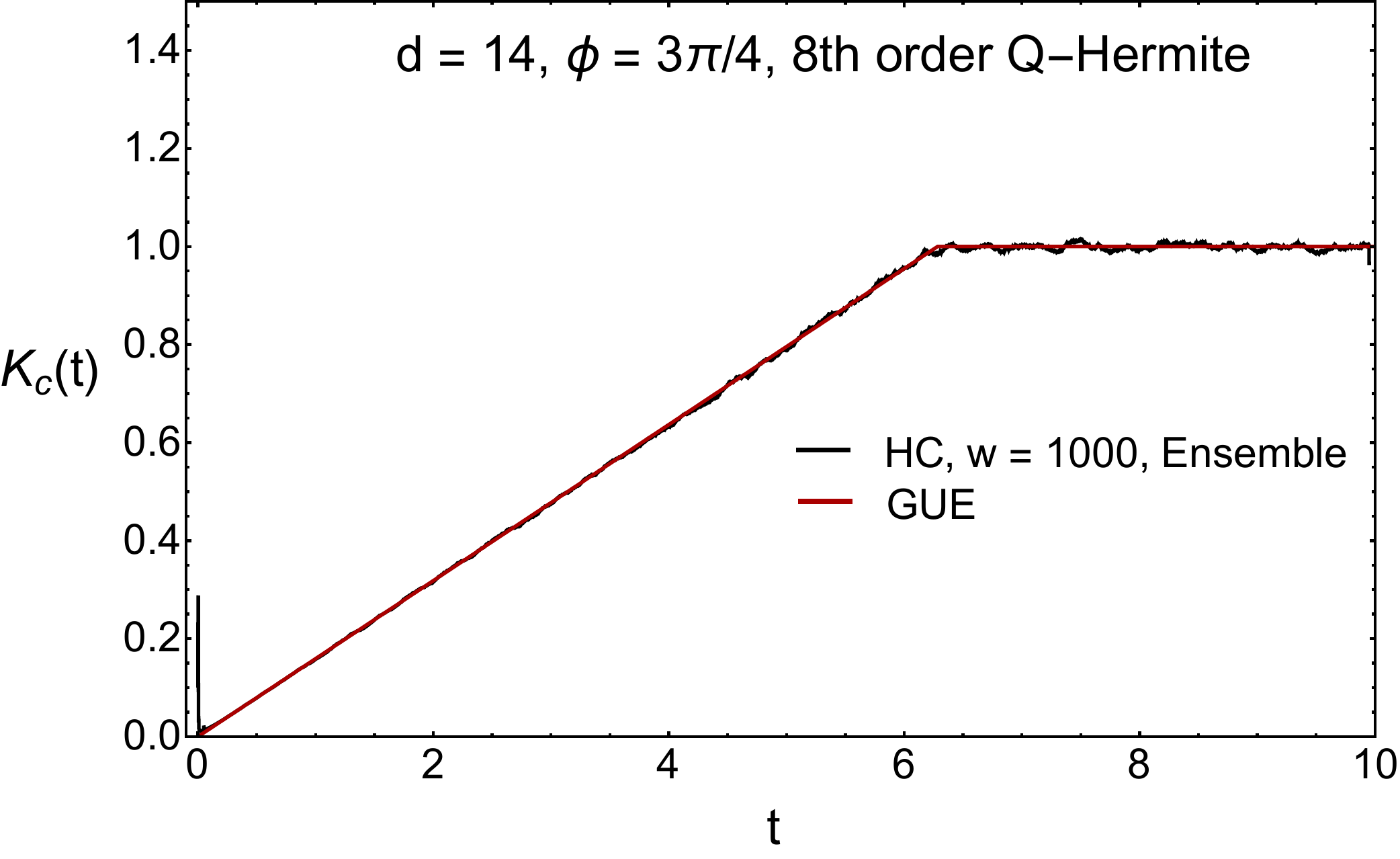}
   \includegraphics[width=8cm]{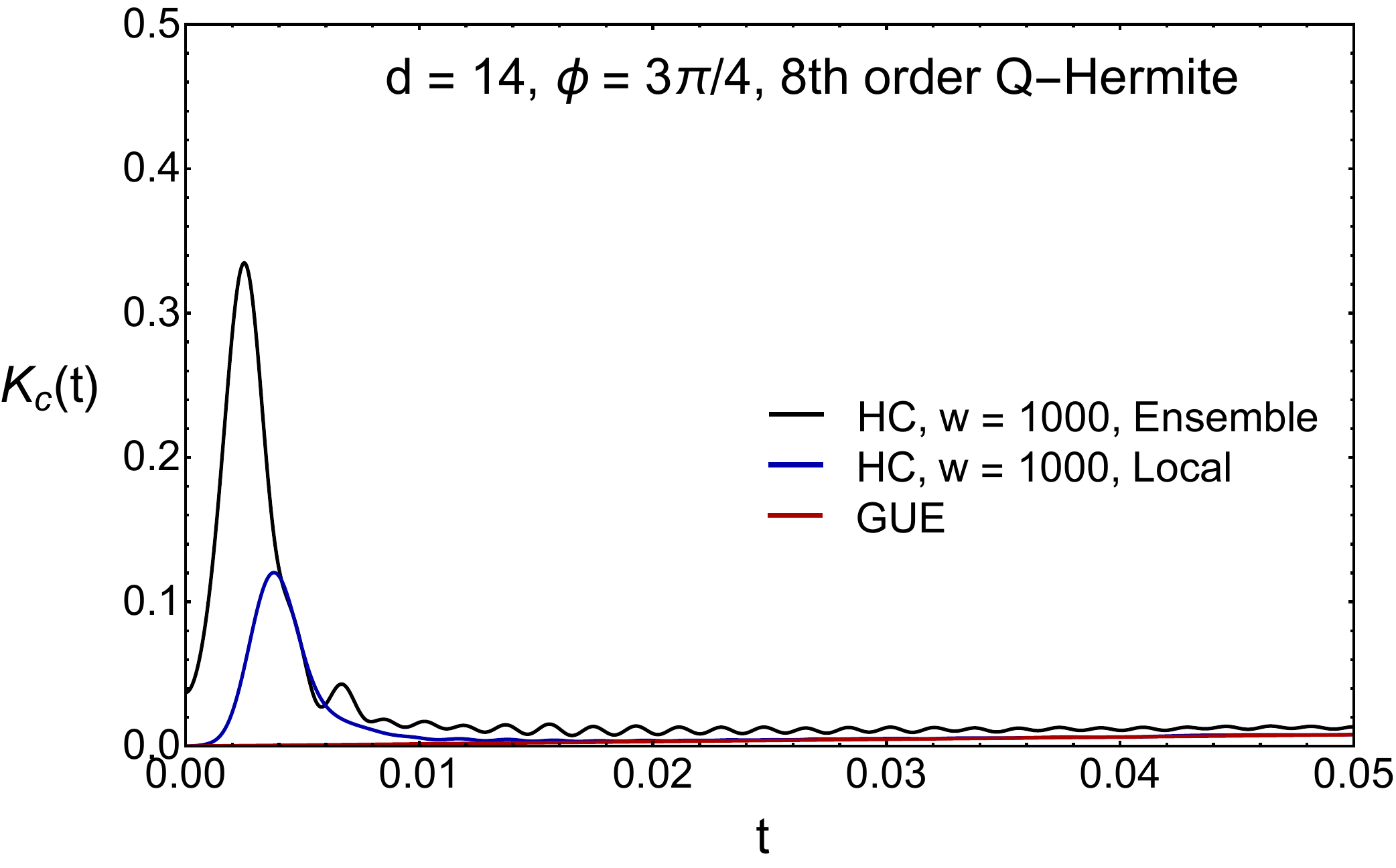}}
  \caption{The ensemble-unfolded spectral form factor for $d=14$ at $\phi = 0.57 \pi/2$, $\phi=0.95 \pi/2$ and $\phi=3\pi/4$ (black curves). The results are compared with the result for the GUE (red curves). In the
  left figure, a careful observer can see a tiny peak at $\tau$ close to zero
  which is responsible for the large deviation of  the number variance from the universal GUE result. This peak is magnified in the right figure (black curve), where we also show the result for local unfolding (blue curve). 
  Ensemble unfolding and local unfolding give almost indistinguishable results for $\phi=0.57 \pi/2$. If the results of local unfolding were plotted in the left figures, the differences with the ensemble unfolding results would not be visible for any of the three figures.}  \label{fig:form57}
  \end{figure}
The difference between the results for ensemble  unfolding and local unfolding is due
to the fluctuations of $a_8=\langle a_8\rangle +\delta a_8$.
Table \ref{tab:one} contains the results for  the simulation parameters of the above figures. We conclude that for $\phi = 0.57 \pi/2$ the collective fluctuations only
contribute a negligible amount to the spectral fluctuations,  while they are important
for $\phi=\arccos(1/13)$ and $\phi=3\pi/4$.

The deviations from the universal RMT result
are barely visible in the spectral form factor (see the left column of figure \ref{fig:form57}), where the results for
the hypercube model (black curve) agree very well with the GUE result (red curve) except for a very narrow peak for $t$ close to zero. To reduce finite
size effects, the spectral form factor is calculated using a Gaussian
window of width 2000 for $\phi=0.57 \pi/2$ and $\phi =\arccos(1/13)$; for $\phi = 3\pi/4$, where the range of the spectrum that can be reliably
unfolded is smaller,
the  width is taken to be $500$. For $ \phi =0.57\pi/2$ local unfolding and
ensemble unfolding give almost identical results (see upper right figure
of figure \ref{fig:form57}), while for the other values of $\phi$ in this
figure, there are significant reductions of the  small time peaks
for local unfolding (blue curves). This suggests the  moments that
are responsible for the early-time peak are much beyond the eighth order,
and more so for $\phi=0.57 \pi/2$ than larger values of $\phi$. Indeed,
as we have shown in section \ref{sec:sumRules}, there is no fluctuation up
to the sixth moment, so that the first moment that can fluctuate
is the eighth moment. In this light it is perhaps not too surprising
that the eighth-order local unfolding does not reduce the fluctuations very significantly.
It is instructive to contrast this phenomenon in the HC model to
its counterpart in the SYK model \cite{Jia:2019orl}, where the eighth-order
local unfolding is quite adequate to remove the early-time peak that is present in the ensemble-unfolded spectral form factor.
The early-time peak  is 
responsible for the deviation from the random matrix result in terms of the number variance. This can be shown
explicitly by calculating the number variance directly from the spectral
form factor with and without this peak 
using the relation \cite{delon-1991} 
\be
\Sigma^2(n)=  \frac {n^2}{2\pi} \int_{-\infty}^\infty dt K(t)
\left(  \frac{\sin(nt/2)}{nt/2}\right )^2.
\ee
Note the derivation of this relation assumes translational invariance of the
spectral correlations which is not the case close to the center of the spectrum for a chirally symmetric spectrum.

\begin{figure}
\centerline{\includegraphics[width=8cm]{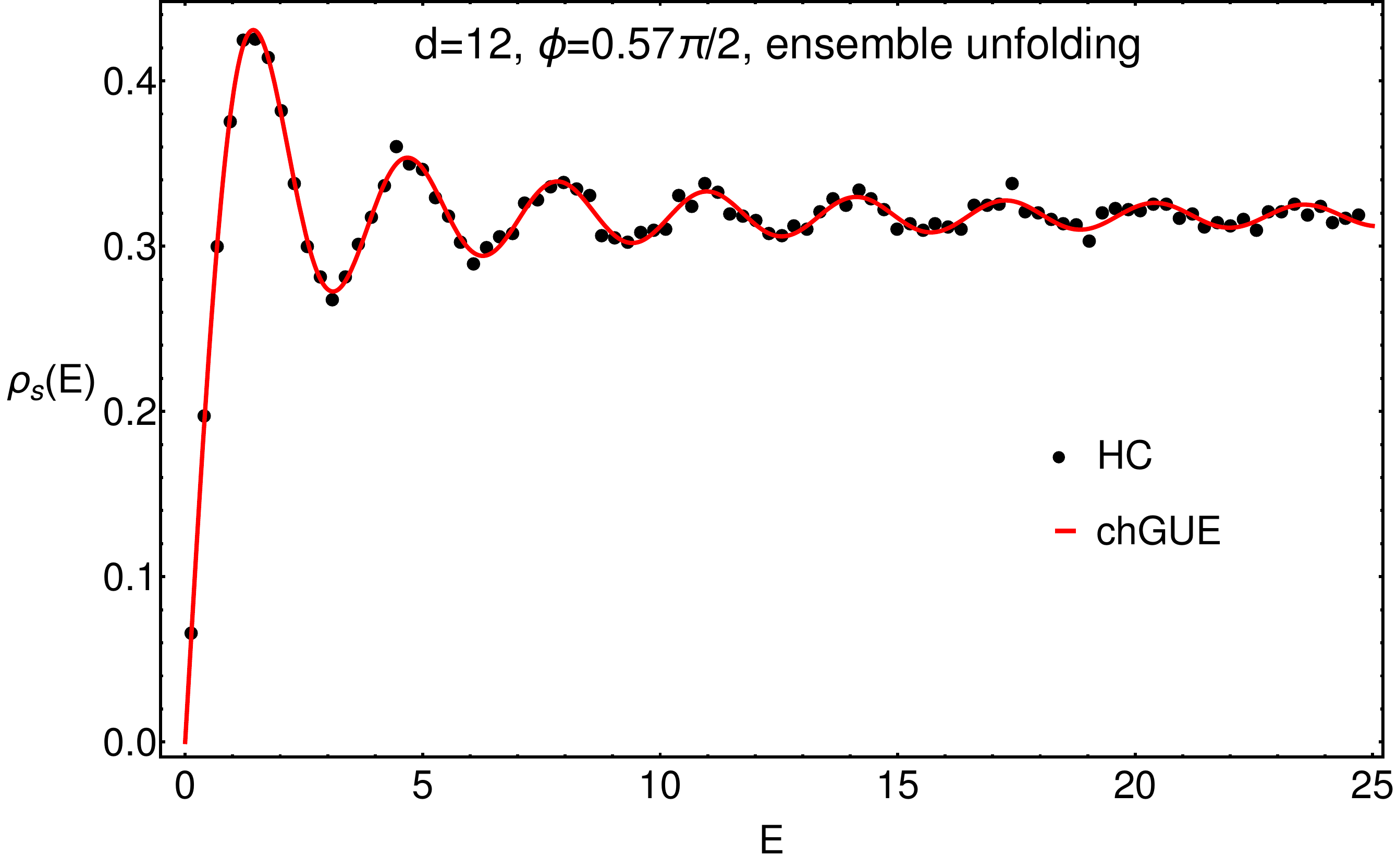}}
\caption{The microscopic spectral densities. Black dots: numerically calculated from 10,000 realizations of the $d=12, \phi = 0.57\pi/2$ Hamiltonian. Red curve: analytically predicted by the chGUE random matrix theory.}
\label{fig:micro}
\end{figure}

Since we deal with a bipartite lattice the Hamiltonian has a chiral symmetry,
and the eigenvalues correlations are in the universality class of chiral
Random Matrix Theory \cite{Verbaarschot:1994qf}, specifically the chiral Gaussian Unitary Ensemble (chGUE) since the system does not have any anti-unitary symmetry. The chGUE ensemble is characterized by an oscillatory structure
in the spectral density near zero on the scale of the average level spacing, and we call the spectral density in this regime the \textit{microscopic spectral density}.
The  microscopic spectral density is defined by \cite{Shuryak:1992pi}
\be
\rho_s(E) = \frac 1{\Sigma N}\rho\left( \frac E{\Sigma N}\right ),
\ee
where\footnote{This $\Sigma$ is not to be confused with the number variance $\Sigma(n)$.}
\be
\Sigma =\lim_{\lambda\to 0}\lim_{N\to \infty}\frac {\pi \rho(\lambda)}N
\ee
and $N$ is a parameter that counts the total number of eigenvalues such as the
size of the random matrix. For an overview of chiral Random Matrix Theory and its applications to lattice QCD
we refer to \cite{Verbaarschot:2000dy}. In the case of hypercube model $N=2^d$ and $\rho(\lambda) = \vev {\rho^{HC}(\lambda)}$.
In figure \ref{fig:micro} we show the microscopic spectral density for an ensemble
of 10,000 Hamiltonians for $d=12$ and $\phi=0.57 \pi/2$ (black dots). The result is compared
with the analytical result for the chGUE microscopic spectral density  
(red curve) \cite{Verbaarschot:1993pm}:
\be
\rho_s(E)=\frac E2 (J_0^2(E) + J_1^2(E)),
\ee
where $J_n(E)$ are the Bessel functions. We remark that there is no fitting and the agreement is excellent.

The chiral symmetry also affects the number variance, but the effects are negligible unless
the intervals for which the number variance is calculated are chosen symmetrically about
zero. The correlations due to the pairing $\pm \lambda_k$ are also visible in the short
time behavior of the form factor. Instead of $K_c(t) \sim t^2$ for the GUE we have
$K_c(t) \sim t^4$ for the chGUE, when $t\to 0$ and the matrices have finite size.
However, the peak near zero in the numerical
results obscures this effect. The number variance of the chGUE is reduced by
a factor 2 (in the domain where $\Sigma^2(n) \sim \log n$) for intervals that are symmetric about zero \cite{Toublan:2000dn}.
However, because we calculate the number variance by
spectral averaging over the spectrum, this has only a small effect except when
$n$ becomes large. In fact the kinks in the number variance for $n > 400$
are due to this effect.

\section{Thermofield double state}\label{sec:tfd}

In this section we construct the ThermoField Double (TFD) state corresponding the ground
state of the hypercubic model. Whether or not the ground state is a TFD state is a basis-dependent
statement, and we have to identify an appropriate basis. Inspired by the Maldacena-Qi
model we use the sum of a left  SYK model and a right SYK model to construct a basis, and in this case 
we illustrate our construction by choosing a two-body Hamiltonian.  We remark
that in the MQ model, ``left'' and ``right'' refer to the two sides of a
worm hole, and quantum mechanically this translates to the fact the elementary
fermion operators factorize into tensor products in a product Hilbert space.
In this paper we do not dwell on the space-time interpretations of the
HC model, so we use the terms simply to  refer to the tensor product structure. General arguments to construct
a TFD state are given in \cite{cottrell:2018ash}, and applications of the TFD state can be
found in \cite{Maldacena:2001kr,delCampo:2017ftn}.
In this section mostly focus on the zero flux case which can be analyzed
analytically. At nonzero flux, the ground state can only be obtained
numerically, and is compared to a TFD state at the end of this section.

The first observation is that the coupling of the Maldacena-Qi model is equivalent to the Parisi Hamiltonian
at zero flux, which can be expressed in terms of the gamma matrices defined in equation \eqref{gamma-hc}. We thus have
\be\label{eqn:HsimilarityTrans}
H(\phi = 0) =i \sum_{k=1}^d \gamma_k^L\gamma_k^R = U H_{MQ}U^{-1}
\ee
with
\be
H_{MQ}=i\sum_{k=1}^d \tilde \gamma_k^L \tilde \gamma_k^R ,
\label{ham-mq}
\ee
where the gamma matrices $\tilde \gamma_k^{L(R)}$ are in a representation that was used in \cite{Garcia-Garcia:2019poj} to prove
that the ground state of the Maldacena-Qi model is a TFD state. Specifically,
   \be\label{eqn:gammaTildeTensor}
   \begin{split}
\tilde \gamma_k^L =&\tilde  \gamma_k \otimes 1, \quad k=1,2,\ldots, d/2,\\
\tilde \gamma_k^R =& \tilde \gamma_c \otimes \tilde \gamma_k,  \quad k=1,2,\ldots, d/2,
\end{split}
\ee
where $\tilde{\gamma}_k$ are Dirac matrices in $d/2$ dimensions and $\tilde{\gamma}_c$ is the corresponding chirality Dirac matrix. For this construction to work we need $d/2$ to be even, namely $d$ is a multiple of $4$. The $\tilde\gamma^L$ and $\tilde\gamma^R$ matrices can be obtained by a permutation of the $\gamma^L$ and $\gamma^R$ matrices in equation \eref{gamma-hc} as follows:
\begin{equation}\label{eqn:gammaTildePermute}
\begin{split}
\tilde{\gamma}^L_{2k-1} & = \gamma^L_k, \ \tilde{\gamma}^L_{2k} = \gamma^R_k,\\
 \tilde{\gamma}^R_{2k-1} & = \gamma^L_{\frac{d}{4}+k}, \ \tilde{\gamma}^R_{2k} = \gamma^R_{\frac{d}{4}+k},
 \end{split}
\end{equation}
for $k=1,2,\ldots, d/4$. Then we can check the $\tilde\gamma_k$ matrices in equation \eqref{eqn:gammaTildeTensor} take the form:
\begin{equation}\label{eqn:gammaTildeExplicit}
\begin{split}
\tilde{\gamma}_{2k-1} &= \overbrace{\sigma_1 \otimes\cdots\otimes\sigma_1}^{k}  \otimes\sigma_{3}\otimes\overbrace{\sigma_0 \otimes\cdots\otimes\sigma_0}^{\frac d2-k-1},\\
\tilde{\gamma}_{2k} &= \overbrace{\sigma_1 \otimes\cdots\otimes\sigma_1}^{k}  \otimes\sigma_{2}\otimes\overbrace{\sigma_0 \otimes\cdots\otimes\sigma_0}^{\frac d2-k-1},
\end{split}
\end{equation}
for $k=1,2,\ldots, d/4$, and
\be\label{gamma-c}
\tilde \gamma^c = \overbrace{\sigma_1 \otimes \cdots\otimes \sigma_1}^{\frac d2}.
\ee
Since both $\{\gamma^{L}_k,\gamma^R_k\}$ and $\{\tilde\gamma^{L}_k,\tilde\gamma^R_k\}$ are Hermitian representations of the Clifford algebra in even dimensions, the similarity transformation $U$ in equation \eqref{eqn:HsimilarityTrans} that relates the two is unitary.
In the Maldacena-Qi model, the basis of the TFD state is constructed from the Hamiltonian
\be
H_{SYK}^R + H_{SYK}^L = \sum_\alpha J_\alpha (\tilde \Gamma_\alpha^R +\tilde\Gamma_\alpha^L),
\label{ham-kmq}
\ee
where $\tilde\Gamma_\alpha^{L(R)}$ is a product of $q$ different $\tilde \gamma^{L (R)}$ matrices, $\alpha$ is the set of $q$ indices of these gamma matrices, and $J_\alpha$ is the Gaussian-random coupling.\footnote{Note again that here $q$ is an integer in the SYK model, independent of the HC model's flux parameter.} It is
important that the left and right Hamiltonian share the same coupling $J_\alpha$. Because of the tensor
structure of the Hamiltonian it is clear that the eigenstates of this Hamiltonian are given by
\be
|m\rangle \otimes |n\rangle
\ee
with eigenvalues $E_m+E_n$. Here, $|m\rangle $ are the eigenstates of $H^L$ projected
onto the left space.
In this basis, the thermofield double state at inverse temperature $\beta$ is given by
\be
|{\rm TDF}\rangle= 2^{-d/4}\sum_m  e^{-\beta E_m/2}|m\rangle
| C^R e^{\frac \pi4 i \tilde \gamma_c} K m\rangle
\label{tfd}
\ee
with $C^R$ the charge conjugation matrix,
   \be 
      {C^R}^\dagger \tilde \gamma_k^R C^R =  \tilde{\gamma}_k^{R*} ,
      \ee
and $K$ the complex conjugation operator.
In a  convention where  gamma matrices $\tilde \gamma_{2k}$ are purely imaginary while
   the $\tilde \gamma_{2k-1}$ are purely real like in equation \eqref{eqn:gammaTildeExplicit}, we have that
      \be
   C^{L(R)} =\prod_{k=1}^{d/4} {\tilde \gamma_{2k-1}^{L(R)}}.
   \ee

   The argument to show that the ground state of the Hamiltonian $H_{MQ}$
   is given by the TFD state at $\beta =0$
does not depend on the details of the Hamiltonian \eqref{ham-kmq} that
determines the basis states \cite{Garcia-Garcia:2019poj}, for example it does not matter if we use a 2-body, 4-body or 6-body SYK model Hamiltonians. This follows
from the expectation value
      \be
     &&2^{-d/2} \sum_{mn}\langle   m|\langle C^Re^{\frac \pi4 i \tilde\gamma_c} K  m| i \sum_k \tilde \gamma_k^L  \tilde \gamma_k^R|n\rangle  |C^Re^{\frac \pi4 i \tilde \gamma_c} Kn \rangle\nn\\ 
        &=& 2^{-d/2}
        \sum_{mn}\langle m|{\tilde \gamma_{k}} \tilde \gamma_c|n\rangle
        \langle K m|e^{-\frac \pi4 i \tilde \gamma_c} {C^R}^\dagger i{\tilde \gamma_{k}}C^R
        e^{\frac \pi4 i \tilde \gamma_c}|K n\rangle\nn\\
         &=& 2^{-d/2}
        \sum_{mn}\langle m|{\tilde \gamma_{k}} \tilde \gamma_c|n\rangle
        \langle K m|\tilde \gamma_c{{\tilde \gamma_{k}}}^* | K n\rangle\nn\\
         &=&  2^{-d/2}
        \sum_{mn} \langle m|{\tilde \gamma_{k}} \tilde \gamma_c|n\rangle
        \langle n| {\tilde \gamma_{k}} \tilde \gamma_c|m\rangle,
        \ee
 where in going from the second line to the third line, we have used the fact that
 \begin{equation}
\tilde \gamma_c^*=\tilde \gamma_c, \quad i e^{-i\frac{\pi}{2}\tilde \gamma_c}=\tilde \gamma_c.
 \end{equation}
Now we can use completeness to do the sum over $n$, and  employ
that the gamma matrices square to 1,  we then see the sum over
        $k$ yields a factor $d$ resulting in 
        \be
        \langle \text{TFD} |  i \sum_{k=1}^d \tilde \gamma_k^L  \tilde \gamma_k^R |\text{TFD} \rangle = -d.
        \ee
        Since $-d$ is the ground state energy and the ground state is nondegenerate, the TFD state must be the ground state.

\begin{figure}[t!]
          \centerline{
            \includegraphics[width=8cm]{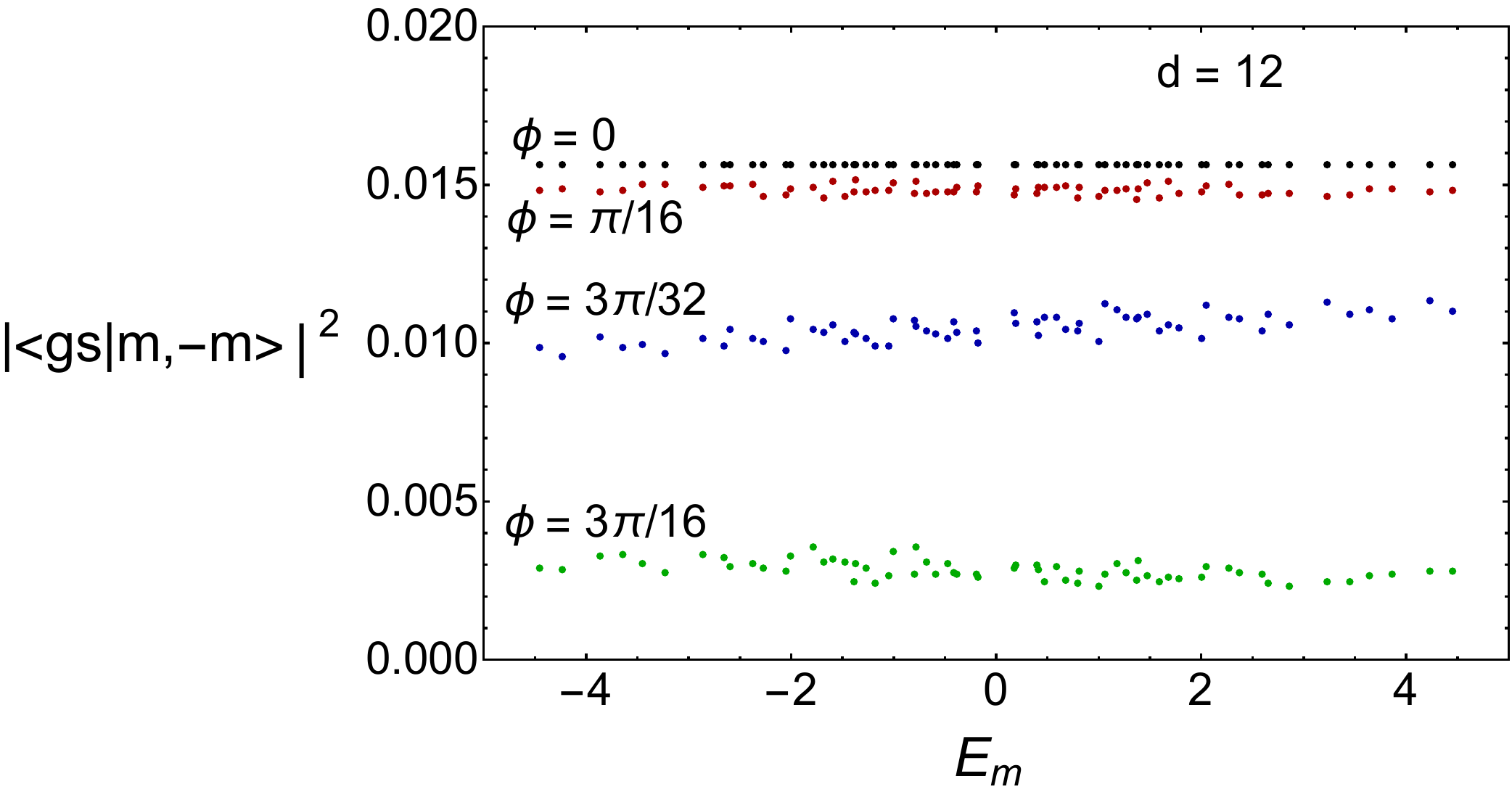}
            \includegraphics[width=8cm]{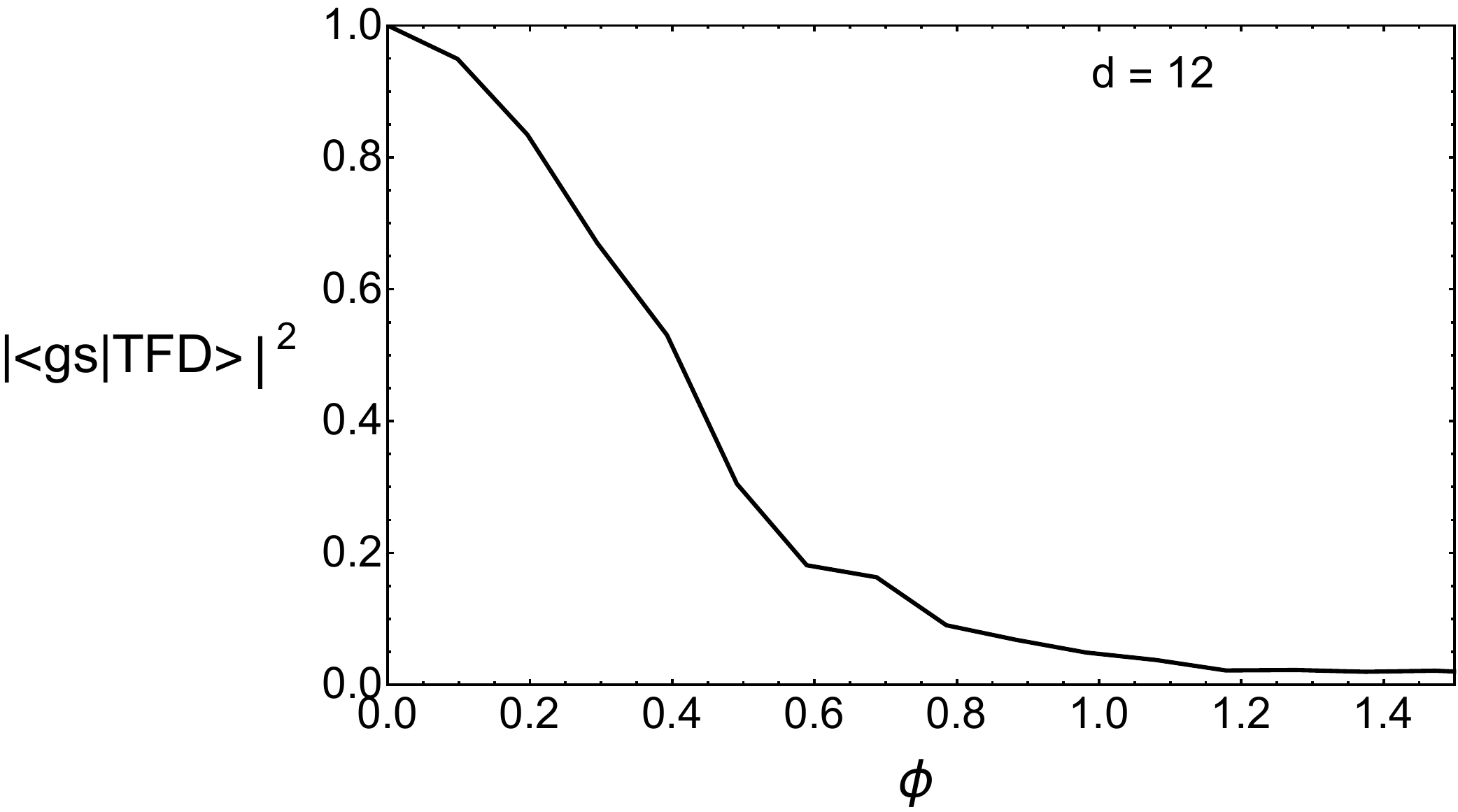}}
            \caption{The overlap of the ground state of the HC model
              with the components of the TFD state (left) for several values
              of the magnetic flux. In the right figure we show the total overlap between the ground state of the HC model and the TFD state.
              Because the zero energy states \eref{ham-lr} are degenerate we have
              a small symmetry breaking term to this Hamiltonian by
              $i\gamma_k^L\gamma_l^L\to i\gamma_k^L\gamma_l^L(1+\epsilon)$. For
              the figures above we used $\epsilon =10^{-4}$.
            }
            \label{fig:tdf}
\end{figure}

        To illustrate the above argument, we choose the two-body SYK Hamiltonian
        \be
        H_{SYK}=\sum_{k<l} J_{kl}\left (i\gamma_k^L\gamma_l^L+i\gamma_k^R\gamma_l^R \right),
\label{ham-lr}
        \ee
        to determine the basis states entering the TDF state and consider the overlap
        with the ground state of
\be
H(\phi=0)=i\sum_k\gamma_k^L\gamma_k^R.
\label{ham-hc}
\ee
The gamma matrices in  both Hamiltonians are in the representation \eref{gamma-hc}.
Since the overlap between states is invariant under a unitary transformation, we can do the unitary transformation $U$ in equation \eqref{eqn:HsimilarityTrans} to transform the Hamiltonians \eref{ham-lr} and \eref{ham-hc}
into the Hamiltonians \eref{ham-kmq} and the coupling matrix in the right-hand side
of \eref{ham-mq}, respectively.
Using the above argument, the ground state of \eref{ham-hc} is given by
\be
U^{-1} |\text{TFD}\rangle=  U^{-1} 2^{-d/4}\sum_m |m\rangle
        | C^R e^{\frac \pi4 i \tilde \gamma_c} K m\rangle.
        \ee
        Since for even $d/2$ the anti-commutator $\{C^R K,H^R_{SYK}\}=0$, if $|m\rangle$ is an eigenstate of $H^R_{SYK}$ with eigenvalue
        $E_m$, then  $C^R K|m\rangle$ is an eigenstate of $H^R_{SYK}$ with eigenvalue $-E_m$. The ground state of
        \eref{ham-hc} is thus a linear combination of the zero energy states of \eref{ham-lr}.
        In figure \ref{fig:tdf} we show the magnitude of the  overlap of the ground state with the $|m\rangle| C^R e^{\frac \pi4 i \gamma_c} K m\rangle $
       (denoted by $|m\rangle|-m\rangle$
       in the figure) for $d=12$. The total strength in this subspace decreases rapidly with
       increasing magnetic flux, but the temperature of the TFD state remains
       infinite.

        There are other possibilities to choose a basis for a TFD state. For example at zero flux, the Hamiltonian
        may be written as
        \be
        H_d(\phi=0) = H_{d/2}(\phi=0) \otimes 1+ 1\otimes H_{d/2}(\phi=0), 
        \ee
        and a TFD state can be constructed out of the eigenstates of $H_{d/2}$. For $\phi\ne 0$
        the Hamiltonian
        \be
        H_{d/2}(\phi) \otimes 1+ 1\otimes H_{d/2}(\phi)
        \ee
        has its nonzero matrix elements in the same position as the ones of $H_d(\phi)$, and also
        at $\phi\ne 0$ the eigenstates of $H_{d/2}$ could be used to construct a TFD state.
        We have explored these and other related possibilities, but they did not give
        a better description of the ground state of the hypercubic Hamiltonian.

        \section{Conclusions and discussions}
        \label{sec:conclusions}
        We have studied the spectral density and the spectral correlations of Parisi's hypercubic model.
        This model is described by the Laplacian  on a hypercube with two lattice points in each dimension and
        U(1) gauge fields on the links such that the magnitude of the
        magnetic flux through each of its faces is constant, but its orientation is chosen to be
        random.
        We have confirmed that the  spectral density of this model
        is given by the density function of the
        Q-Hermite polynomials. This has the important implication that the spectral
        density above the ground state $E_0$ behaves as $\sinh\sqrt{c(E-E_0)}$. However, contrary
        to the SYK model, the ground state of the hypercubic model is separated from the rest of the spectrum by
        a gap. In this respect, the hypercubic model resembles the Maldacena-Qi model, and we
        expect it to have a similar phase diagram with a first order phase transition as
        a function of the temperature. We hope to address this point in a future
        publication. Remarkably, at zero flux the Hamiltonian of this model coincides
        with the coupling term of the Maldacena-Qi model. We have constructed a basis such that
        in the zero-flux case the ground state is given by a thermofield double state. Contrary to
        the Maldacena-Qi model,
        at nonzero
        flux the overlap with the TFD state rapidly decreases. Since the hypercubic
        Hamiltonian at nonzero flux is not the sum of a left and a right Hamiltonian,
        this did not come as a surprise.

       Though not explicitly stated in the main text, the initial analysis
        of the spectral correlations of this model
        led to
        the observation that they are described by the superposition of
        two Gaussian Unitary Ensembles. This resulted in the discovery of a discrete
        symmetry that we later identified as a magnetic inversion symmetry which is
        analogous to magnetic translation symmetries studied in the literature. Since
        this operator is related to space inversion (which is the same as a translation
        mod 2 on a hypercubic lattice), it squares to unity and its eigenvalues are $\pm 1$.
        We have analyzed the correlations of the eigenvalues of the hypercubic Hamiltonian for
        fixed quantum number of this symmetry and found that they are correlated according
        to the GUE. Since this model is determined by $d^2$ random numbers, the fluctuations
        of the number of eigenvalues in an interval containing $n$ eigenvalues on average
        behave as $\delta n/ n \sim 1/d$, and hence the number variance for large $n$ behaves
        as $\Sigma^2(n) \sim n^2/d^2$ resulting in a ``Thouless energy scale'' of order $d$.
        This is in qualitative  agreement with our numerical results. In the spectral form factor,
        this deviation is visible as a peak close to zero time with area $\sim 1/d^2$, which is only         apparent in plots
        of the {\it connected} form factor (which we always plot). 

        Because of the sublattice symmetry, the Hamiltonian has a chiral
        symmetry with eigenvalues occurring in pairs $\pm \lambda_k$ so that the eigenvalues
        are correlated according to the chiral Gaussian Unitary  Ensemble (chGUE). Indeed
        we have shown that the microscopic spectral density exhibits the universal oscillations
        characteristic for this ensemble. If the number variance is calculated for an interval
        that is symmetric about zero, the chiral symmetry reduces the variance by a factor
        two. Since we calculate the number variance by averaging over the spectrum, this effect
        only affects large values of $n$ where the number variance is dominated by the $n^2$
        correction.

        The traces of powers of the hypercubic Hamiltonian are given by the Wilson
        loops of closed paths on the hypercube. We have extended (in appendix \ref{app:chords}) Parisi's work
        on a one-to-one mapping 
        between these paths and the chord diagrams that occur in the  calculation
        of the moments of both the SYK model and the hypercubic model. This explains why in both cases
        the spectral density  is given by the density function
        of the Q-Hermitian polynomials. This suggests that the low-energy effective
        partition function of the hypercubic Hamiltonian can also be expressed in terms
        of a Schwarzian action. We hope to address this point in a future publication. Moreover, in appendix \ref{app:chords} we developed three chord diagram representations of the subleading moments of the Parisi model. Remarkably, one of the representations (the averaged scheme) coincides with the chord diagram representation of the subleading moments of the sparse SYK model \cite{garcagarca2020sparse, xu2020sparse}, up to an overall factor of three. We end up with the surprising relation 
                \begin{equation}
\text{(sparse SYK moments)}_\text{QH} = \text{Parisi leading} + \frac{1}{k N}\times 3 \times \text{Parisi subleading} +O(1/N^2), 
\end{equation}
where the subscript ``QH'' denotes Q-Hermite approximation,  $N$ is the number of Majorana fermions in the sparse SYK model and $k$ indicates sparseness (smaller $k$ means more sparseness). We note however this relation does not hold to higher orders.
        
        Our work confirms the power of random matrix universality. Although the model is very
        different from a random matrix theory, and in the tensor product representation of section \ref{sec:tensorRep} it describes a many-body theory with a sparse
        Hamiltonian, the level correlations are still very well described by the corresponding
        Random Matrix Theory. This further supports the paradigm, going back to the first
        applications of random matrix theory to the nuclear many-body problem,
        that generically spectra of many-body systems are chaotic. Our chord diagram analysis of the leading and subleading moments reveals a surprising connection between the Parisi's model and the sparse SYK model, which is worthy of further investigation.
Moreover, it makes the  polynomials in $q$ of the subleading moments of Parisi hypercubic model  and the sparse SYK model a more luring mathematical problem -- a complete analysis of those polynomials will deepen our understanding of both  models in one strike.   

        \acknowledgments{
We  acknowledge partial support from U.S. DOE Grant
No. DE-FAG-88FR40388. We acknowledge useful comments by Antonio Garc\'ia-Garc\'ia and Dario Rosa, and their critical readings of the manuscript. We also acknowledge Peter van Nieuwenhuizen for a useful discussion on representations of
gamma matrices.}

     \newpage

 \appendix

 \section{Disorder independence of the fourth and sixth moments in the tensor product representation}\label{app:m4m6Tensor}
\subsection{$\Tr H^4$} \label{app:m4Tensor}

  In this section we calculate  $\Tr H^4$ in the tensor representation of the Hamiltonian. We obtain an explicit expression for the fourth moment.
 In agreement with the geometric picture in the main text, it 
 only depends on the magnitude of the magnetic flux through the faces of
 the hypercube and
is independent of its random orientations.
 
To facilitate the discussion, we define (Here, $\sigma_0$ is the $2\times 2$ identity
matrix, and we refer to equation \eref{rhodef} for the
definition $\rho_x$.)
\begin{equation}
h_{d,\mu}:=\overbrace{\sigma_0\otimes\sigma_0\cdots\otimes\sigma_0}^{d-\mu}\otimes \sum_{x_1,\ldots,x_{\mu-1}} \sigma_{\mu,\vec{x}} \otimes \rho_{x_1}\cdots\otimes\rho_{x_{\mu-1}},
\end{equation}
so that the Hamiltonian of the hypercubic model is given by (see equation \eref{eqn:tensorHami})
\begin{equation}
H_d =\sum_{\mu=1}^d h_{d,\mu} \ .
\end{equation}
The fourth moment can be expressed as
\begin{equation}
\Tr H_d^4 =\sum_{\mu\nu\kappa\omega} \Tr \left(h_{d,\mu}h_{d,\nu}h_{d,\kappa}h_{d,\omega}\right).
\end{equation}
Since each $h_{d,\mu}$ has only one off-diagonal $2\times 2$ matrix in the tensor product, and its position is labeled by $\mu$, the only nonzero traces are of the forms $\Tr h_{d,\mu}^4$, $\Tr \left(h_{d,\mu}h_{d,\mu}h_{d,\nu}h_{d,\nu}\right)$ with $\mu>\nu$, $\Tr \left(h_{d,\mu}h_{d,\nu}h_{d,\mu}h_{d,\nu}\right)$ with $\mu>\nu$, and their cyclic permutations. It is clear that
\begin{align}
\Tr\; h_{d,\mu}^4=&2^d,\label{eqn:fourthMom1stTerm}\\
\Tr \left(h_{d,\mu}h_{d,\mu}h_{d,\nu}h_{d,\nu}\right) =&2^d\label{eqn:fourthMom2ndTerm}.
\end{align}

We now consider $\Tr \left(h_{d,\mu}h_{d,\nu}h_{d,\mu}h_{d,\nu}\right)$ with $\mu>\nu$. Let us first work out $h_{d,\mu}h_{d,\nu}$, it is given by
  \begin{equation}
  \begin{split}
& \sum_{\{x\},\{y\}}\overbrace{\sigma_0\otimes\cdots\sigma_0}^{d-\mu}\otimes \sigma_{\mu,\vec x} \otimes \overbrace{\rho_{x_1}\otimes \cdots \otimes\rho_{x_{\mu-\nu-1}}}^{\mu-\nu-1}\otimes\rho_{x_{\mu-\nu}}\sigma_{\nu,\vec y}\otimes\overbrace{\rho_{x_{\mu-\nu+1}}\rho_{y_1}\otimes\cdots\otimes\rho_{x_{\mu-1}}\rho_{y_{\nu-1}}}^{\nu-1} \\
=&\sum_{\{x\}}\overbrace{\sigma_0\otimes\cdots\sigma_0}^{d-\mu}\otimes \sigma_{\mu,\vec x} \otimes \overbrace{\rho_{x_1}\otimes \cdots \otimes\rho_{x_{\mu-\nu-1}}}^{\mu-\nu-1}\otimes\rho_{x_{\mu-\nu}}\sigma_{\nu,\vec y(\vec x)}\otimes\overbrace{\rho_{x_{\mu-\nu+1}}\otimes\cdots\otimes\rho_{x_{\mu-1}}}^{\nu-1},
 \end{split}
  \end{equation}
where 
\begin{equation}
\sum_{\{x\}}:=\sum_{x_1,\ldots,x_{\mu-1}},\quad \sum_{\{y\}}:=\sum_{y_1,\ldots,y_{\nu-1}},
\end{equation}
and in the second line we have used that $y_k=x_{\mu-\nu+k}$ for $k =1,\cdots, \nu-1$
so that
\begin{equation}
\vec{y}(\vec x) :=(x_{\mu-\nu+1},x_{\mu-\nu+2},\ldots,x_{\mu-1},y_{\nu}).
\end{equation}
Hence we can write $h_{d,\mu}h_{d,\nu}h_{d,\mu}h_{d,\nu}$ as 
\begin{equation}
\begin{split}
&\sum_{\{x\},\{x'\}}\overbrace{\sigma_0\otimes\cdots\sigma_0}^{d-\mu}\otimes \sigma_{\mu,\vec x}\sigma_{\mu,\vec x'}  \otimes \overbrace{\rho_{x_1}\rho_{x'_1}\otimes \cdots \otimes\rho_{x_{\mu-\nu-1}}\rho_{x'_{\mu-\nu-1}}}^{\mu-\nu-1}\\
&\qquad\quad\otimes\rho_{x_{\mu-\nu}}\rho_{x'^{c}_{\mu-\nu}}\sigma_{\nu,\vec y(\vec x)}\sigma_{\nu,\vec y(\vec x')}\otimes\overbrace{\rho_{x_{\mu-\nu+1}}\rho_{x'_{\mu-\nu+1}}\otimes\cdots\otimes\rho_{x_{\mu-1}}\rho_{x'_{\mu-1}}}^{\nu-1},
\end{split}
\end{equation}
where we have used $\sigma_{\mu,\vec y(\vec x)}\rho_{x'_{\mu-\nu}}=\rho_{x'^{c}_{\mu-\nu}}\sigma_{\mu,\vec y(\vec x)}$. It is clear the only nonzero terms are those with 
\begin{equation}
x'_{\kappa} =x_{\kappa}, \quad \text{if $\kappa\neq \mu-\nu$}; \quad x'_{\mu-\nu} =x^c_{\mu-\nu}=1-x_{\mu-\nu}.
\end{equation}
Under this condition we see (note that $ \sigma_{\nu,\vec y(\vec x)}$ does
not depend on the last component of $\vec y(\vec x)$)
\begin{equation}
\sigma_{\nu,\vec y(\vec x)}\sigma_{\nu,\vec y(\vec x')}  = \sigma_0,
\end{equation} 
\begin{equation}
\sigma_{\mu,\vec x}\sigma_{\mu,\vec x'}  =\begin{pmatrix}
e^{i \phi S_{\mu,\mu-\nu}(x_{\mu-\nu}-x_{\mu-\nu}^c)}&0\\0&e^{-i \phi S_{\mu,\mu-\nu}(x_{\mu-\nu}-x_{\mu-\nu}^c)}
\end{pmatrix},
\end{equation} 
whose trace is 
\begin{equation}
\Tr \sigma_{\mu,\vec x}\sigma_{\mu,\vec x'} =2\cos (\phi S_{\mu,\mu-\nu} )=2\cos \phi.
\end{equation}
Now we can perform the sum over $x$ explicitly and obtain 
\begin{equation}\label{eqn:fourthMom3rdTerm}
\Tr h_{d,\mu}h_{d,\nu}h_{d,\mu}h_{d,\nu} =2^d \cos\left(\phi S_{\mu\nu}\right) =2^d\cos \phi, 
\end{equation}
which is independent of $S_{\mu\nu}$. Combining equations \eqref{eqn:fourthMom1stTerm}, \eqref{eqn:fourthMom2ndTerm} and \eqref{eqn:fourthMom3rdTerm}, we obtain the total fourth moment
\be
 \Tr H^4 = 2^d \left[ d +4\binom{d}{2}+ 2\binom{d}{2}\cos\phi\right]= 2^d \left[d(d-1)(2+\cos\phi)+d\right],
\ee
                        and the normalized fourth moment is equal to
                        \be
                        \frac {2^{-d}\Tr H^4}{[2^{-d}\Tr H^2]^2}=   \frac{d-1}{d}(2+\cos \phi)+\frac{1}{d},                                             
                        \ee
                        which is in agreement with the averaged fourth moment $\langle\Tr H^4\rangle$ first obtained in \cite{Marinari:1995jwr}.

 \subsection{$\Tr H^6$}\label{app:m6Tensor}
                       
In this section we show that  $\Tr H^6$ does not depend on the disorder realization.

 Most contribution to $\Tr H^6$ can be reduced to combinations occurring in $\Tr H^4$. We have two new combinations:  $ \Tr h_\mu h_{\nu}h_\mu h_{\omega} h_{\nu} h_{\omega}$  and
                        $ \Tr h_\mu h_{\nu}h_{\omega}h_\mu h_{\nu}h_{\omega}$ with $\mu>\nu>\omega$. For notational clarity we focus on the cases of  $ \Tr h_d h_{d-1}h_d h_{d-2} h_{d-1} h_{d-2}$  and $ \Tr h_d h_{d-1}h_{d-2}  h_d h_{d-1}h_{d-2}$, so that we only need to use 
                         \be
                        h_d &=& \sum_{x_1,\ldots , x_{d-1}} \sigma_{d,\vec x}\otimes \rho_{x_1}\otimes
                        \rho_{x_2}\otimes \cdots,\\
                          h_{d-1}& = &\sigma_0\otimes\sum_{x'_1,\ldots , x'_{d-2}} \sigma_{d-1,\vec{x}'}\otimes
                          \rho_{x_1'}\otimes  \cdots,\\
                          h_{d-2} &= &\sigma_0\otimes\sigma_0\otimes\sum_{x''_1,\ldots , x''_{d-3}}  \sigma_{d-2,\vec{x}''}\otimes \cdots
                          \ee
               with the $\cdots$ representing   $d-3$ additional factors $\rho_x$.  Each $h$ appears two times in the traces, and we use another set of dummy indices $y$, $y'$ and $y''$ to be the summation indices for their second appearances.   From the multiplication of the last $d-3$ factors, we know the sum only receive contributions from 
              \begin{equation}\label{eqn:sixthMomSumCondition}
              \begin{split}
                 x''_1=y''_1=x'_2&=y'_2=x_3=y_3, \\
                 x''_2=y''_2=x'_3&=y'_3=x_4=y_4, \\
                 &\ \vdots \\
                  x''_{d-3}=y''_{d-3}=x'_{d-2}&=y'_{d-2}=x_{d-1}=y_{d-1},
                 \end{split}
\end{equation}            
and the summation symbol simplifies accordingly:
\begin{equation}
\sum_{x_1,\ldots,x_{d-1}}\sum_{y_1,\ldots,y_{d-1}}\sum_{x'_1,\ldots,x'_{d-2}}\sum_{y'_1,\ldots,y'_{d-2}}\sum_{x''_1,\ldots,x''_{d-3}}\sum_{y''_1,\ldots,y''_{d-3}} \rightarrow \sum_{x_3,\ldots,x_{d-1}} \sum_{x_1,x_2, x_1'}\sum_{y_1,y_2,y_1'}. 
\end{equation}
We now work out $\Tr (h_d h_{d-1}h_d h_{d-2} h_{d-1} h_{d-2})$. The nontrivial part of the trace is 
                        \be\label{eqn:sixthmom1stTrace}
                        &&\sum_{x_1,x_2, x_1'}\sum_{y_1,y_2,y_1'}\Tr \left(\sigma_{d,\vec x}
                        \sigma_{d,\vec y}\right)
                        \Tr \left(\rho_{x_1} \sigma_{d-1,\vec x'} \rho_{y_1}\sigma_{d-1,\vec y'} \right)\Tr \left(\rho_{x_2}
                          \rho_{x_1'}\rho_{y_2} \sigma_{d-2,\vec x''}\rho_{y'_1}\sigma_{d-2,\vec y''}\right)\nn\\
                          &=&\sum_{x_1,x_2, x_1'}\sum_{y_1,y_2,y_1'}\Tr \left(\sigma_{d,\vec x}
                        \sigma_{d,\vec y}\right)
                        \Tr \left(\rho_{x_1}\rho_{y^c_1} \sigma_{d-1,\vec x'} \sigma_{d-1,\vec y'} \right)\Tr \left(\rho_{x_2}
                          \rho_{x_1'}\rho_{y_2}\rho_{{y'_1}^c } \sigma_{d-2,\vec x''}\sigma_{d-2,\vec y''}\right)\nn\\
                          &=&\sum_{x_1,x_2, x_1'}\sum_{y_1,y_2,y_1'}\Tr \left(\sigma_{d,\vec x}
                        \sigma_{d,\vec y}\right)
                        \Tr \left(\rho_{x_1}\rho_{y^c_1} \sigma_{d-1,\vec x'} \sigma_{d-1,\vec y'} \right)\Tr \left(\rho_{x_2}
                          \rho_{x_1'}\rho_{y_2}\rho_{{y'_1}^c } \right),
                        \ee
where in the last equality we used $\sigma_{d-2,\vec x''}\sigma_{d-2,\vec y''} =\sigma_{d-2,\vec x''}^2=\sigma_0$. The nonzero traces are those with the extra conditions 
\begin{equation}
x_1=y_1^c, \quad x_2=x_1' =y_2 = {y'_1}^c
\end{equation}
on top of the conditions \eqref{eqn:sixthMomSumCondition}. With these conditions we get
\begin{equation}
\begin{split}
\Tr \left(\sigma_{d,\vec x}\sigma_{d,\vec y}\right)=&2\cos(\phi S_{d,1}),\\
\sum_{x_1}\Tr \left(\rho_{x_1}\rho_{y_1^c} \sigma_{d-1,\vec x'} \sigma_{d-1,\vec y'} \right)=&\Tr\left(\sigma_{d-1,\vec x'} \sigma_{d-1,\vec y'}\right)=2\cos(\phi S_{d-1,1}),\\
 \Tr \left(\rho_{x_2} \rho_{x_1'}\rho_{y_2}\rho_{{y'_1}^c } \right) =& \Tr\left(\rho_{x_2}\right)=1.
\end{split}
\end{equation}
Taking the trace over the remaining $\rho_{x_3}\otimes\cdots \rho_{x_{d-1}}$ and sum over the remaining indices $x_2,x_3,\ldots,x_{d-1}$, we finally arrive at 
\be
                        \Tr h_d h_{d-1}h_d h_{d-2} h_{d-1} h_{d-2}= 2^d \cos \left(\phi S_{d,1}\right) \cos\left(\phi S_{d-1,1}\right)=2^d (\cos\phi)^2,
\ee
which is independent of disorder realizations.\\
Next we proceed to $\Tr h_d h_{d-1}h_{d-2} h_{d} h_{d-1} h_{d-2}$. The nontrivial part is given by
                        \be
                        \begin{split}
                        &\sum_{x_1,x_2, x_1'}\sum_{y_1,y_2,y_1'}\Tr \left(\sigma_{d,\vec x}
                        \sigma_{d,\vec y}\right)
                        \Tr \left(\rho_{x_1} \sigma_{d-1,\vec x'} \rho_{y_1}\sigma_{d-1,\vec y'} \right)\Tr \left(\rho_{x_2}
                          \rho_{x_1'}\sigma_{d-2,\vec x''}\rho_{y_2}\rho_{y'_1} \sigma_{d-2,\vec y''}\right)    \\
                          =& \sum_{x_1,x_2, x_1'}\sum_{y_1,y_2,y_1'}\Tr \left(\sigma_{d,\vec x}
                        \sigma_{d,\vec y}\right)
                        \Tr \left(\rho_{x_1}\rho_{y^c_1} \sigma_{d-1,\vec x'} \sigma_{d-1,\vec y'} \right)\Tr \left(\rho_{x_2}
                          \rho_{x_1'}\rho_{y_2^c}\rho_{{y'_1}^c }\right),
                           \end{split} \label{h6}
                         \ee
 Note only the last factor is different from that of equation \eqref{eqn:sixthmom1stTrace}, and the extra conditions enforced this time are
\begin{equation}
x_1=y_1^c, \quad x_2=x_1' =y_2^c= {y'_1}^c.
\end{equation} 
So equation \eqref{h6} reduces to 
\begin{align}\label{eqn:sixthmom2ndLast}
&\sum_{x_1,x_2}\Tr \left(\sigma_{d,\vec x}
                        \sigma_{d,\vec y}\right)
                        \Tr \left( \rho_{x_1}\sigma_{d-1,\vec x'} \sigma_{d-1,\vec y'} \right)\nn\\                        =& \sum_{x_1,x_2}2\cos\left[\phi(x_1-x_1^c)S_{d,1}+\phi(x_2-x_2^c)S_{d,2}\right] \Tr \left[\rho_{x_1} \begin{pmatrix}
                       e^{i\phi(x_2-x_2^c)S_{d-1,1}}&0\\
                        0&e^{-i\phi(x_2-x_2^c)S_{d-1,1}}
                        \end{pmatrix}\right]\nn\\
                        =&8\cos\left(\phi S_{d,1}\right)\cos\left(\phi S_{d,2}\right)\cos\left(\phi S_{d-1,1}\right)+8i\sin\left(\phi S_{d,1}\right)\sin\left(\phi S_{d,2}\right)\sin\left(\phi S_{d-1,1}\right). 
\end{align}
Note this does depend on disorder realizations of $S_{\mu\nu}$ due to the sine terms. However, in the expansion of $\Tr H^6$, the $\Tr h_d h_{d-1}h_{d-2} h_{d} h_{d-1} h_{d-2}$ can always be paired with a reverse-ordered term, namely $\Tr h_{d-2} h_{d-1} h_d h_{d-2} h_{d-1} h_{d}$. The previous calculations can be repeated easily, with the only change being a reverse ordering of matrices, and instead of equation  \eqref{eqn:sixthmom2ndLast} we now have
\begin{align}\label{eqn:sixthmom2ndLastReverse}
&\sum_{x_1,x_2}\Tr \left(\sigma_{d,\vec x}
                        \sigma_{d,\vec y}\right)
                        \Tr \left( \rho_{x_1} \sigma_{d-1,\vec y'}\sigma_{d-1,\vec x'}\right)\nn\\                        =& \sum_{x_1,x_2}2\cos\left[\phi(x_1-x_1^c)S_{d,1}+\phi(x_2-x_2^c)S_{d,2}\right] \Tr \left[\rho_{x_1} \begin{pmatrix}
                       e^{-i\phi(x_2-x_2^c)S_{d-1,1}}&0\\
                        0&e^{i\phi(x_2-x_2^c)S_{d-1,1}}
                        \end{pmatrix}\right]\nn\\
                        =&8\cos\left(\phi S_{d,1}\right)\cos\left(\phi S_{d,2}\right)\cos\left(\phi S_{d-1,1}\right)-8i\sin\left(\phi S_{d,1}\right)\sin\left(\phi S_{d,2}\right)\sin\left(\phi S_{d-1,1}\right). 
\end{align}
Taking the sum of \eqref{eqn:sixthmom2ndLast} and \eqref{eqn:sixthmom2ndLastReverse} we notice the sine terms cancel, thus the result no longer depends on $S_{\mu\nu}$. After performing the sum over $x_3,\ldots, x_{d-1}$, we conclude 
\begin{equation}
\begin{split}
&\Tr h_d h_{d-1}h_{d-2} h_{d} h_{d-1} h_{d-2}+\Tr h_{d-2} h_{d-1} h_d h_{d-2} h_{d-1} h_{d}\\
=& 2^{d+1}\cos\left(\phi S_{d,1}\right)\cos\left(\phi S_{d,2}\right)\cos\left(\phi S_{d-1,1}\right)\\
=& 2^{d+1}\left(\cos \phi\right)^3,
\end{split}
\end{equation}
which readily generalizes to the generic cases $ \Tr h_\mu h_{\nu}h_{\omega}h_\mu h_{\nu}h_{\omega}+\Tr h_{\omega}h_\nu h_{\mu}h_{\omega}h_\nu h_{\mu}$ with $\mu>\nu>\omega$.


 \section{Moments, words, chord diagrams and intersection graphs}\label{app:chords}

In this appendix, we will discuss how the leading and subleading large $d$ contributions to moments can be obtained through chord diagrams. For the leading contributions Parisi's original paper \cite{Parisi:1994jg} already has a comprehensive discussion, so we will briefly rephrase his work. In a follow-up work \cite{Marinari:1995jwr}, Marinari, Parisi and Ritort explicitly listed the subleading contributions up to the eighteenth moment, without giving a chord diagram interpretation of the results. We find in fact there is a nice correspondence between subleading contributions and the leading-contribution chord diagrams through a deletion procedure, and we will discuss it at some length.

\subsection{Leading contributions}
\begin{figure}
\begin{center}
 \begin{tikzpicture}
 \node at (-2,3) {Word};
 \node at (1,3) {Lattice path};
 \node at (5.5,3) {Chord diagram};
  \node at (9.5,3) {Intersection graph};

 \node at (-2,1) {$abab$};
\draw[fill=black] (0,0) circle (1.5pt);
\draw[fill=black] (0,2) circle (1.5pt);
\draw[fill=black] (2,0) circle (1.5pt);
\draw[fill=black] (2,2) circle (1.5pt);
\begin{scope}[thick,decoration={
    markings,
    mark=at position 0.5 with {\arrow{>}}}
    ] 
    \draw[postaction={decorate}] (0,0)--(2,0);
    \draw[postaction={decorate}] (2,0)--(2,2);
    \draw[postaction={decorate}] (2,2)--(0,2);
    \draw[postaction={decorate}] (0,2)--(0,0);
\end{scope}
\node at (1,-0.25) {$a$};
\node at (2.25,1) {$b$};
\node at (1,2.25) {$a$};
\node at (-0.25,1) {$b$};

\draw [] (4,0)--(4,1.5)--(6,1.5)--(6,0);
\draw [] (5,0)--(5,1)--(7,1)--(7,0);
\node at (4,-0.2) {$a$};
\node at (5,-0.2) {$b$};
\node at (6,-0.2) {$a$};
\node at (7,-0.2) {$b$};

\draw[fill=black] (9,0.5) circle (1pt);
\draw[fill=black] (10,0.5) circle (1pt);
\draw[] (9,0.5)--(10,0.5);
\node at (8.9,0.3) {$a$};
\node at (10.1,0.3) {$b$};

 \node at (-2,-2.5) {$abcacb$};
 
\draw[fill=black] (0,-3.5) circle (1.5pt);
\draw[fill=black] (1.5,-3.5) circle (1.5pt);
\draw[fill=black] (0.5,-3) circle (1.5pt);
\draw[fill=black] (0.5,-1.5) circle (1.5pt);
\draw[fill=black] (2,-3) circle (1.5pt);
\draw[fill=black] (2,-1.5) circle (1.5pt);
\draw[fill=black] (0,-2) circle (1.5pt);
\draw[fill=black] (1.5,-2) circle (1.5pt);

\begin{scope}[thick,decoration={
    markings,
    mark=at position 0.5 with {\arrow{>}}}
    ] 
    \draw[postaction={decorate}] (0,-3.5)--(1.5,-3.5);
    \draw[postaction={decorate}] (1.5,-3.5)--(2,-3);
    \draw[postaction={decorate}] (2,-3)--(2,-1.5);
    \draw[postaction={decorate}] (2,-1.5)--(0.5,-1.5);
      \draw[postaction={decorate}] (0.5,-1.5)--(0.5,-3);
        \draw[postaction={decorate}] (0.5,-3)--(0,-3.5);
\end{scope}
\node at (0.75,-3.75) {$a$};
\node at (1.9,-3.35) {$b$};
\node at (2.25,-2.25) {$c$};
\node at (1.25,-1.3) {$a$};
\node at (0.7,-2.25) {$c$};
\node at (0.43,-3.25) {$b$};

\draw[dashed] (0.5,-3)--(2,-3);
\draw[dashed] (0.5,-1.5)--(0,-2)--(0,-3.5);
\draw[dashed] (0,-2)--(1.5,-2)--(2,-1.5);
\draw[dashed] (1.5,-3.5)--(1.5,-2);

\draw [] (4,-3.5)--(4,-2)--(5.8, -2)--(5.8,-3.5);
\draw [] (4.6,-3.5)--(4.6,-2.5)--(7,-2.5)--(7,-3.5);
\draw [] (5.2,-3.5)--(5.2,-3)--(6.4,-3)--(6.4,-3.5);

\node at (4,-3.7) {$a$};
\node at (4.6,-3.7) {$b$};
\node at (5.2,-3.7) {$c$};
\node at (5.8,-3.7) {$a$};
\node at (6.4,-3.7) {$c$};
\node at (7,-3.7) {$b$};

\draw[fill=black] (9,-3.2) circle (1pt);
\draw[fill=black] (10,-3.2) circle (1pt);
\draw[fill=black] (9.5,-2.334) circle (1pt);

\draw[] (9.5,-2.334)--(9,-3.2)--(10,-3.2);
\node at (8.85,-3.35) {$a$};
\node at (10.15,-3.35) {$b$};
\node at (9.5,-2.15) {$c$};

 \node at (-2,-6) {$abcabc$};
 
\draw[fill=black] (0,-7) circle (1.5pt);
\draw[fill=black] (1.5,-7) circle (1.5pt);
\draw[fill=black] (0.5,-6.5) circle (1.5pt);
\draw[fill=black] (0.5,-5) circle (1.5pt);
\draw[fill=black] (2,-6.5) circle (1.5pt);
\draw[fill=black] (2,-5) circle (1.5pt);
\draw[fill=black] (0,-5.5) circle (1.5pt);
\draw[fill=black] (1.5,-5.5) circle (1.5pt);

\begin{scope}[thick,decoration={
    markings,
    mark=at position 0.5 with {\arrow{>}}}
    ] 
    \draw[postaction={decorate}] (0,-7)--(1.5,-7);
    \draw[postaction={decorate}] (1.5,-7)--(2,-6.5);
    \draw[postaction={decorate}] (2,-6.5)--(2,-5);
    \draw[postaction={decorate}] (2,-5)--(0.5,-5);
      \draw[postaction={decorate}] (0.5,-5)--(0,-5.5);
        \draw[postaction={decorate}] (0,-5.5)--(0,-7);
\end{scope}
\node at (0.75,-7.25) {$a$};
\node at (1.9,-6.85) {$b$};
\node at (2.25,-5.75) {$c$};
\node at (1.25,-4.8) {$a$};
\node at (0.13,-5.15) {$b$};
\node at (-0.2,-6.25) {$c$};

\draw[dashed] (0.5,-5)--(0.5,-6.5)--(2,-6.5);
\draw[dashed] (0.5,-6.5)--(0,-7);
\draw[dashed] (2,-5)--(1.5,-5.5)--(0,-5.5);
\draw[dashed] (1.5,-5.5)--(1.5,-7);

\draw [] (4,-7)--(4,-5.5)--(5.8, -5.5)--(5.8,-7);
\draw [] (4.6,-7)--(4.6,-6)--(6.4,-6)--(6.4,-7);
\draw [] (5.2,-7)--(5.2,-6.5)--(7,-6.5)--(7,-7);

\node at (4,-7.2) {$a$};
\node at (4.6,-7.2) {$b$};
\node at (5.2,-7.2) {$c$};
\node at (5.8,-7.2) {$a$};
\node at (6.4,-7.2) {$b$};
\node at (7,-7.2) {$c$};

\draw[fill=black] (9,-6.7) circle (1pt);
\draw[fill=black] (10,-6.7) circle (1pt);
\draw[fill=black] (9.5,-5.834) circle (1pt);

\draw[] (9,-6.7)--(10,-6.7)--(9.5,-5.834)--(9,-6.7);
\node at (8.85,-6.85) {$a$};
\node at (10.15,-6.85) {$b$};
\node at (9.5,-5.65) {$c$};
\end{tikzpicture} 
\end{center}
\caption{Three examples of loop paths on hypercube and their representations in terms of words, chord diagrams and intersection graphs.}\label{fig:loopWords}
\end{figure}
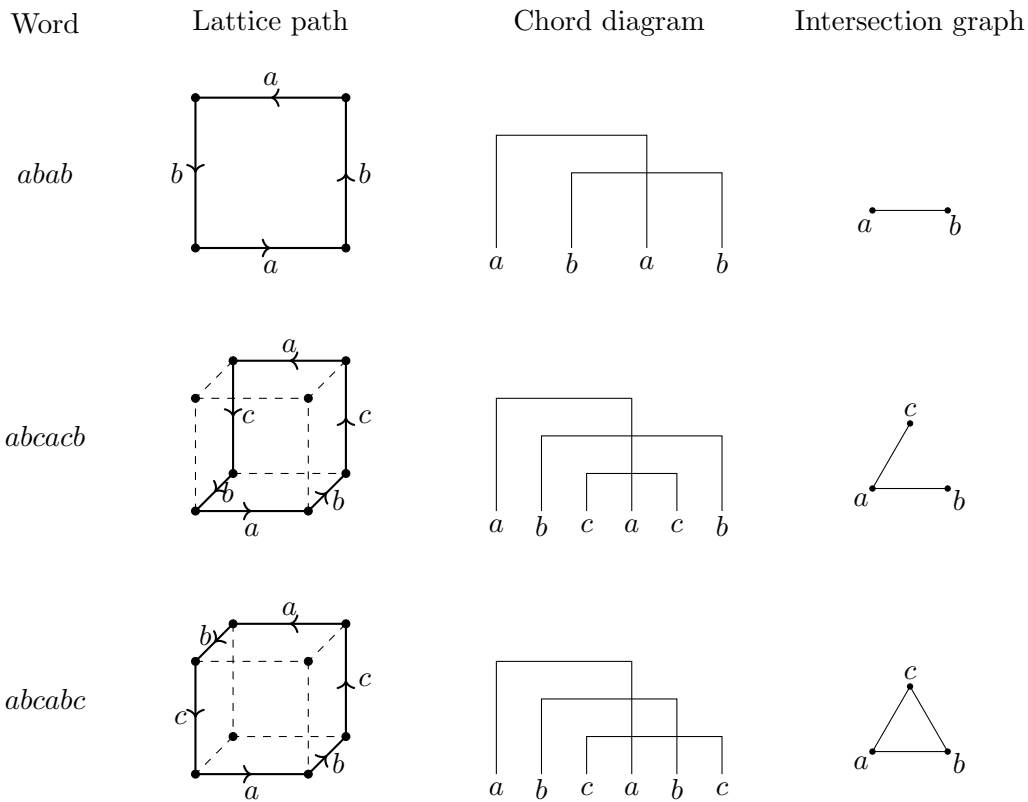

The $2p$-th moment $\langle \Tr H^{2p} \rangle$ is given by the sum of all $2p$-step Wilson loops:
\begin{equation}\label{eqn:momentsFromWilsonLoop}
\langle \Tr H^{2p} \rangle = \sum_{\mathcal{C},|\mathcal{C}|=2p } \langle W(\mathcal{C})\rangle.
\end{equation}
We will classify all the $2p$-step Wilson loops into groups by the total number of Euclidean dimensions they traverse. Since the $2p$ steps need to form a loop, at most they can traverse $p$ different dimensions.   If we follow the path of a $2p$-step loop, each time a new step is taken along a dimension that has not been traversed, we pick up a multiplicity factor counting the remaining dimensions. For example, the first step of any loop can freely choose any of the $d$ dimensions; the nearest next step that takes a different dimension has the remaining $d-1$ dimensions to choose from, and so on. By this reasoning we see if a loop traverses $k$ dimensions, the multiplicity factor from this effect alone is $d(d-1)\cdots (d-k+1)\sim d^k$. Since $k\leq p$, the leading large $d$ contributions will come from those loops that traverse $p$ different dimensions, having a multiplicity factor of  $d(d-1)\cdots (d-p+1)$,
and each of the $p$ chosen dimensions is traversed twice, forward and backward, so that a loop can be formed in the end. We can use an alphabet of $p$ different letters to represent the $p$ different dimensions, and use a $2p$-letter word with each alphabet letter appearing twice to represent a loop:  we read the $2p$ letters in the word from left to right, and we traverse the dimension that is represented by the letter. To avoid double counting we should demand that the first appearances of the letters in a word must be ordered as they are in the alphabet. As a few  examples, $aabb$ is a permissible word but $bbaa$ is not; $abcacb$ is permitted but $acbabc$ is not; $abbacc$ is permitted but $caacbb$ is not. It is easy to see there are $(2p-1)!!$ different words we can form by having $p$ letters each appearing twice.
If we connect the same letters in a word with lines in the upper half plane, we form what is called a \textit{chord diagram}, and the lines are called the \textit{chords}. See figure \ref{fig:loopWords} for a few examples. 

The Wilson loop value can be calculated by decomposing a loop into elementary plaquettes, namely the faces of our hypercube. We can project the path into all the $\binom{d}{2}$ coordinate axis planes, and if the projection into the $\mu\nu$ plane has a plaquette shape, we pick up phase of $e^{i \phi S_{\mu\nu}}$; if the projection is a backtracking path which has zero area, then the contribution is just $1$. Multiplying the contributions from all projections gives the value of the Wilson loop. It is important to note for  loops of the leading contributions the projections do not loop around the same plaquette twice, because there are only two steps along each dimension: a step forward and a step back. This means we cannot pick up phases like $e^{2 i \phi S_{\mu\nu}}$ from the $\mu\nu$ plane, and the disorder average over $S_{\mu\nu}$ on each face results in a $\cos \phi$ for each projection that loops around a face, and contributions from different plaquettes multiply. Hence we have the following formula:
\begin{equation}\label{eqn:WilsonLoopFromLatticePath}
\langle W(\mathcal{C})\rangle = q^{A(\mathcal{C})}, 
\end{equation}
with
\begin{equation}
\begin{split}
q:=& \cos\phi,\\
A(\mathcal{C}):=& \sum_{\mu<\nu} A_{\mu\nu}(\mathcal{C}),
\end{split}
\end{equation}
where $A_{\mu\nu}(\mathcal{C})$ is the area of the loop's projection into $\mu\nu$ plane which takes value of either $0$ or $1$. 

In the word representation of lattice paths, if we want to study the loop projection into a particular plane, we only need to focus on the two alphabet letters that represent the plane. Suppose the $\mu$ and $\nu$ dimensions are represented by the letters $a$ and $b$, respectively. To study the projection into $\mu\nu$ plane, we can temporarily forget letters other than $a$ and $b$. With regard to $a$ and $b$, there are only three scenarios:
\begin{align}
 &\ldots a\ldots a \ldots b\ldots b\ldots, \\
  &\ldots a\ldots b \ldots b\ldots a\ldots,\\  &\ldots a\ldots b \ldots a\ldots b\ldots.
\end{align}

It is clear that the first two cases have zero-area projections in the $\mu\nu$ plane and the third case has an area-one projection. In terms of chord diagrams, the first two have zero intersections between chords $a$ and $b$, whereas the third has one intersection. Now we can synthesize equations \eqref{eqn:momentsFromWilsonLoop} and \eqref{eqn:WilsonLoopFromLatticePath} for the leading contributions as
\begin{equation}\label{eqn:momentsFromChordDiagrams}
\langle \Tr H^{2p} \rangle_{\text{leading}} = d(d-1)\cdots(d-p+1) \sum_{C(p \text{-chord})} q^{\# \text{ of intersections in } C},
\end{equation}
where $q=\cos\phi$ and $C$ denotes chord diagrams with $p$ chords. In other words the leading moments are the generating functions of chord intersections. The moments calculated by \eqref{eqn:momentsFromChordDiagrams} also appear in the Sachdev-Ye-Kitaev
model \cite{Erdos:2014a,Garcia-Garcia:2017pzl,Cotler2016,Garcia-Garcia:2018kvh}. The sum on the right-hand side of \eqref{eqn:momentsFromChordDiagrams} has an interesting solution: in his original paper Parisi \cite{Parisi:1994jg}
already suggested mapping the sum to the vacuum expectation values of some observables in the $q$-deformed harmonic oscillator system. This approach was further elaborated in \cite{Marinari:1995jwr}. In fact, much earlier this chord diagram sum was studied by Touchard \cite{Touchard:1952a} and Riordan \cite{Riordan:1975a} in a more combinatorial vein, which led to the Riordan-Touchard formula:
\begin{equation}
\sum_{C(p \text{-chord})} q^{\# \text{ of intersections in } C}=\frac{1}{(1-q)^p}\sum\limits_{k=-p}^{p}(-1)^k q^{k(k-1)/2}\binom{2p}{p+k}.
\end{equation}
Using this formula we can effortlessly generate 
\begin{align}
\langle \Tr H^{2} \rangle_{\text{leading}}=&d,\\
\langle \Tr H^{4} \rangle_{\text{leading}}=&d(d-1)\left[2+q\right],\\
\langle \Tr H^{6} \rangle_{\text{leading}}=&d(d-1)(d-2)\left[5 + 6 q + 3 q^2 + q^3\right],\label{eqn:sixthRT}\\
\langle \Tr H^{8} \rangle_{\text{leading}}=&d(d-1)(d-2)(d-3)\left[14 + 28 q + 28 q^2 + 20 q^3 + 10 q^4 + 4 q^5 + q^6\right],\label{eqn:eighthRT}
\end{align}
and so on.
\subsection{Subleading contributions}
The goal of this section is to give a chord diagram interpretation of the subleading contributions to moments. To be clear, there are already contributions subleading in $d$ included in equation \eqref{eqn:momentsFromChordDiagrams} due to the multiplicity factor $d(d-1)\cdots(d-p+1)$. However, there are still subleading contributions from $2p$-step loops that traverse only $p-1$ dimensions which gives a multiplicity factor $d(d-1)\cdots(d-p+2)$.
This section will be about such loops. We first demonstrate that there is a bijection between subleading words and certain structures of the leading words. The choice for this bijection is not unique, different choices lead to different schemes of calculating the subleading contributions, and unsurprisingly all schemes give the same result. 
\subsubsection{The interlace scheme}
As already discussed, the leading words with $2p$ letters are words with $p$ different pairs of alphabet letters. By the previous discussion it is clear the subleading words with $2p$ letters have one alphabet letter appearing four times, and $p-2$ other alphabet letters each appearing twice. This reflects the fact that the subleading loops discussed at the beginning of last section must traverse a dimension four times and other remaining dimensions two times each. For examples $aaaabb$, $abbbab$ and $abbabb$ are some subleading words for $p=3$. From the general formula\footnote{We can write down the general formula after some thought. The total number of $2p$-letter words formed by $k$ alphabet letters (each can appear even number of times) is
\begin{equation}
\sum_{[m_1,m_2,\ldots,m_k]_p }\binom{2p-1}{2m_1-1}\binom{2p-2m_1-1}{2m_2-1}\cdots \binom{2p-2\sum_{l=1}^{k-1} m_l -1}{2m_k-1},
\end{equation} where $[m_1,m_2,\ldots,m_k]_p$ denotes a \textit{composition} of $p$, that is, an ordered $k$-tuple $(m_1,m_2,\ldots,m_k)$ such that $\sum_{l=1}^k m_l=p$.}
it is clear that we can form $\frac{1}{3}\binom{p}{2}(2p-1)!!$ subleading words of length $2p$.
 
 The following map is a bijection between subleading words and interlacing structures of leading words:
 \begin{equation}\label{eqn:leadingSubleadingBijection}
 \ldots a\ldots a \ldots a\ldots a \ldots \mapsto \ldots \underline{a}\ldots \underline b \ldots \underline a\ldots \underline b \ldots,
 \end{equation}
where the $\cdots$ part remains unchanged after the mapping. We added underlines on the right--hand side to emphasize the map is toward an interlacing structure, instead of the leading word that contains this interlacing structure. In the context of this mapping, it is convenient for us to adopt a ``jump an alphabet letter'' convention for subleading words: we jump over the alphabet letter that immediately follows (in the alphabet) the letter that appears four times in the word. For example, $aaaabb$ and $aaaacc$ are equivalent words, but we prefer the second representation because it is mapped to  $ababcc$ without changing the letter $c$. Let us also see an example of the inverse mapping. The leading word $abcacb$ has two interlacing structures, each will be mapped to a subleading word:
\begin{align}
\underline{a}\underline{b}c\underline{a}c\underline{b}&\mapsto aacaca,\\
\underline{a} b\underline{cac}b &\mapsto  abaaab.
\end{align}
It is clear that the mappings in both directions are injective and hence bijective. Note each interlacing structure in a leading word corresponds to an intersection in the corresponding chord diagram, so we may also say there is a bijection between subleading words and the intersections of the leading chord diagrams.\footnote{A byproduct of this discussion is that we just completed a bijective proof of the following statement: the total number of intersections among all chord diagrams with $p$ chord is $\frac{1}{3}\binom{p}{2}(2p-1)!!$. We can easily generalize the proof to other intersection structures. Other proofs of this statement already exist, see for example \cite{Flajolet:1997a,Jia:2018ccl}.}

We have demonstrated that the bijection \eqref{eqn:leadingSubleadingBijection} allows us to use the interlacing structures in leading words to represent the subleading Wilson loops. The remaining question is how to read off the values of the Wilson loops from the leading word interlacing structures. Let us recall that for a leading Wilson loop, each interlacing structure in its word representation represents a projection of the path that loops around a plaquette.  Obviously, after the mapping  \eqref{eqn:leadingSubleadingBijection}, this particular interlacing structure is removed, and the leading Wilson loop becomes a subleading Wilson loop in which this plaquette projection gets squashed to a zero-area projection. However, this is not the end of the story: it is conceivable that the removal of one interlacing structure interferes with other interlacing structures in the same word, so that more plaquette-shaped projections get squashed as a result. We are faced with three possibilities:
\begin{enumerate}
\item The other interlacing structure that might be interfered with by the removal of $\underline{a}\underline b \underline a \underline b$  is formed by two other alphabet letters, so we have
 \begin{equation}\label{eqn:interlace1stCase}
  \underline{a}\ldots \underline b \ldots \underline a\ldots \underline b  \ldots c\ldots d \ldots c \ldots d \ldots  \mapsto a\ldots a \ldots a\ldots a \ldots c\ldots d \ldots c \ldots d \ldots.
\end{equation}
In this case the plaquette projection represented by $cdcd$ cannot be affected because the hypercube dimensions represented by $c$ and $d$ are in the orthogonal complement of $a$ and $b$.

\item The other interlacing structure that might be interfered with by the removal of $\underline{a}\underline b \underline a \underline b$ is formed by one other alphabet letter interlacing with one of $a$ or $b$ but not both, so we have 
  \begin{equation}\label{eqn:interlace2ndCase}
  \ldots \underline{a}\ldots \underline b \ldots c\ldots \underline a\ldots c \ldots \underline b \ldots  \mapsto \ldots a\ldots a \ldots c\ldots a \ldots c \ldots a \ldots.
\end{equation}
The effect on the corresponding Wilson loops can be read off visually:
\begin{equation}
\begin{tikzpicture}
\draw[fill=black] (0,-3.5) circle (1.5pt);
\draw[fill=black] (1.5,-3.5) circle (1.5pt);
\draw[fill=black] (0.5,-3) circle (1.5pt);
\draw[fill=black] (0.5,-1.5) circle (1.5pt);
\draw[fill=black] (2,-3) circle (1.5pt);
\draw[fill=black] (2,-1.5) circle (1.5pt);
\draw[fill=black] (0,-2) circle (1.5pt);
\draw[fill=black] (1.5,-2) circle (1.5pt);

\begin{scope}[thick,decoration={
    markings,
    mark=at position 0.5 with {\arrow{>}}}
    ] 
    \draw[postaction={decorate}] (0,-3.5)--(1.5,-3.5);
    \draw[postaction={decorate}] (1.5,-3.5)--(2,-3);
    \draw[postaction={decorate}] (2,-3)--(2,-1.5);
    \draw[postaction={decorate}] (2,-1.5)--(0.5,-1.5);
      \draw[postaction={decorate}] (0.5,-1.5)--(0.5,-3);
        \draw[postaction={decorate}] (0.5,-3)--(0,-3.5);
\end{scope}
\node at (0.75,-3.75) {$a$};
\node at (1.9,-3.35) {$b$};
\node at (2.25,-2.25) {$c$};
\node at (1.25,-1.3) {$a$};
\node at (0.7,-2.25) {$c$};
\node at (0.43,-3.25) {$b$};

\draw[dashed] (0.5,-3)--(2,-3);
\draw[dashed] (0.5,-1.5)--(0,-2)--(0,-3.5);
\draw[dashed] (0,-2)--(1.5,-2)--(2,-1.5);
\draw[dashed] (1.5,-3.5)--(1.5,-2);

\node at (3,-2.45) {$\mapsto$};

\draw[fill=black] (4,-3.5) circle (1.5pt);
\draw[fill=black] (5.5,-3.5) circle (1.5pt);
\draw[fill=black] (4.5,-3) circle (1.5pt);
\draw[fill=black] (4.5,-1.5) circle (1.5pt);
\draw[fill=black] (6,-3) circle (1.5pt);
\draw[fill=black] (6,-1.5) circle (1.5pt);
\draw[fill=black] (4,-2) circle (1.5pt);
\draw[fill=black] (5.5,-2) circle (1.5pt);

\begin{scope}[thick,decoration={
    markings,
    mark=at position 0.5 with {\arrow{>}}}
    ] 
    \draw[postaction={decorate}] (4,-3.5)--(5.5,-3.5);
     \draw[postaction={decorate}] (5.45,-3.65)--(3.95,-3.65);
      \draw[postaction={decorate}] (3.95,-3.65)--(3.95,-2);
       \draw[postaction={decorate}] (3.95,-2)--(5.5,-2);
       \draw[postaction={decorate}] (5.5,-2)--(5.5,-3.35);
        \draw[postaction={decorate}] (5.5,-3.35)--(4,-3.35);
\end{scope}
\node at (4.75,-3.8) {$a$};
\node at (4.95,-3.5) {$a$};
\node at (4.6,-3.3) {$a$};
\node at (3.8,-2.75) {$c$};
\node at (4.75,-1.85) {$a$};
\node at (5.3,-2.75) {$c$};

\draw[thick] (5.5,-3.5)--(5.45,-3.65);
\draw[thick] (4,-3.35)--(4,-3.5);
\draw[dashed] (4.5,-3)--(6,-3);
\draw[dashed] (4,-3.5)--(4.5,-3);
\draw[dashed] (5.5,-3.5)--(6,-3)--(6,-1.5)--(4.5,-1.5)--(4.5,-3);
\draw[dashed] (4.5,-1.5)--(4,-2);
\draw[dashed] (4,-2)--(5.5,-2)--(6,-1.5);
\end{tikzpicture}
\end{equation}
and we see the projected area in the $ac$ plane remains one. 
\item Another alphabet letter $c$ interlaces with both $a$ and $b$, and we want to investigate what happens to the projected areas represented by the two new interlacing structures  $acac$ and $bcbc$. The word map is 
\begin{equation}\label{eqn:interlace3rdCase}
  \ldots \underline{a}\ldots \underline b \ldots c\ldots \underline a \ldots \underline b \ldots c\ldots  \mapsto \ldots a\ldots a \ldots c\ldots a \ldots a \ldots c \ldots.
\end{equation}
The corresponding Wilson loop transforms as 
\begin{equation}
\begin{tikzpicture}
\draw[fill=black] (0,-7) circle (1.5pt);
\draw[fill=black] (1.5,-7) circle (1.5pt);
\draw[fill=black] (0.5,-6.5) circle (1.5pt);
\draw[fill=black] (0.5,-5) circle (1.5pt);
\draw[fill=black] (2,-6.5) circle (1.5pt);
\draw[fill=black] (2,-5) circle (1.5pt);
\draw[fill=black] (0,-5.5) circle (1.5pt);
\draw[fill=black] (1.5,-5.5) circle (1.5pt);

\begin{scope}[thick,decoration={
    markings,
    mark=at position 0.5 with {\arrow{>}}}
    ] 
    \draw[postaction={decorate}] (0,-7)--(1.5,-7);
    \draw[postaction={decorate}] (1.5,-7)--(2,-6.5);
    \draw[postaction={decorate}] (2,-6.5)--(2,-5);
    \draw[postaction={decorate}] (2,-5)--(0.5,-5);
      \draw[postaction={decorate}] (0.5,-5)--(0,-5.5);
        \draw[postaction={decorate}] (0,-5.5)--(0,-7);
\end{scope}
\node at (0.75,-7.25) {$a$};
\node at (1.9,-6.85) {$b$};
\node at (2.25,-5.75) {$c$};
\node at (1.25,-4.8) {$a$};
\node at (0.13,-5.15) {$b$};
\node at (-0.2,-6.25) {$c$};

\draw[dashed] (0.5,-5)--(0.5,-6.5)--(2,-6.5);
\draw[dashed] (0.5,-6.5)--(0,-7);
\draw[dashed] (2,-5)--(1.5,-5.5)--(0,-5.5);
\draw[dashed] (1.5,-5.5)--(1.5,-7);

\node at (3,-6) {$\mapsto$};

\draw[fill=black] (4,-7) circle (1.5pt);
\draw[fill=black] (5.5,-7) circle (1.5pt);
\draw[fill=black] (4.5,-6.5) circle (1.5pt);
\draw[fill=black] (4.5,-5) circle (1.5pt);
\draw[fill=black] (6,-6.5) circle (1.5pt);
\draw[fill=black] (6,-5) circle (1.5pt);
\draw[fill=black] (4,-5.5) circle (1.5pt);
\draw[fill=black] (5.5,-5.5) circle (1.5pt);

\begin{scope}[thick,decoration={
    markings,
    mark=at position 0.5 with {\arrow{>}}}
    ] 
    \draw[postaction={decorate}] (4,-7)--(5.5,-7);
     \draw[postaction={decorate}] (5.4,-7.15)--(3.9,-7.15);
     \draw[postaction={decorate}](3.9,-7.15)--(3.9,-5.4);
     \draw[postaction={decorate}](3.9,-5.4)--(5.5,-5.4);
     \draw[postaction={decorate}](5.5,-5.5)--(4,-5.5);
     \draw[postaction={decorate}](4,-5.5)--(4,-7);
\end{scope}

\draw[thick] (5.5,-5.4)--(5.5,-5.5);
\draw[thick] (5.5,-7)--(5.4,-7.15);
\draw[dashed] (4,-7)--(4.5,-6.5)--(4.5,-5)--(6,-5)--(6,-6.5)--(5.5,-7)--(5.5,-5.5)--(6,-5);
\draw[dashed] (4,-5.5)--(4.5,-5);
\draw[dashed] (4.5,-6.5)--(6,-6.5);

\node at (4.7,-7.3) {$a$};
\node at (4.8,-6.85) {$a$};
\node at (4.9,-5.7) {$a$};
\node at (5,-5.3) {$a$};
\node at (3.75,-6.35) {$c$};
\node at (4.15,-6.15) {$c$};
\end{tikzpicture}
\end{equation}
and we see the all three plaquettes in the Wilson loop  before the mapping collapse to
zero area after the mapping.
\end{enumerate}
We can summarize the above three cases as the following: 
\textit{for any subleading Wilson loop represented by the leading word interlacing structure $\ldots \underline{a}\ldots \underline b \ldots \underline a\ldots \underline b \ldots$,  this subleading Wilson loop has the value
\begin{equation}\label{eqn:subleadingRuleWords}
q^{\text{\# of interlacing structures in this word} - 2(\text{\# of triangular structures containing $\underline{abab}$)}-1},
\end{equation}
where  ``triangular structures containing $\underline{abab}$'' are structures like $\ldots \underline{a}\ldots \underline b \ldots c\ldots \underline a \ldots \underline b \ldots c\ldots $.}
For example, if the leading word is $abcdabcd$ and the interlacing structure we are interested in $\underline{ab}cd\underline{ab}cd$, then there are two triangular structures containing $\underline{abab}$, namely $\underline{ab}c\cdot\underline{ab}c\cdot$ and  $\underline{ab}\cdot d\underline{ab}\cdot d$.
There are six interlacing structures in the word $abcdabcd$, so the subleading Wilson loop has the value $q^{6-(2\times 2+1)}=q$.

We can obtain a rather compact and visual representation of rule \eqref{eqn:subleadingRuleWords} if we introduce the notion of \textit{intersection graphs}. To obtain the intersection graph of a leading word, we first draw its chord diagram. The intersection graph is then obtained by the following two steps:
\begin{enumerate}
\item represent every chord by a vertex,
\item connect two vertices if and only if the chords they represent intersect each other. 
\end{enumerate}
\begin{table}[t!]
  \begin{center}
\begin{tabular}{|c|c|c|c|c|}
\hline Intersection graph & \begin{tikzpicture} \draw[fill=black] (0,0) circle (1pt); \draw[fill=black] (0.4,0) circle (1pt);\draw[fill=black] (0.2,0.346) circle (1pt); \node at (0.2,0.35) {};\end{tikzpicture} & \begin{tikzpicture} \draw[fill=black] (0,0) circle (1pt); \draw[fill=black] (0.4,0) circle (1pt);\draw[fill=black] (0.2,0.346) circle (1pt);
\draw (0,0)--(0.4,0); \end{tikzpicture} & \begin{tikzpicture} \draw[fill=black] (0,0) circle (1pt); \draw[fill=black] (0.4,0) circle (1pt);\draw[fill=black] (0.2,0.346) circle (1pt);
\draw (0,0)--(0.4,0)--(0.2,0.346); \end{tikzpicture} & \begin{tikzpicture} \draw[fill=black] (0,0) circle (1pt); \draw[fill=black] (0.4,0) circle (1pt);\draw[fill=black] (0.2,0.346) circle (1pt);
\draw (0,0)--(0.4,0)--(0.2,0.346)--(0,0); \end{tikzpicture} \\ \hline
Leading value & 1 & $q$ & $q^2$ & $q^3$\\ \hline
Subleading value & 0 & $1$ & $2q$ & $3$\\ \hline
Multiplicity & 5 & 6 & 3 & 1\\ \hline
\end{tabular}
\caption{All the intersection graphs for the sixth moment.}\label{tab:6thMomIntersecGraphs}
\end{center}
\end{table}
\begin{table}
  \begin{center}
\begin{tabular}{|c|c|c|c|c|c|c|c|c|c|c|c|}
\hline Intersection graph & \begin{tikzpicture} \draw[fill=black] (0,0) circle (1pt); \draw[fill=black] (0.4,0) circle (1pt);\draw[fill=black] (0.4,0.4) circle (1pt);\draw[fill=black] (0,0.4) circle (1pt); \node at (0.2,0.44) {};\end{tikzpicture} & \begin{tikzpicture}\draw[fill=black] (0,0) circle (1pt); \draw[fill=black] (0.4,0) circle (1pt);\draw[fill=black] (0.4,0.4) circle (1pt);\draw[fill=black] (0,0.4) circle (1pt);
\draw (0,0)--(0.4,0); \end{tikzpicture} & \begin{tikzpicture} \draw[fill=black] (0,0) circle (1pt); \draw[fill=black] (0.4,0) circle (1pt);\draw[fill=black] (0.4,0.4) circle (1pt);\draw[fill=black] (0,0.4) circle (1pt);
\draw (0,0)--(0.4,0);\draw (0,0.4)--(0.4,0.4); \end{tikzpicture} & \begin{tikzpicture} \draw[fill=black] (0,0) circle (1pt); \draw[fill=black] (0.4,0) circle (1pt);\draw[fill=black] (0.4,0.4) circle (1pt);\draw[fill=black] (0,0.4) circle (1pt);
\draw (0,0)--(0.4,0)--(0.4,0.4); \end{tikzpicture} & \begin{tikzpicture} \draw[fill=black] (0,0) circle (1pt); \draw[fill=black] (0.4,0) circle (1pt);\draw[fill=black] (0.4,0.4) circle (1pt);\draw[fill=black] (0,0.4) circle (1pt);
\draw (0,0)--(0.4,0)--(0.4,0.4);\draw (0.4,0)--(0,0.4); \end{tikzpicture}& \begin{tikzpicture} \draw[fill=black] (0,0) circle (1pt); \draw[fill=black] (0.4,0) circle (1pt);\draw[fill=black] (0.4,0.4) circle (1pt);\draw[fill=black] (0,0.4) circle (1pt);
\draw (0,0)--(0.4,0)--(0.4,0.4)--(0,0.4); \end{tikzpicture}& \begin{tikzpicture} \draw[fill=black] (0,0) circle (1pt); \draw[fill=black] (0.4,0) circle (1pt);\draw[fill=black] (0.4,0.4) circle (1pt);\draw[fill=black] (0,0.4) circle (1pt);
\draw (0,0)--(0.4,0)--(0.4,0.4)--(0,0); \end{tikzpicture}& \begin{tikzpicture} \draw[fill=black] (0,0) circle (1pt); \draw[fill=black] (0.4,0) circle (1pt);\draw[fill=black] (0.4,0.4) circle (1pt);\draw[fill=black] (0,0.4) circle (1pt);
\draw (0,0)--(0.4,0)--(0.4,0.4)--(0,0)--(0,0.4); \end{tikzpicture}& \begin{tikzpicture} \draw[fill=black] (0,0) circle (1pt); \draw[fill=black] (0.4,0) circle (1pt);\draw[fill=black] (0.4,0.4) circle (1pt);\draw[fill=black] (0,0.4) circle (1pt);
\draw (0,0)--(0.4,0)--(0.4,0.4)--(0,0.4)--(0,0); \end{tikzpicture}& \begin{tikzpicture} \draw[fill=black] (0,0) circle (1pt); \draw[fill=black] (0.4,0) circle (1pt);\draw[fill=black] (0.4,0.4) circle (1pt);\draw[fill=black] (0,0.4) circle (1pt);
\draw (0,0)--(0.4,0)--(0.4,0.4)--(0,0.4)--(0,0)--(0.4,0.4); \end{tikzpicture}& \begin{tikzpicture} \draw[fill=black] (0,0) circle (1pt); \draw[fill=black] (0.4,0) circle (1pt);\draw[fill=black] (0.4,0.4) circle (1pt);\draw[fill=black] (0,0.4) circle (1pt);
\draw (0,0)--(0.4,0)--(0.4,0.4)--(0,0.4)--(0,0)--(0.4,0.4);\draw (0,0.4)--(0.4,0); \end{tikzpicture}\\ \hline
Leading value & 1 & $q$ & $q^2$ & $q^2$ & $q^3$ & $q^3$ &$q^3$ & $q^4$ & $q^4$& $q^5$ & $q^6$\\ \hline
Subleading value & 0 & $1$ & $2q$ & $2q$ & $3q^2$ & $3q^2$ &$3$ & $q^3+3q$ & $4q^3$& $4q^2+1$ & $6q$\\ \hline
Multiplicity & 14 & 28 & 4 & 24 & 4 & 8 & 8& 8& 2 & 4 &1\\ \hline
\end{tabular}
\caption{All the intersection graphs for the eighth moment.}\label{tab:8thMomIntersecGraphs}
\end{center}
\end{table}
We refer readers to figure \ref{fig:loopWords} for a few examples. In the intersection graph language, the leading moments \eqref{eqn:momentsFromChordDiagrams} can be written as 
\begin{equation}\label{eqn:leadingMomentsFromIntersectionGraph}
\langle \Tr H^{2p} \rangle_{\text{leading}} = d(d-1)\cdots(d-p+1) \sum_{G(p \text{-vertex})} q^{E_G},
\end{equation}
where the sum is over all the $(2p-1)!!$ intersection graphs $G$, and $E_G$ denotes the total number of edges in $G$. And from formula \eqref{eqn:subleadingRuleWords}, the subleading moments can be written as 
\begin{equation}\label{eqn:subleadingMomentsFromIntersectionGraph}
\langle \Tr H^{2p} \rangle_{\text{subleading}} = d(d-1)\cdots(d-p+2) \sum_{G(p \text{-vertex})}\sum_{e \in G} q^{E_G-(2T_e +1)},
\end{equation}
where $e$ denotes edges in $G$ and $T_e$ is the number of triangles that has $e$ as one of its sides. Notice in intersection graphs, the triangular structures in words literally become triangles. So equation \eqref{eqn:subleadingMomentsFromIntersectionGraph} is telling us to go through all the edges of the intersection graphs one by one, delete the edge we are looking at and all the triangles that has it as a side,
then count the number of edges of the remaining graph, and that is the power we raise $q$ to. Let us work out how equation \eqref{eqn:subleadingMomentsFromIntersectionGraph} for low-order moments: in tables \ref{tab:6thMomIntersecGraphs} and \ref{tab:8thMomIntersecGraphs}, all the intersection graphs contributing to the sixth and the eighth moment are respectively listed. The leading-contribution values they represent are just $q$ raised to the powers being the numbers of edges of those graphs.  The subleading-contribution values are obtained by the edge and triangle deletion procedure just described. After summing over all graphs we can check the total leading contributions are just those given by equations \eqref{eqn:sixthRT} and \eqref{eqn:eighthRT}; the subleading contributions are 
\begin{align}
\langle \Tr H^{6} \rangle_{\text{subleading}}=&d(d-1)\left[9+6q\right],\\
\langle \Tr H^{8} \rangle_{\text{subleading}}=&d(d-1)(d-2)\left[56 + 86 q + 52 q^2 + 16 q^3\right],
\end{align}
which are consistent with the results of Marinari, Parisi and Ritort\cite{Marinari:1995jwr}. For subleading contributions of higher moments, we refer readers to the same reference.

 It would be very useful to develop a Riordan-Touchard-like formula for subleading moments \eqref{eqn:subleadingMomentsFromIntersectionGraph}, but we have not found one yet.
\subsubsection{The nest scheme and the alignment scheme}
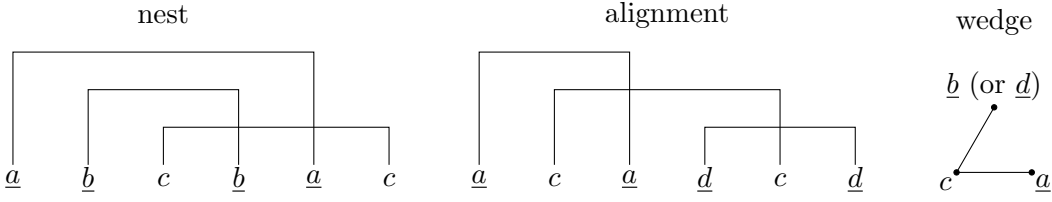
\begin{figure}
\begin{center}

\begin{tikzpicture}
\draw [] (4,0)--(4,1.5)--(8,1.5)--(8,0);
\draw [] (5,0)--(5,1)--(7,1)--(7,0);
\draw [] (6,0)--(6,0.5)--(9,0.5)--(9,0);

\node at (4,-0.2) {$\underline a$};
\node at (5,-0.2) {$\underline b$};
\node at (6,-0.2) {$c$};
\node at (7,-0.2) {$\underline b$};
\node at (8,-0.2) {$\underline a$};
\node at (9,-0.2) {$c$};

\node at  (6,2) {nest};
\end{tikzpicture}
\hspace{0.5cm}
\begin{tikzpicture}
\draw [] (4,0)--(4,1.5)--(6,1.5)--(6,0);
\draw [] (5,0)--(5,1)--(8,1)--(8,0);
\draw [] (7,0)--(7,0.5)--(9,0.5)--(9,0);

\node at (4,-0.2) {$\underline a$};
\node at (5,-0.2) {$c$};
\node at (6,-0.2) {$\underline a$};
\node at (7,-0.2) {$\underline d$};
\node at (8,-0.2) {$c$};
\node at (9,-0.2) {$\underline d$};
\node at (6.5,2) {alignment};

\end{tikzpicture}
\hspace{0.5cm}
\begin{tikzpicture}
\draw[fill=black] (9,-6.7) circle (1pt);
\draw[fill=black] (10,-6.7) circle (1pt);
\draw[fill=black] (9.5,-5.834) circle (1pt);

\draw[] (10,-6.7)--(9,-6.7)--(9.5,-5.834);
\node at (8.85,-6.85) {$c$};
\node at (10.15,-6.85) {$\underline{a}$};
\node at (9.5,-5.55) {$\underline{b}$ (or $\underline{d})$ };
\node at (9.5,-4.7) {wedge};
\end{tikzpicture}
\end{center}
\caption{The two left figures show a chord $c$ intersecting both chords in a nest structure and in an alignment structure. Note only the underlined letters represent the nest or alignment structures. The right figure shows the intersection graph of such scenarios.} \label{fig:nestAlignmentIntersect}
\end{figure}

The readers may have noticed that we can easily form two other bijections similar to equation \eqref{eqn:leadingSubleadingBijection}, namely:
 \begin{equation}\label{eqn:leadingSubleadingBijection2}
 \ldots a\ldots a \ldots a\ldots a \ldots \mapsto \ldots \underline{a}\ldots \underline b \ldots \underline b\ldots \underline a \ldots
 \end{equation}
 or
 \begin{equation}
 \label{eqn:leadingSubleadingBijection3}
 \ldots a\ldots a \ldots a\ldots a \ldots \mapsto \ldots \underline{a}\ldots \underline a \ldots \underline b\ldots \underline b \ldots.
 \end{equation}
 In some literature \cite{kim:2015a} the $abba$ structure is called a \textit{nest} and the $aabb$ structure is called an \textit{alignment}. Hence we will call the calculations based on the former the \textit{nest scheme} and the latter the \textit{alignment scheme}. By the same reasoning in the interlace scheme section, we know there are exactly the same total number of interlaces, nests and alignments when all the chord diagrams with $p$ chords are counted, which is  $\frac{1}{3}\binom {p}{2} (2p-1)!!$. We can do a hypercube Wilson loop analysis similar to that of the interlace scheme, namely the analysis wrapping around equations \eqref{eqn:interlace1stCase}-\eqref{eqn:interlace3rdCase} and see what the reduction to subleading words does to the power of $q$. The end result is the following:
for both the nest scheme and the alignment scheme when there is a third chord (in terms of chord diagrams) intersecting both chords represented by $\underline a \ldots \underline b \ldots \underline b \ldots \underline a$ (nest scheme) or $\underline a \ldots \underline a\ldots \underline b \ldots \underline b$ (alignment scheme), the power of $q$ reduces by two. In all other scenarios the power of $q$ remains the same. That is,
\begin{enumerate}
\item The nest scheme: 
\begin{equation}
\begin{split}
\ldots \underline{a}\ldots \underline b \ldots c \ldots\underline b\ldots \underline a \ldots c &\mapsto  \ldots \underline{a}\ldots \underline a \ldots c \ldots\underline a\ldots \underline a \ldots c \ldots\\
\implies q^{\#}&\mapsto   \ q^{\#-2},
\end{split}
\end{equation}
otherwise $q^{\#} \mapsto   \ q^{\#}.$
\item The alignment scheme\footnote{Here we temporarily use $\underline{adad}$ instead of $\underline{abab}$, so that upon the insertion of $c$, we still comply with the convention that what comes earlier in the alphabet comes earlier in the word. } 
\begin{equation}
\begin{split}
\ldots \underline{a}\ldots  c \ldots \underline a \ldots\underline d \ldots c \ldots \underline{d} \ldots &\mapsto  \ldots \underline{a}\ldots c \ldots\underline a  \ldots\underline a\ldots c \ldots\underline a \ldots \\
\implies q^{\#}&\mapsto   \ q^{\#-2},
\end{split}
\end{equation}
otherwise $q^{\#} \mapsto   \ q^{\#}.$
\end{enumerate}
In terms intersection graphs, we cannot distinguish a nest from an alignment because both are represented by a pair of vertices not connected by any edge. Hence in terms of intersection graphs we can at best give a prescription in terms of the sum of the nest scheme and the alignment scheme, which gives two times the subleading contribution. A chord that intersects both chords of a nest or an alignment translates to a ``wedge'' structure in intersections graphs, see figure \ref{fig:nestAlignmentIntersect}. Therefore, the prescription for the sum of the nest scheme and alignment scheme is this: \textit{for all the pairs of the vertices that are not connected by any edge in an intersection graph, delete all the ``wedges'' that connect the two vertices. The number of edges in the resulting graph is the power on $q$. The sum of all such resulting graphs from all leading intersection graphs gives two times the subleading coefficients of moments.}
\subsubsection{The averaged scheme}
The interlace scheme picks all the edges in the leading intersection graphs, whereas the nest and the alignment schemes pick all the pairs of vertices not connected by any edge. All three schemes give the same contribution, so we can average over all three schemes and get a prescription that picks all pairs of vertices in intersection graphs, regardless of whether the pairs are connected by any edge or not. It is clear we can combine all the scheme prescriptions into the following one: \textit{for every pair of vertices in a leading intersection graph, delete the edge that connects the two vertices if there is one, and delete all the wedges that has the two vertices as the two ends.\footnote{For example $\underline{a}$ and $\underline{b}$ are two ends of the wedge in the third figure in figure \ref{fig:nestAlignmentIntersect}, whereas $c$ is not an end of the wedge.} Raise $q$ to power of the number of edges of the resulting graph and sum them over all such graphs. The subleading coefficient is one third of this sum.} 

\begin{figure}
\begin{center}
\begin{tikzpicture}
\draw[fill=black] (1,0) circle (1pt);
\draw[fill=black] (2,0) circle (1pt);
\draw[fill=black] (1.5,0.866) circle (1pt);
\draw[] (1.5,0.866)  -- (1,0) -- (2,0);

\node at (1,-0.2) {$a$};
\node at (2,-0.2) {$b$};
\node at (1.5,1.066) {$c$};

\node at (2.5,0.43) {$\rightarrow$};

\draw[fill=black] (3,0) circle (1pt);
\draw[fill=black] (3.5,0.866) circle (1pt);
\draw[] (1.5,0.866)  -- (1,0) -- (2,0);

\node at (3.2,0) {$a$};
\node at (3.5,1.066) {$c$};
\draw[] (3.5,0.866)  -- (3,0);
\draw [] plot [smooth cycle] coordinates {(3,0) (3.2,-0.4)(2.8,-0.4)};

\node at (4.5,0.43) {$\rightarrow$};

\draw[fill=black] (5,0) circle (1pt);
\draw[fill=black] (5.5,0.866) circle (1pt);
\draw[] (5.5,0.866)  -- (5,0);

\node at (5,-0.2) {$a$};
\node at (5.5,1.066) {$c$};

\node at (-1,0.43) {merge $ab:$};
\end{tikzpicture}
\end{center}
\begin{center}
\begin{tikzpicture}
\draw[fill=black] (1,0) circle (1pt);
\draw[fill=black] (2,0) circle (1pt);
\draw[fill=black] (1.5,0.866) circle (1pt);
\draw[] (1.5,0.866)  -- (1,0) -- (2,0);

\node at (1,-0.2) {$a$};
\node at (2,-0.2) {$b$};
\node at (1.5,1.066) {$c$};

\node at (2.5,0.43) {$\rightarrow$};

\draw[fill=black] (3,0) circle (1pt);
\draw[fill=black] (3.5,0.866) circle (1pt);

\node at (3,-0.2) {$a$};
\node at (3.5,1.066) {$c$};

\draw [] plot [smooth cycle] coordinates {(3,0) (3.3,0.4)(3.5,0.866)(3.1,0.5)};

\node at (4.5,0.43) {$\rightarrow$};

\draw[fill=black] (5,0) circle (1pt);
\draw[fill=black] (5.5,0.866) circle (1pt);
\node at (5,-0.2) {$a$};
\node at (5.5,1.066) {$c$};

\node at (-1,0.43) {merge $bc:$};
\end{tikzpicture}
\end{center}

\begin{center}
\begin{tikzpicture}
\draw[fill=black] (1,0) circle (1pt);
\draw[fill=black] (2,0) circle (1pt);
\draw[fill=black] (1.5,0.866) circle (1pt);
\draw[] (1.5,0.866)  -- (1,0) -- (2,0)--(1.5,0.866);

\node at (1,-0.2) {$a$};
\node at (2,-0.2) {$b$};
\node at (1.5,1.066) {$c$};

\node at (2.5,0.43) {$\rightarrow$};

\draw[fill=black] (3,0) circle (1pt);
\draw[fill=black] (3.5,0.866) circle (1pt);

\node at (3,-0.2) {$a$};
\node at (3.5,1.066) {$c$};

\draw [] plot [smooth cycle] coordinates {(3,0) (3.3,0.4)(3.5,0.866)(3.1,0.5)};
\draw [] plot [smooth cycle] coordinates {(3.5,0.866)(3.8,0.9)(3.8,0.732)};

\node at (4.5,0.43) {$\rightarrow$};

\draw[fill=black] (5,0) circle (1pt);
\draw[fill=black] (5.5,0.866) circle (1pt);
\node at (5.2,0) {$a$};
\node at (5.5,1.066) {$c$};

\node at (-1,0.43) {merge $bc:$};
\end{tikzpicture}
\end{center}
\caption{Three examples of the ``merge and delete'' prescription. In the first figure, a loop is formed after merging and then deleted, in the end $q^2$ is reduced to $q$; in the second figure, a double edge is formed after merging and then deleted, and $q^2$ is reduced to $1$; in the third figure, a loop and a double edge are formed and deleted, as a result $q^3$ is reduced to $1$.}\label{fig:mergeAndDelete}
\end{figure}
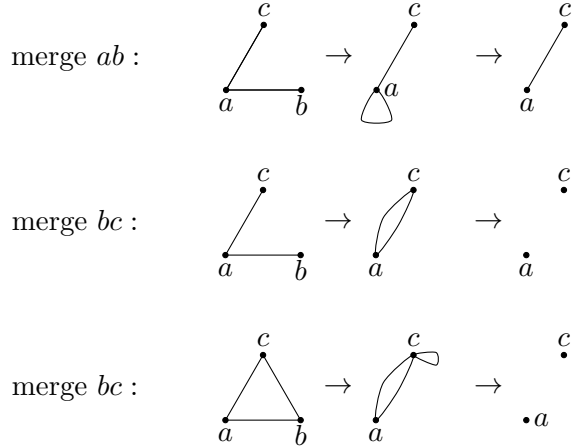

There is a more graphical way to describe the edge and wedge deletion prescription. That is, we take a pair of vertices and merge them into one vertex, and all the edges before merging are inherited. However, loops (edge that connects a vertex to itself) and double edges (two edges connecting the same two vertices) may appear after merging,  and we delete all the loops and double edges to form a subleading intersection graph. Note that deleting a loop is equivalent to deleting the edge connecting a chosen pair in the language of the last paragraph, and deleting a double edge is equivalent to deleting the wedge. Figure \ref{fig:mergeAndDelete} demonstrates a few examples of such ``merge and delete'' process. We can summarize the averaged scheme  into one formula:
\begin{equation}\label{eqn:subleadingMomentsAveScheme}
\frac{\langle \Tr H^{2p} \rangle_{\text{subleading}}}{d(d-1)\cdots(d-p+2)} = \frac{1}{3}\sum_{G}\sum_{\{v_1,v_2\} \subset v(G)} q^{E\left[G_{(v_1,v_2)}\right]},
\end{equation}
where $G$'s are all the intersection graphs formed by all the chord diagrams with $p$ chords; $v_1, v_2$ are any two vertices of $G$ and $v(G)$ denotes the vertex set of $G$; $G_{(v_1,v_2)}$ is the graph formed by the ``merge and delete'' procedure applied to $G$ with respect to $v_1$ and $v_2$ (namely, $v_1$ and $v_2$ are merged), and finally $E\left[G_{(v_1,v_2)}\right]$ is the number of edges in $G_{(v_1,v_2)}$.

Quite remarkably, the ``merge and delete'' prescription exactly coincides with the prescription to calculate the subleading moments of the sparse SYK model \cite{garcagarca2020sparse, xu2020sparse}, except that in sparse SYK model we do not have to divide by three. In the sparse SYK model, the values associated with the intersection graphs are $q^{E(G)}$ if a Q-Hermite approximation is applied. Hence, the coefficients of the sparse SYK subleading moments (after Q-Hermite approximation) are three times those of the Parisi subleading moments. We have already seen in this paper that at leading order the SYK moments after Q-Hermite approximation coincide with the Parisi moments, and at leading order the sparse SYK moments are the same as the SYK moments by construction, and so they coincide with Parisi moments as well. With the ``merge and delete '' prescription, we see that even their subleading moments are related.  So we arrive at 
\begin{equation}
\text{(sparse SYK moments)}_\text{QH} = \text{Parisi leading} + \frac{1}{k N}\times 3 \times \text{Parisi subleading} +O(1/N^2), 
\end{equation}
where the subscript ``QH'' denotes Q-Hermite approximation,  $N$ is the number of Majorana fermions in the sparse SYK model and $k$ indicates sparseness (smaller $k$ means more sparseness). We note however this relation does not hold to higher orders.

 \bibliographystyle{unsrt}

\end{document}